\documentclass[english, 12pt, a4paper]{article}
\usepackage{lmodern}
\usepackage[utf8]{inputenc}
\usepackage[T1]{fontenc}
\usepackage[english]{babel}
\usepackage[top=1in, bottom=1in, left=1in, right=1in]{geometry}
\usepackage[pdftex]{graphicx}
\usepackage{verbatim}
\usepackage[width=0.8\textwidth]{caption}
\usepackage{amsmath,amssymb,amsopn,dsfont,mathrsfs,bm,mathtools}
\usepackage{float,placeins,flafter,longtable,array,booktabs}
\usepackage{enumitem}
\usepackage{color}
\usepackage{natbib}
\usepackage{xr}
\usepackage[hidelinks]{hyperref}
\usepackage{subcaption}
\usepackage{tikz}
\usetikzlibrary{calc}
\graphicspath{ {./HistogramsAER/} }

\makeatletter

\newtheorem{theorem}{Theorem}

\newtheorem{lemma}{Lemma}
\newtheorem{hyp}{Assumption}

\newtheorem{example}{Example}

\newcommand{\norm}[1]{\left\|#1\right\|}

\newcommand{\set}[1]{\left\{#1\right\}}
\newcommand{\ind}[1]{\mathds{1}\left\{#1\right\}}
\newcommand{\Id}{\mathrm{I}}
\newcommand{\Prob}{P}
\renewcommand{\P}{\mathcal{P}}

\newcommand{\E}{E}
\newcommand{\RP}{\mathcal{Q}}
\newcommand{\V}{V}
\newcommand{\Cov}{\text{Cov}}
\newcommand{\W}{\Upsilon}
\newcommand{\Wh}{\Upsilon_n}
\newcommand{\Cinf}{\underline{C}}
\newcommand{\se}{\text{se}}

\newcommand{\K}{\mathcal K}
\newcommand{\R}{\mathbb R}
\newcommand{\N}{\mathbb N}

\newcommand{\eps}{\varepsilon}
\renewcommand{\k}{\bm{k}}

\newcommand{\zero}{\bm{0}}

\newcommand{\deriv}[2]{\partial #1/\partial #2}
\newcommand{\Deriv}[2]{\frac{\partial #1}{\partial #2}}
\newcommand{\tr}{\text{tr}}

\newcommand{\indep}{\perp \!\!\! \perp}
\newcommand{\convP}{\stackrel{p}{\longrightarrow}}
\newcommand{\convD}{\stackrel{d}{\longrightarrow}}
\newcommand{\convNor}[1]{\stackrel{d}{\longrightarrow} \mathcal{N}\left(0,#1\right)}
\newcommand{\convAS}{\underset{n\to +\infty}{\overset{a.s.}{\longrightarrow}}}
\newcommand{\convLtwo}{\underset{}{\overset{\ell^2}{\longrightarrow}}}

\newcommand{\CR}[1]{\text{CR}^{\,#1}_{1-\alpha}}

\newcommand{\Rb}{\mathbb{R}}

\renewcommand{\section}{\@startsection{section}{2}{0mm}{-1\baselineskip}{1\baselineskip}{\normalfont\large\bfseries}}
\renewcommand{\subsection}{\@startsection{subsection}{2}{0mm}{-1\baselineskip}{1\baselineskip}{\normalfont\normalsize\bfseries}}
\renewcommand{\subsubsection}{\@startsection{subsubsection}{3}{0mm}{-0.8\baselineskip}{0.4\baselineskip}{\normalfont\normalsize\itshape}}
\allowdisplaybreaks

\newcommand*{\storecounter}[2]{%
  \edef\@currentlabel{\the\value{#1}}
  \label{#2}
}

\date{}

\begin{document}

	\title{Analytic inference with two-way clustering\thanks{We gratefully acknowledge financial support from the research grant Ricode (ANR-23-CE26-0009). We would like to thank Yanis Zhegal for superb research assistance. We are also thankful to Aureo de Paula, Davide Viviano and seminar and workshop participants at Aarhus, 
Bologna, Bristol, Harvard-MIT, Leuven, Manchester, TSE and CREST for their helpful comments. We gratefully acknowledge financial support from the research grants Ricode (ANR-AAP CE26).}}
	\author{Laurent Davezies\thanks{CREST-ENSAE, laurent.davezies@ensae.fr} \and Xavier D'Haultf\oe uille\thanks{CREST-ENSAE, xavier.dhaultfoeuille@ensae.fr.} \and Yannick Guyonvarch\thanks{INRAE, yannick.guyonvarch@inrae.fr.}}
	
	\maketitle

\begin{abstract}
This paper studies analytic inference along two dimensions of clustering. In such setups, the commonly used approach has two drawbacks. First, the corresponding variance estimator is not necessarily positive. Second, inference is invalid in non-Gaussian regimes, namely when the estimator of the parameter of interest is not asymptotically Gaussian. We consider a simple fix that addresses both issues. In Gaussian regimes, the corresponding tests are asymptotically exact and equivalent to usual ones; otherwise, they are asymptotically conservative. We also establish uniform validity under suitable restrictions. Finally, we compare our approach with existing ones through simulations. 

			
	\medskip
	\textbf{Keywords:} two-way clustering, inference, GMM. \\
	\noindent\textbf{JEL Codes:} C1, C12.
\end{abstract}
		
\newpage

\section{Introduction} 
	\label{sec:introduction}
	
Applied researchers are often reluctant to assume independence between units, because these units may be affected by common shocks. Moreover, these shocks may be of different nature. For instance, the wages of two individuals could be correlated either because these individuals belong to the same industry or because they live in the same area. This case is referred to as two-way clustering because clustering occurs along two dimensions, industry and geographical area in this example. To account for such possible dependence, researchers routinely apply the variance estimator of \citeauthor{miglioretti2007marginal} (\citeyear{ChenRao2007}, MH hereafter), \citeauthor{cameron2011} (2011, CGM hereafter) and \cite{thompson2011}, denoted by $\widehat{V}_u$ below.
	
\medskip
However, these ``usual'' variance estimators have a major drawback, namely, they may not be positive, neither in finite samples nor even asymptotically. Similarly, variance matrices may not be semidefinite positive. We find evidence of this in applied work (see Appendix \ref{sec:aer_meta} for details). CGM propose a fix to this issue, but the corresponding inference in linear regressions is not invariant to affine transforms of the variables anymore. In simulations, we find that the level of a test can vary from zero to one as we change the location and scale of a regressor. 

\medskip
Theoretically speaking, \cite{menzel2021bootstrap} showed that in a particular data generating process (DGP), standard variance estimators are negative with probability almost 40\% asymptotically, which leads to severely distorted inference, even if we use CGM's fix. Beyond this particular example, failure of usual inference occurs whenever estimators are not asymptotically Gaussian, a situation that arises naturally with multiway clustering, as illustrated in \cite{menzel2021bootstrap} and discussed multiple times below.

\medskip
The aim of this paper is to suggest an elementary fix for inference, which eliminates the issue of negative variances, remains invariant to affine transforms of the covariates in linear regressions and is asymptotically valid.\footnote{We develop the Stata package \texttt{twc\_inf}, available on SSC, which implements this method for linear, probit, logit and poisson regressions.} Consider a univariate equality test. Let $\widehat{V}_1$ and $\widehat{V}_2$ be the variance estimators obtained assuming that only one of the two dimensions of clustering matters, and let $\se_1$ and $\se_2$ be the associated standard errors. Then, we suggest to use as a standard error the maximum between $\se_1$, $\se_2$ and $\se_u$, where the latter is the standard error associated to $\widehat{V}_u$, with the understanding that $\se_u=0$ if $\widehat{V}_u$ is negative. This modification has also been proposed by \cite{mackinnon2024jackknife}, though they do not establish its validity in cases where the usual method fails. We suggest a similar construction for joint hypothesis testing, which is new to our best knowledge.

\medskip
We first focus on equality tests for univariate expectations and establish the asymptotic validity of our procedure both in a pointwise and uniform sense. To do so, we model the data as a dissociated, separately exchangeable array, following in particular \cite{ddg2021} and \cite{menzel2021bootstrap}. Then, we rely on results for such arrays, in particular the so-called Aldous-Hoover-Kallenberg representation \citep[see][]{Aldous1981,Hoover1979,kallenberg1989}. Our main insight is that even if the sample mean may not be asymptotically Gaussian and $\widehat{V}_1$ or $\widehat{V}_2$ may remain random asymptotically (once properly normalized), the distribution of our $t$-statistic is asymptotically more concentrated than a standard Gaussian distribution. As a result, the pointwise validity of our univariate tests holds under no further restriction on the DGP. Moreover, we show that our test is equivalent to the usual test whenever the usual $t$-statistic is asymptotically standard Gaussian. Hence, our method does not lead to any power loss asymptotically in cases where usual inference is justified.
	
\medskip
The results on univariate means extend to functionals of GMM estimators, as long as a condition 
that has been overlooked so far holds. Specifically, if linear combinations of the empirical moments under consideration, evaluated at the true parameter, do not all converge at the same rate, the GMM estimator may not be close to the average of the ``standard'' influence functions, namely the influence functions we use for i.i.d. data. We illustrate this with a simple linear regression example. This issue has consequences for any analytic inference method, ours and the usual one included.\footnote{In our simple linear regression example, usual inference is highly distorted while our method is not, though our theoretical results do not cover this case.} On the other hand, if all linear combinations of the empirical moments converge at the same rate, this peculiar phenomemon disappears and we show the validity of our testing procedure.

\medskip
We also consider equality tests for multivariate expectations. Unlike in the i.i.d. setup, this extension is not straightforward, however, since the properly normalized matrices $\widehat{V}_1$ and $\widehat{V}_2$ may converge to random and singular matrices, an issue that also affects standard inference and has not been identified yet, to the best of our knowledge. To handle these challenges, we impose sufficient and (partly) necessary conditions on the DGP and resort to a new result on Gaussian matrices that is of independent interest (see Lemma \ref{lem:invertible_QF} in Appendix \ref{sec:multiv_extension}).
	
\medskip
Finally, we compare in simulations our method with the usual one and the bootstrap method of \cite{menzel2021bootstrap}. We show in particular that usual inference can be very distorted, while ours seems to perform well even in cases not covered by our theory. 	

\paragraph{Related literature.}
	
First and foremost, our paper contributes to the literature on analytic inference under multiway clustering. As mentioned above, the variance estimator $\widehat{V}_u$ was proposed by \cite{miglioretti2007marginal}, \cite{cameron2011} and \cite{thompson2011}. These papers do not show the validity of the corresponding inference. The latter is established by \cite{menzel2021bootstrap} for sample means of univariate variables if such means are asymptotically Gaussian. \cite{menzel2021bootstrap} also shows that if sample means are not asymptotically Gaussian, inference based on $\widehat{V}_u$ may not be valid. \cite{chiang2023using} extend Menzel's results by showing asymptotic Gaussianity for specific drifting sequences of DGPs. Another extension of Menzel's result, to large $T$ panel data where temporal shocks can be dependent both over time and across individuals, is considered by \cite{Chiang2024}. \cite{Yap2025_1}	 shows the validity of usual inference under the same independence structure as here, but without exchangeability. Compared to these papers, we show that a simple modification of inference based on $\widehat{V}_u$ solely also works in non-Gaussian cases, while being equivalent to it in Gaussian cases. To our knowledge, this is the first analytic inference method for which validity is established in non-Gaussian cases. 

\medskip
Several papers also consider resampling-based inference, and here we just mention a few of them. \cite{ddg2021} show the validity of the so-called pigeonhole bootstrap, and a multiplier bootstrap, for ``non-degenerate'' DGPs, for which the estimator under consideration converges at a slow rate. \cite{mackinnon2021wild} show the validity of a certain wild bootstrap method in some Gaussian regimes. \cite{menzel2021bootstrap} develop other wild bootstrap schemes and show that one of them is pointwise valid both in Gaussian and non-Gaussian regimes, while another one controls size over a large set of DGPs but is possibly conservative \citep[see also][for an extension to panel data]{juodis2021}. Our paper complements Menzel's by showing that to some extent, adaptivity is also possible with analytic inference in this set-up. Our approach also has the advantage of being computationally very cheap and not requiring any tuning parameter. 
	
\paragraph{Organization of the paper.} 
\label{par:organization_of_the_paper}
	
Section \ref{sec:scalar_expecs} introduces the setup and the tests on univariate expectations we propose, and presents our pointwise and uniform results for these tests. Section \ref{sec:GMM} extends our approach to the GMM case. Section \ref{sec:monte_carlo_simulations} analyses differences between our method and others, in particular $\widehat{V}_u$, in simulations. Section \ref{sec:conclusion} concludes. The appendix gathers some extensions, such as equality tests for multivariate expectations, and most of the proofs. The remaining proofs and supporting lemmas can be found in the Online Appendix.
	
	
	
\section{Inference on scalar expectations}	
\label{sec:scalar_expecs}

\subsection{Set-up and inference method} 
\label{sec:set_up}
	
We have access to the observed random variables $(Y_{ij})_{1\le i\le C_1,1\le j\le C_2}$. The two indices $i$ and $j$ correspond to the two dimensions of ``clustering'', with a dependence structure that will be clarified below. For instance, these dimensions may correspond to industries and geographical areas.
We would like to make inference on a scalar parameter $\theta_0 := E[\overline{Y}]$, where, for any array of random variables $(D_{ij})_{1\leq i \leq C_1, 1 \leq j \leq C_2}$, we let $\overline{D}:=(C_1C_2)^{-1}\sum_{i=1}^{C_1}\sum_{j=1}^{C_2} D_{ij}$. Note that for simplicity, we assume in this section to observe a single random variable in each ``cell'' $(i,j)$. If, instead, we observe $(Y_{ij\ell})_{\ell=1,...,N_{ij}}$ in cell $(i,j)$ and if $\theta_0=\E[(C_1C_2)^{-1}\sum_{i,j}\sum_{\ell=1}^{N_{ij}}Y_{ij\ell}]$, we can always define $Z_{ij}$ as $\sum_{\ell=1}^{N_{ij}} Y_{ij\ell}$ and $\theta_0$ as $\E[\overline{Z}]$. For other parameters of interest (e.g., $\theta_0=\E[\overline{Z}]/\E[\overline{N}]$, the estimand corresponding to $\sum_{i,j}\sum_{\ell}^{N_{ij}}Y_{ij\ell}/\sum_{i,j} N_{ij}$), assuming a single observation per cell is not without loss of generality, and we do allow for this case in our GMM setup in Section \ref{sec:GMM}.  

\medskip
Hereafter, we mostly consider tests of nominal level $\alpha\in (0,1)$ of the null hypothesis that $\theta_0 = \theta$, against $\theta_0\ne \theta$; we also briefly discuss unilateral tests, as well as confidence intervals. As those proposed by MH and CGM, our tests rely on the following three variance estimators:
\begin{align}
		\widehat{V}_1 & := \frac{1}{C_1^2}\sum_{i=1}^{C_1}\left(\frac{1}{C_2}\sum_{j=1}^{C_2}Y_{ij} - \overline{Y} \right)^2, \notag \\
		\widehat{V}_2 & := \frac{1}{C_2^2}\sum_{j=1}^{C_2}\left(\frac{1}{C_1}\sum_{i=1}^{C_1} Y_{ij} - \overline{Y} \right)^2, \label{eq:def_Vhat} \\
		\widehat{V}_{12} & := \frac{1}{(C_1C_2)^2}\sum_{i=1}^{C_1}\sum_{j=1}^{C_2} \left(Y_{ij} - \overline{Y}\right)^2. \notag 	
\end{align}

For simplicity, and since they do not matter asymptotically, we do not consider the degrees-of-freedom corrections suggested by CGM. 

\medskip
We first present the test proposed by MH and CGM. Let $\widehat{V}_u:=\widehat{V}_1+\widehat{V}_2-\widehat{V}_{12}$, where the index ``u'' refers to ``usual''. Then, MH and CGM consider the test $\phi_{u,\alpha} := \ind{|t_u|>z_{1-\alpha/2}}$, where $z_{1-\alpha/2}$ is the quantile of order $1-\alpha/2$ of a standard normal distribution and
$$t_u := \frac{\overline{Y}-\theta}{\widehat{V}_u^{1/2}}.$$
This approach has one major drawback, namely, $\widehat{V}_u$  can be negative, in which case the test above is not defined. This may happen in finite samples even in DGPs for which this test is asymptotically valid and thus $P(\widehat{V}_u>0)$ tends to one. There are also cases for which the latter condition does not hold and the test is asymptotically invalid. Suppose for instance that $Y_{ij}=\theta_0+U_{i0}U_{0j}$ with $(U_{i0})_{i\ge 1}$ and $(U_{0j})_{j\ge 1}$ two independent sequences of centered, i.i.d. variables. \cite{menzel2021bootstrap} shows that in this example, $P(\widehat{V}_u<0)\to 39.3\%$. This multiplicative structure is a special case of a factor model (in fact, similar results would hold if we replaced $U_{i0}U_{0j}$ by $\sum_{k=1}^K U_{i0,k}U_{0j,k}$) and thus seems relevant in economic contexts.

\medskip
To solve these issues, we propose the following modification, which was suggested before by \cite{mackinnon2024jackknife}. Let $\se_k:=\widehat{V}_k^{1/2}$ for $k\in\{1,2\}$, $\se_u:=\max(0,\widehat{V}_u)^{1/2}$  and let $\se:=\max(\se_1,\se_2,\se_u)$. Then, consider the test 
\begin{equation}\phi_{\alpha} := \ind{|t|>z_{1-\alpha/2}},\text{ where }
t := \frac{\overline{Y}-\theta}{\se},\label{eq:def_test}
\end{equation}
with the convention that $\phi_{\alpha}=1$ when $\se=0$. Remark that we simply replace the usual standard error $\se_u$ with the maximum of $\se_u$, $ \se_1$ and $\se_2$. An intuition behind this test is that the second (resp., the first) dimension of clustering may not matter. In such a case, it would be more natural to consider $\se_1$ (resp. $\se_2$) rather than $\se_u$. We then take a conservative approach by picking the maximum of these three standard errors. It turns out, however, that when $\overline{Y}$ is asymptotically Gaussian, our test is not asymptotically conservative. Our test is (potentially) conservative only in non-Gaussian cases, for which  $\phi_{u,\alpha}$ may be asymptotically invalid. In the rest of paper, we do not discuss unilateral tests nor confidence intervals on $\theta_0$ (given by $[\overline{Y} \pm \, z_{1-\alpha/2} \, \se]$), but our results on bilateral tests below directly extend to them.

\paragraph{Multivariate extension.} Suppose that $Y_{ij}\in\R^d$, and we want to test $E[\overline{Y}]=\theta$. In Appendix \ref{sec:multiv_extension}, we consider a test in a same spirit as above. Specifically, we basically take the minimum of three $F$-tests, with an adjustment in cases some of the involved variance matrices are singular. To our knowledge, this test is new.


\subsection{Assumptions}
	\label{ssub:a_result_for_sequences_of_dgps}
	
We obtain our results below under two conditions, which put restrictions on the data generating process and the asymptotic framework. The first assumption clarifies the dependence structure underlying two-way clustering and imposes minimal moment conditions:	
		
\begin{hyp}\label{as:DGP}
We observe a sample $(Y_{ij})_{1\le i\le C_1,1\le j\le C_2}$, which is extracted from the dissociated and separately exchangeable array $\bm{Y}:=(Y_{ij})_{(i,j)\in \N^{*2}}$. Namely, $\bm{Y}$ satisfies:
	\begin{enumerate}
			\item (dissociation) for any $(E_1,F_1,E_2,F_2) \subset \N^{*4}$ such that $E_1\cap E_2=F_1\cap F_2=\emptyset$, $(Y_{ij})_{(i,j)\in E_1\times F_1}$ is independent of $(Y_{ij})_{(i,j)\in E_2\times F_2}$.
			\item (separate exchangeability) for any pair of permutations $(\pi_1,\pi_2)$ on $\mathbb{N}^*$,
			$$(Y_{ij})_{(i,j)\in \N^{*2}}\overset{d}{=}(Y_{\pi_1(i)\pi_2(j)})_{(i,j)\in \N^{*2}}.$$
	\end{enumerate}
	Note that the distribution of $\bm{Y}$ may depend on $(C_1,C_2)$ and satisfies for every $(C_1,C_2)$ $E[\|Y_{11}\|^2]<\infty$ and $\lambda_{\min}\left(V(Y_{11})\right)>0$.\footnote{Here $||\cdot||$ denotes the Euclidean norm and $\lambda_{\min}(A)$ is the smallest eigenvalue of a square matrix $A$. We introduce multivariate notation here to avoid duplicating this assumption when presenting multivariate extensions and proofs in the Appendix.}
\end{hyp}

The first condition implies that two subsets of $\bm{Y}$ sharing no common cluster are independent. On the other hand, this condition does not impose any restriction on the dependence between $Y_{ij}$ and $Y_{ij'}$ or between $Y_{ij}$ and $Y_{i'j}$. The second condition states that the labels $i$ and $j$ do not carry any information: replacing them by any other labelling (through permutations) leads to the same distribution of the array. This implies in particular that the variables $(Y_{ij})_{i,j \ge 1}$  are identically distributed. Imposing finite second moments and non-zero variance allows us to rule out pathological situations without any variation in the observed data and is required to derive the asymptotic properties of our tests. Finally, allowing the distribution of $\bm{Y}$ to depend on $(C_1,C_2)$ is essential when studying the (asymptotic) uniform validity of our inference method, as in this case we must study the asymptotic behavior of our test under sequences of DGPs rather than a fixed DGP. 

\medskip
Our second assumption pertains to the asymptotic framework. We suppose hereafter that both $C_1$ and $C_2$ tend to infinity, but without restricting their respective rates of convergence.
	
\begin{hyp}
	\label{as:asym}
There exists $n\in \N$ and increasing functions $g_1$ and $g_2$ from $\N$ to $\N$ such that $C_k=g_k(n) \rightarrow\infty$ as $n \rightarrow\infty$ for $k=1,2$.
\end{hyp}
	

\subsection{A useful decomposition} 
\label{sub:a_useful_decomposition}

Before showing our results, we present a useful decomposition. We assume that Assumption~\ref{as:DGP} holds and introduce the variable $W_{ij} := Y_{ij}-\theta_0$ and vector $\bm{W} := (Y_{ij}-\theta_0)_{(i,j)\in \N^{*2}}$. First remark that as a dissociated and separately exchangeable array, $\bm{W}$ satisfies a Aldous-Hoover-Kallenberg (AHK for short) representation, see \cite{Aldous1981}, \cite{Hoover1979} and \cite{kallenberg1989}. Namely, there exist i.i.d. continuously distributed random variables $(U_{i0},U_{0j},U_{ij})_{i,j \ge 1}$ and a function $\tau$ such that almost surely,
	\begin{equation}
		W_{ij}=\tau(U_{i0},U_{0j},U_{ij}).
		\label{eq:AHK}
	\end{equation}
We can assume without loss of generality (wlog) that $U_{i0}$, $U_{0j}$ and $U_{ij}$ are centered and admit second-order moments. The variables $U_{i0}$ and $U_{0j}$ may be seen as row and column shocks, respectively, while $U_{ij}$ can be interpreted as a ``cell''-specific shock. Then, we consider a similar decomposition as that in~\cite{menzel2021bootstrap}. Specifically, let us define
\begin{align*}
	\alpha_i & :=E[W_{ij}|U_{i0}] , \\
	\beta_j & :=E[W_{ij}|U_{0j}] , \\
	\gamma_{ij} & :=E[W_{ij}|U_{i0},U_{0j}]  - \alpha_i -\beta_j, \\
	\eps_{ij} & :=W_{ij}  - \alpha_i - \beta_j - \gamma_{ij}.	
\end{align*}
Observe that by construction,
\begin{equation}
W_{ij} = \alpha_i + \beta_j + \gamma_{ij} + \eps_{ij}.	
	\label{eq:decomp}
\end{equation}
Finally, we define $\Omega_1:=V(\alpha_1)$, $\Omega_2:=V(\beta_1)$, $\Omega_3:=V(\gamma_{11})$ and $\Omega_4:=V(\eps_{11})$. Because the AHK decomposition is not unique, it may seem that $(\alpha_i,\beta_j,\gamma_{ij}, \eps_{ij})_{i,j \ge 1}$ and the $(\Omega_k)_{k=1,...,4}$ depend on the choice of the variables $(U_{i0},U_{0j},U_{ij})_{i,j \ge 1}$. The following lemma shows that this is not the case. Let $\mathcal{S}_{1,n}:=\sigma(W_{ij}:j>n, i\geq 1)$, $\mathcal{S}_{2,n}:=\sigma(W_{ij}:i>n, j\geq 1)$ and $\mathcal{S}_{12,n}:=\sigma(W_{ij}:\max(i,j)>n)$ and $\mathcal{S}_1:=\bigcap_{n\geq 1}\mathcal{S}_{1,n}$, $\mathcal{S}_2:=\bigcap_{n\geq 1}\mathcal{S}_{2,n}$ and $\mathcal{S}_{12}:=\bigcap_{n\geq 1}\mathcal{S}_{12,n}$.

\begin{lemma}\label{lem:decomp}
	We have:
\begin{align*}
	\alpha_i & =E[W_{ij}|\mathcal{S}_1] , \\
	\beta_j & =E[W_{ij}|\mathcal{S}_2] , \\
	\gamma_{ij} & =E[W_{ij}|\mathcal{S}_{12}]  - \alpha_i -\beta_j, \\
	\eps_{ij} & =W_{ij}  - \alpha_i - \beta_j - \gamma_{ij}.	
\end{align*}
Moreover, $\Omega_1=\Cov(W_{11},W_{12})$, $\Omega_2=\Cov(W_{11},W_{21})$ and $\Omega_1+\Omega_2+\Omega_3+\Omega_4=V(W_{11})$.
\end{lemma}

To our knowledge, there is no simple expression for $\Omega_3$, though we can still express it as a function of $\bm{W}$ only through the following equality
$$\Omega_3 =  V\left\{E[W_{11}|\mathcal{S}_{12}] - E[W_{11}|\mathcal{S}_1] - E[W_{11}|\mathcal{S}_2]\right\}.$$

	\begin{example}
		Consider the DGP $Y_{ij}=\theta_0 + U_{i0} + U_{0j} + U_{i0}U_{0j} + U_{ij}$, where the $(U_{ij})_{i,j\ge 0}$ are as in \eqref{eq:AHK}. Then $\alpha_i=U_{i0}$, $\beta_j=U_{0j}$, $\gamma_{ij}=U_{i0}U_{0j}$ and $\eps_{ij}=U_{ij}$.
		\label{ex:ex1}
	\end{example}


\subsection{Pointwise results} 
\label{sub:main_results}

We now state and discuss validity results for our test when the probability distribution of $\bm{Y}$ does not vary with $n$.

\begin{theorem}
\label{thm:point_univ}
Suppose that $Y_{ij}\in\R$, $P_{\bm{Y}}$ does not depend on $n$, Assumptions \ref{as:DGP}-\ref{as:asym} hold, $\phi_\alpha$ be defined according to \eqref{eq:def_test} and $\theta=\theta_0$. Then, for every $\alpha\in(0,1)$,
\begin{equation}
\limsup_{n \to \infty} \E[\phi_\alpha]\le \alpha. \label{eq:pointwise_valid}
\end{equation}
Moreover, if either $\Omega_1+\Omega_2>0$ or $\Omega_3=0$,
\begin{equation}
	\lim_{n \to \infty} E[\phi_\alpha] = \alpha.
	\label{eq:pointwise_exact}
\end{equation}
\end{theorem}

Even if Theorem \ref{thm:point_univ} follows from our uniform result below (Theorem \ref{thm:unif_univ}), let us give some intuition on its proof. Assume first that $\Omega_1+\Omega_2>0$. In that case, we show that $\overline{Y}-\theta_0=O_p\left((\Omega_1/C_1+\Omega_2/C_2)^{1/2}\right)$ and
$$\frac{\overline{Y}-\theta_0}{\se_u}=\left[\frac{\overline{\alpha}+\overline{\beta}}{(\Omega_1/C_1 + \Omega_2/C_2)^{1/2}}+\frac{\overline{\gamma}+\overline{\eps}}{(\Omega_1/C_1 + \Omega_2/C_2)^{1/2}}\right] + o_P(1).$$
By the central limit theorem, the first fraction on the right-hand side converges to a standard normal distribution. Also, observing that $\Cov(\gamma_{ij},\gamma_{i'j'})=V(\gamma_{11})$ $\times \ind{i=i',j=j'}$ and $\Cov(\eps_{ij},\eps_{i'j'})=V(\eps_{11})\ind{i=i',j=j'}$, we prove that $\overline{\gamma}+\overline{\eps}=O_p((C_1C_2)^{-1/2})$. As a result,
$$\frac{\overline{Y}-\theta_0}{\se_u} \convD \mathcal{N}(0,1).$$
This proves the asymptotic validity of usual inference, as well as asymptotic normality of $\overline{Y}$, if $\Omega_1+\Omega_2>0$. Moreover, we show that $\se/\se_u\convP 1$ (see Eq. \eqref{eq:equiv_usual} in the appendix),
which implies that our test is also asymptotically valid in this case, and in fact equivalent to the usual test.

\medskip
Next, assume that $\Omega_1+\Omega_2=0$. Then, $\overline{\alpha}=\overline{\beta}=0$. Since we still have $\overline{\gamma}+\overline{\eps}=O_p((C_1C_2)^{-1/2})$, we obtain
\begin{equation}
\overline{Y}-\theta_0 =\overline{\gamma}+\overline{\eps}  + o_P\left((C_1C_2)^{1/2}\right).	
	\label{eq:approx_nong}
\end{equation}
Equation \eqref{eq:approx_nong} implies the estimator converges at a faster rate when $\Omega_1+\Omega_1=0$. If $\Omega_3=0$, then $\overline{\gamma}=0$ and $(C_1C_2 /\Omega_4)^{1/2} \overline{\eps}\convNor{1}$ ensuring that $\overline{Y}$ is again asymptotically normal. Moreover, we establish  that $(C_1C_2/\Omega_4)^{-1} \se \convP 1$ and $\se_u/\se\convP 1$. Thus, in this case again, the usual test and ours are equivalent. Moreover, they are both asymptotically valid and non-conservative.

\medskip
Finally, if $\Omega_3>0$, two complications occur. First, $\overline{\gamma}$ is not asymptotically normal and second, the standard errors remain random asymptotically. The key point we establish is that conditional on $(U_{0j})_{j\ge 1}$ (say), we have
\begin{equation}
\frac{\overline{\gamma}+\overline{\eps}}{\se_1} \convD \mathcal{N}(0,1).	
	\label{eq:conv_nong}
\end{equation}
Since the limit (Gaussian) distribution in \eqref{eq:conv_nong} does not depend on the $(U_{0j})_{j\ge 1}$, we obtain unconditional convergence as well. Combined with \eqref{eq:approx_nong}, this yields
$$\frac{\overline{Y}-\theta_0}{\se_1} \convD \mathcal{N}(0,1).$$
We finally obtain \eqref{eq:pointwise_valid} using the fact that $\se\ge \se_1$ and that $\theta_0=\theta$ under the null hypothesis.

\paragraph{Asymptotically exact tests.} Our test $\phi_{\alpha}$ is conservative in non-Gaussian regimes. It is actually possible to consider an asymptotically exact test. To understand how, remark that $t_u$ is asymptotically exact when $\Omega_1+\Omega_2>0$, in which case $\overline{Y}$ has a slow rate of convergence, whereas in view of \eqref{eq:approx_nong} and \eqref{eq:conv_nong}, the test based on $t_1:=(\overline{Y}-\theta)/\se_1$ is asymptotically exact when $\Omega_1+\Omega_2=0$. Moreover, we show in the proof of Theorem \ref{thm:point_univ} that $(\widehat{V}_1+\widehat{V}_2)/\widehat{V}_{12}$ converges to infinity when $\Omega_1+\Omega_2>0$, whereas $(\widehat{V}_1+\widehat{V}_2)/\widehat{V}_{12} =O_P(1)$ when $\Omega_1+\Omega_2=0$. Now, consider
	\begin{equation*}
		t_a := t_u \ind{\widehat{V}_1+\widehat{V}_2 > s_{\Cinf} \widehat{V}_{12}} + t_1 \ind{\widehat{V}_1+\widehat{V}_2 \le s_{\Cinf} \widehat{V}_{12}},
	\end{equation*}
	where $\Cinf:=\min(C_1,C_2)$ and $s_{\Cinf}$ is such that $s_{\Cinf}\to \infty$ and $s_{\Cinf}/\Cinf \to 0$. Such conditions ensure that one selects the statistic that is asymptotically exact with probability approaching one.\footnote{In this sense, this construction is related to Menzel's bootstrap with selection.} As a result, the corresponding test is also asymptotically exact. Remark also that to treat the dimensions of clustering symmetrically, one could replace $t_1$ by $t_{j_{\textrm{max}}}$, with $j_{\textrm{max}}:=\arg\max_{k=1,2}C_k$. However, this test suffers from at least two drawbacks. First, the precise choice of the tuning parameter $s_{\underline{C}}$ remains unclear. Second, the test associated with $t_a$ does not have uniform guarantees, contrary to $t$. 

\paragraph{Power loss.} Related to the previous point, we explore in Appendix \ref{app:power_loss} to what extent our test $\phi_\alpha$ is conservative, by computing the average increase in confidence intervals we obtain when using $\se$ instead of $\se_1$ on asymptotically non-Gaussian DGPs. Across multiple draws of possible DGPs, we obtain an average increase of the length of around 9\%, with a maximum of around 25\%.

\paragraph{Multivariate extension.} We derive in Appendix \ref{sec:multiv_extension} the asymptotic validity of the joint test mentioned above. We obtain therein a similar result as Theorem \ref{thm:point_univ}, under an additional condition ensuring that the limit of estimated variance matrices have full rank. Note that this latter issue is specific to the multivariate set-up.

\subsection{Uniform results} 
\label{sub:uniform}

We now consider a uniform version of Theorem \ref{thm:point_univ}. In this context, we have to make some of our previous conditions uniform. Let $\P$ denote the set of probability distributions such that Assumption \ref{as:DGP} holds. By Theorem \ref{thm:point_univ}, we automatically obtain ``uniform'' asymptotic validity on $\P$ as long as it is finite; but additional restrictions have to be imposed otherwise.

\medskip
To introduce these restrictions, we index relevant objects such as expectation signs or $\Omega_k$ by $P$. For any $\tau_P$ that satisfies Equation~\eqref{eq:AHK}, let us define\footnote{Here and in the proofs, the variables $(U_{ij})_{i,j\ge 0}$ are supposed wlog to be uniformly distributed on $[0,1]$.}
\begin{equation}
\begin{array}{rrcl} \tau_{1P}: & [0,1]^3 & \to & \R \\
 & (u_1,u_2,u_3) & \mapsto & (\Omega_{1P} + \mathds{1}\{\Omega_{1P}=0\})^{-1/2}E_P[Y_{11}-\theta_{0P} \mid U_{10} = u_1]\end{array}.	
	\label{eq:def_tau1}
\end{equation}
 We define $\tau_{2P}$ similarly, just replacing $\Omega_{1P}$ and $U_{1,0}=u_1$ by $\Omega_{2P}$ and $U_{0,1}=u_2$.  For any $m>0$ and $H$ compact subset of $L_2([0,1]^3,\R)$, let us introduce
\begin{align*}
	\mathcal{P}_{m,H} := & \bigg\{ P\in \P: \; V_P(Y_{11}) \ge m,  \exists \tau_P \in H \text{ satisfying  Eq. \eqref{eq:AHK}} \\
	& \quad \text{and such that  $\tau_{kP}\in H$ for $k=1,2$},  \\
	& \quad \text{either } \Omega_{1P} \wedge \Omega_{2P} = 0 \text{ or } \Omega_{3P} \le m^{-1} \left(\Omega_{1P}+\Omega_{2P}\right) \bigg\}, \\
	\mathcal{P}_{m,H}^G := & \bigg\{P \in \mathcal{P}_{m,H} \; : \; \Omega_{3P} \leq m^{-1}(\Omega_{1P} + \Omega_{2P}) \bigg\}.	
\end{align*}
The compactness restriction states that we can approximate elements of $H$ uniformly well by elements of a finite-dimensional space. We comment on the other restrictions in $\mathcal{P}_{m,H}$ and $\mathcal{P}_{m,H}^G$ below.

	\begin{theorem}
		\label{thm:unif_univ}
Fix $m>0$ and $H$ a compact subset of $L_2([0,1]^3,\R)$. If Assumptions~\ref{as:DGP}-\ref{as:asym} hold, $\phi_\alpha$ is defined according to \eqref{eq:def_test} and $\theta=\theta_{0P}$, we have for any $\alpha \in (0,1)$:
		\begin{equation}
			\limsup_{n \to \infty} \sup_{P \in \mathcal{P}_{m,H}}  \E_P[\phi_\alpha] \le \alpha. 			\label{eq:valid_unif} \\
		\end{equation}
		Moreover, if $\mathcal{P}_{m,H}^G\ne \emptyset$:
		\begin{equation}
			\limsup_{n \to \infty} \sup_{P \in \mathcal{P}_{m,H}^G} \E_P[\phi_\alpha] = \liminf_{n \to \infty} \inf_{P \in \mathcal{P}_{m,H}^G} \E_P[\phi_\alpha]= \alpha.
			\label{eq:exactness_unif}
		\end{equation}
	\end{theorem}	

Let us sketch the proof of Theorem \ref{thm:unif_univ}. First, we show that it suffices to establish the result for any sequence of DGPs $(P_n)_{n\ge 1}$ in $\mathcal{P}_{m,H}$ (or in $\mathcal{P}_{m,H}^G$). The difficulty, then, is that for such a sequence, the four terms in the decomposition \eqref{eq:decomp} may matter asymptotically. To illustrate this, consider the following sequence of DGPs:
	\begin{equation}
		Y_{ij} = \theta_{0P} + \frac{b_1 U_{i0} + b_2 U_{0j}}{n^{1/2}} + b_3 U_{i0} U_{0j} + b_4 U_{ij},	
		\label{eq:illustr_issues_unif}
	\end{equation}
	where $C_1=C_2=n$, $(b_1,...,b_4)\in\R^4$ and the $(U_{ij})_{i,j\ge 0}$ are i.i.d., mean-zero variables. By dividing $b_1 U_{i0} + b_2 U_{0j}$ by $n^{1/2}$, we make the four terms of the decomposition ($\overline{\alpha}$, $\overline{\beta}$, $\overline{\gamma}$ and $\overline{\eps}$) converge at the same rate, namely $n=(C_1C_2)^{1/2}$. The term $\overline{\alpha}+\overline{\beta}+\overline{\eps}$ is asymptotically normal but $\overline{\gamma}$ is not, and it is not asymptotically independent of the first term. The general asymptotic distribution of $\overline{Y}$ for such sequences of DGPs is complicated and given by Lemma \ref{lem:weak_conv_means} in Appendix \ref{ssub:weak_conv_seq}. Still, we can explain the logic of our results in the simple example given by \eqref{eq:illustr_issues_unif}. Specifically, Lemma \ref{lem:weak_conv_means} implies that
$$\bigg[(\overline{Y}-\theta_{0P})/\sqrt{V(\overline{Y})}, (\widehat{V}_1, \widehat{V}_2,  \widehat{V}_{12})/V(\overline{Y})\bigg] \convD (L, V_1, V_2, V_{12}),$$
where, letting $(Z_1, Z_2, Z_4)$ be three i.i.d. standard normal variables,
\begin{align*}
	L & := c_1 Z_1 + c_2 Z_2 + c_3 Z_1 Z_2 + c_4 Z_4, \\
	V_1  & := c_4^2 + (c_1 + c_3 Z_2)^2, \\
	V_2 & :=  c_4^2 + (c_2 + c_3 Z_1)^2, \\
	V_{12} &:= c_4^2 + c_3^2,
\end{align*}	
with $(c_1,...,c_4)$ a vector that is related to $(\Omega_{1P},...,\Omega_{4P})$ (see Lemmas \ref{lem:mu_nu} and \ref{lem:weak_conv_means} for details) and also to $(b_1,...,b_4)$. In particular, if $b_2=0$, which implies $\Omega_{2P}=0$, we have $c_2=0$. Then, $L|Z_2 \sim \mathcal{N}(0, c_4^2 + (c_1 + c_3 Z_2)^2)$. As a result, $L/V_1^{1/2} \sim \mathcal{N}(0,1)$ and thus, as in the non-normal, pointwise case,
\begin{equation}
\frac{\overline{Y}-\theta_{0P}}{\se_1} \convNor{1}.	
	\label{eq:conv_with_se1}
\end{equation}
Similarly, if $b_1=0$, so that $\Omega_{1P}=0$, we can show that $(\overline{Y}-\theta_{0P})/\se_2 \convNor{1}$. The conclusion on \eqref{eq:valid_unif} follows as in the pointwise case.

\medskip
To what extent are the conditions in $\mathcal{P}_{m,H}$ necessary? Without fully answering this question, we can at least ascertain that the last condition in $\mathcal{P}_{m,H}$, namely $\Omega_{1P} \wedge \Omega_{2P} = 0$ or $\Omega_{3P} \le m^{-1} \left(\Omega_{1P}+\Omega_{2P}\right)$, cannnot be omitted. To see this, let us consider the following particular case of \eqref{eq:illustr_issues_unif}:
	\begin{equation}
		Y_{ij} = \theta_{0P} + \frac{U_{i0} + U_{0j}}{n^{1/2}} + \zeta\; U_{i0} U_{0j},	
		\label{eq:counterex}
	\end{equation}
	for some $\zeta\in\R$ and an i.i.d. sequence $(U_{ij})_{i,j\ge 0}$ with standard normal distribution. 
	Remark that no $\mathcal{P}_{m,H}$ includes the full sequence $(P_n)_{n\ge 1}$, since for all $n\ge 1$ $\Omega_{1P_n} \wedge \Omega_{2P_n} > 0$, $\Omega_{3P_n}=\zeta$ and $\Omega_{1P_n}+\Omega_{2P_n}\to 0$. Now, using Lemma \ref{lem:weak_conv_means}, we are able to simulate the asymptotic distribution of the test statistic $t$ in this case, for any $\zeta\in\R$. It appears that the test is not asymptotically valid for $\zeta\in(0,1.16]$, with an asymptotic level peaking at around $11\%$ for $\zeta\simeq 0.65$. 
	Mathematically, the expressions for $L$ and $V_1$ above show that $L|Z_2 \sim \mathcal{N}(c_2 Z_2, V_1)$. Moreover, $c_2\ne 0$ and thus we do not obtain \eqref{eq:conv_with_se1} anymore.


\section{Inference based on GMM estimators} 
\label{sec:GMM}

We extend our results on scalar expectations to tests based on smooth functionals of GMM estimators. Because we  consider possibly nonlinear estimators, we have to explicitly account for the fact that several units may be observed in each cell $(i,j)$. We thus assume to observe $(A_{ij\ell})_{1\le i\le C_1,1\le j\le C_2, 1 \le \ell \le N_{ij}}$, where $N_{ij}\in \N$ is a random variable denoting the number of units observed in cell $(i,j)$. We are interested in 
testing $\theta_0=\theta$ against $\theta_0\neq \theta$, where $\theta_0:=\varphi(\beta_0)\in \R$ for some function $\varphi$ and $\beta_0\in \Theta \subseteq \R^p$ satisfies
	\begin{equation}
		\label{eq:gmm_identification}
		\E\left[\frac{1}{C_1C_2} \sum_{i=1}^{C_1}\sum_{j=1}^{C_2}\sum_{\ell=1}^{N_{ij}}\psi(A_{ij\ell},\beta_0)\right] = 0, 
	\end{equation}
	with $\psi(a,\beta) \in \R^q$ $(q\ge p)$, and we use the convention $\sum_{\ell=1}^0 t_{\ell}=0$ for any sequence $(t_{\ell})_{\ell\ge 1}$. We let $\widehat{\theta}:=\varphi(\widehat{\beta})$, where $\widehat{\beta}$ is the GMM estimator:
	\begin{equation*}
		\widehat{\beta} \in \arg\min_{\beta\in\Theta} \left(\frac{1}{C_1C_2}\sum_{i=1}^{C_1}\sum_{j=1}^{C_2}\psi_{ij}(\beta)\right)' \Wh \left(\frac{1}{C_1C_2}\sum_{i=1}^{C_1}\sum_{j=1}^{C_2}\psi_{ij}(\beta)\right),
	\end{equation*}
for some symmetric, positive matrix $\Wh$ and with $\psi_{ij}(\beta) :=\sum_{\ell=1}^{N_{ij}} \psi(A_{ij\ell},\beta)$. To build our test, let $J:=\E\left[\deriv{\psi_{11}(\beta_0)}{\beta}\right]'$ and  $\Psi_{ij} := -\deriv{\varphi(\beta_0)}{\beta}'\left(J'\W J \right)^{-1} J'\W$ $\psi_{ij}(\beta_0)$ (we assume below the existence of the derivatives). As with i.i.d. data, we expect to have
\begin{equation}
\widehat{\theta} - \theta_0 \simeq \overline{\Psi}.	
	\label{eq:approx_theta}
\end{equation}
Then, let $\widehat{\Psi}_{ij} := -\deriv{\varphi(\widehat{\beta})}{\beta}'\left(\widehat{J}'\Wh \widehat{J} \right)^+ \widehat{J}'\Wh \psi_{ij}(\widehat{\beta})$, with $A^+$ the Moore-Penrose inverse of $A$ and $\widehat{J} := (C_1C_2)^{-1}\sum_{i=1}^{C_1}\sum_{j=1}^{C_2}\deriv{\psi_{ij}(\widehat{\beta})}{\beta}'$. We consider a similar test as in \eqref{eq:def_test}, namely
\begin{equation}
\phi_{\alpha} := \ind{|t|>z_{1-\alpha/2}},\text{ where }
t := \frac{\widehat{\theta} - \theta}{\se}.
\label{eq:def_test_GMM}
\end{equation}
We compute $\se$ as in Section \ref{sec:scalar_expecs} except that in view of \eqref{eq:approx_theta}, we replace $Y_{ij}$ with $\widehat{\Psi}_{ij}$ in $\widehat{V}_1, \widehat{V}_2$ and $\widehat{V}_{12}$.

\medskip
The validity of this test is obtained under the following conditions:
	
\begin{hyp}\label{as:DGP_GMMs}
We observe a sample $(A_{ij\ell})_{1\le i\le C_1,1\le j\le C_2, 1 \le \ell \le N_{ij}}$, which is extracted from the dissociated and separately exchangeable array $\bm{A}^\infty:=(A^\infty_{ij})_{(i,j)\in \N^{*2}}$, where $A^\infty_{ij}:=((A_{ij\ell})_{\ell\ge 1}, N_{ij})$.\footnote{Namely, $\bm{A}^\infty$ satisfies Conditions 1 and 2 in Assumption \ref{as:DGP}.}
\end{hyp}
	
	\begin{hyp}
		~
		\label{as:GMM_regul}
		\begin{enumerate}[label=(\roman*)]
			\item The parameter space $\Theta$ is a compact subset of $\R^p$ and $\beta_0$ lies in $\overset{\circ}{\Theta}$, the interior of $\Theta$.
			\item Eq.~\eqref{eq:gmm_identification} holds and $\forall \eps>0$,  $\inf_{\|\beta-\beta_0\|\ge\eps}\|\E[\psi_{11}(\beta)]\|>0$.
			\item $E[||\psi_{11}(\beta_0)||^2] < \infty$ and $V(\psi_{11}(\beta_0))$ is invertible.
			\item For every $a \in \R^{d_a}$, $\beta \mapsto \psi(a,\beta)$ is twice continuously differentiable on $\overset{\circ}{\Theta}$.
			\item $ \E\left[\sup_{\beta\in\overset{\circ}{\Theta}} \|\psi_{11}(\beta)\|^2 +\sup_{\beta\in\overset{\circ}{\Theta}} \|\Deriv{\psi_{11}(\beta)}{\beta}\|^2 +\sup_{\beta\in\overset{\circ}{\Theta}} \|\Deriv{^{2}\psi_{11}(\beta)}{\beta\partial \beta'}\|\right]<\infty$.
			\item $J:=\E\left[\deriv{\psi_{11}(\beta_0)}{\beta}\right]'$ is such that $J'J$ is invertible.
			\item $\Wh \convP \W$, an invertible deterministic matrix.
			\item The map $\beta \mapsto \varphi(\beta)$ is twice differentiable on $\overset{\circ}{\Theta}$ and $\sup_{\beta\in\overset{\circ}{\Theta}} \|\Deriv{^{2}\varphi(\beta)}{\beta\partial \beta'}\| < \infty$. We further have $\|\Deriv{\varphi(\beta_0)}{\beta}\| > 0$.
		\end{enumerate}
	\end{hyp}

	\begin{hyp}
		~\\
		\label{as:GMM_main}
			$\lambda_{\max}\left(\V\left(\sum_{i=1}^{C_1}\sum_{j=1}^{C_2} \psi_{ij}(\beta_0)\right)\right)/ \lambda_{\min}\left(\V\left(\sum_{i=1}^{C_1}\sum_{j=1}^{C_2} \psi_{ij}(\beta_0)\right)\right)=O(1)$.
	\end{hyp}
	
Assumption~\ref{as:DGP_GMMs} is simply Assumption~\ref{as:DGP} (without moment constraints) imposed on $\bm{A}^\infty$. Assumption~\ref{as:GMM_regul} includes classical regularity conditions that are not specific to our setup with multiway clustering. 
But we also impose Assumption~\ref{as:GMM_main}, which is specific to our setup: it automatically holds with i.i.d. data under  Assumption~\ref{as:GMM_regul}-(iii). We discuss this condition further below. 

\begin{theorem}\label{thm:GMM}
	Suppose that $P_{\bm{A}^\infty}$ does not depend on $n$, Assumptions \ref{as:asym}-\ref{as:GMM_main} hold, $\phi_\alpha$ be defined according to \eqref{eq:def_test_GMM} and $\theta=\theta_0$. Then, for every $\alpha\in(0,1)$,
	\begin{equation*}
			\limsup_{n \to \infty} \E[\phi_\alpha]\le \alpha,
	\end{equation*}
	with equality if the asymptotic distribution of $\widehat{\theta}$ is Gaussian.
\end{theorem}
	
The proof of Theorem \ref{thm:GMM} can be found in the Supplemental Appendix. In line with Section~\ref{sec:scalar_expecs}, we could strengthen our pointwise results on GMMs to uniform ones, by basically imposing uniform versions of Assumptions~\ref{as:GMM_regul} and~\ref{as:GMM_main}.
	
\medskip
While Assumption \ref{as:GMM_regul} is a standard regularity assumption, Theorem \ref{thm:GMM} also relies on Assumption \ref{as:GMM_main}. The following example illustrates that without this assumption, the usual linear approximation \eqref{eq:approx_theta} may not be valid with two-way clustered data. Note that this issue affects $\widehat{\theta}$ and is thus not specific to our inference method.
	
\begin{example}
	Consider a simple linear regression where $\psi(A_{ij},\beta)= (1, X_{ij})'(Y_{ij}- \alpha - X_{ij} \theta)$, with $A_{ij}=(Y_{ij}, X_{ij})'$ and $\beta=(\alpha,\theta)'$. Assume that $X_{ij}=U_{i0}$, $\eps_{ij}=U_{0j} + U_{ij}$ and $Y_{ij} = \alpha_0 + X_{ij}\theta_0 +\eps_{ij}$, with $(U_{ij})_{i,j\ge 0}$ i.i.d. random variables. With this DGP, the two components of  $\overline{\psi}:=(\overline{\eps},\overline{X\eps})$ do not converge at the same rate, so that Assumption \ref{as:GMM_main} fails. Actually, in this example, we can show that the appropriate $\Psi_{ij}$ is $X_{ij}U_{ij}/ V(X_{11})$ rather than the usual $\Psi_{ij}=X_{ij}\eps_{ij}/V(X_{11})$.
\end{example}	

\paragraph{Other tests.} With GMMs, the naive test based on $\widehat{V}_u=\widehat{V}_1+\widehat{V}_2-\widehat{V}_{12}$ has the same drawbacks as with sample means: $\widehat{V}_u$ can be negative, and inference may be invalid asymptotically because $\widehat{\theta}$ may not be asymptotically Gaussian. The first issue may actually be common in practice: it arises in 9 out of 15 papers published in the {\it American Economic Review} and that use two-way clustering (see Appendix \ref{sec:aer_meta} for details). To solve this issue, CGM consider the following alternative estimator. First, they estimate the asymptotic variance of $\widehat{\beta}$ by the matrix $\widehat{V}^\beta_u$, obtained as $\widehat{V}_u$ but with the $((\widehat{J}'\Wh \widehat{J})^+\widehat{J}'\Wh$ $\psi_{ij}(\widehat{\beta}))_{1\le i\le C_1, 1\le j\le C_2}$ instead of the $(\widehat{\Psi}_{ij})_{1\le i\le C_1, 1\le j\le C_2}$ as inputs. Then, they consider the eigendecomposition of $\widehat{V}^\beta_u$, $P'\Delta P$, and replace the negative eigenvalues in $\Delta$ by 0, and let $\widetilde{V}^\beta_u$ denote the corresponding matrix. Finally, their variance estimator for $\widehat{\theta}$ is $[\deriv{\varphi(\widehat{\beta})}{\beta}]' \,\widetilde{V}^\beta_u\, [\deriv{\varphi(\widehat{\beta})}{\beta}]$. 

\medskip
This solution has two drawbacks, however. First, it does not restore valid inference when $\widehat{\theta}$ is not asymptotically Gaussian. Second, the test is not invariant to affine transforms of variables in linear regressions, contrary to ours. For instance, we show in Subsection \ref{sub:lin_reg} that adding a constant or changing the scale of a regressor can make the rejection rate vary from 0 to 1. Similarly, changing the reference of a binary regressor affects inference.


			
\section{Monte Carlo simulations} 
\label{sec:monte_carlo_simulations}

We illustrate the performance of our test in two cases: univariate means and linear regressions. In these two cases, we let $C_1=C_2=n \in\{10, 20, 40\}$.
	
\subsection{Univariate sample means} 
\label{sub:sample_means}
	
	We first consider $\theta_0= E[Y_{11}]$ estimated by the sample mean $\overline{Y}$, where
	$$Y_{ij}=\delta_{1n} U_{i0} + \delta_{2n} U_{0j} + U_{i0} U_{0j} + \frac{1}{2} U_{ij},$$
	and the $(U_{ij})_{i,j\ge 0}$ are all independent, standard normal variables and $(\delta_{1n}, \delta_{2n})$ are possibly varying with $n$. We consider four DGPs, depending on the values of $(\delta_{1n}, \delta_{2n})$:
	\begin{enumerate}
		\item $\delta_{1n}=\delta_{2n}=1$. This DGP is fixed (independent of $n$) and non-degenerate. Theorem \ref{thm:point_univ} applies and our test is asymptotically exact;
		\item $\delta_{1n}=\delta_{2n}=0$. This DGP is fixed and degenerate. Theorem \ref{thm:point_univ} still applies but our test is expected to be asymptotically conservative;
		\item $\delta_{1n}=1/\sqrt{n}$, $\delta_{2n}=0$. Since $\Omega_{2P_n}=0$, the corresponding  DGPs all belong to $\mathcal{P}_{m,H}$ for some appropriate $m>0$, compact set $H$ and $n$ large enough. Thus, our uniformity result (Theorem \ref{thm:unif_univ}) applies, but our test is expected to be asymptotically conservative;
		\item $\delta_{1n}=\delta_{2n}=1/\sqrt{n}$. Since $\Omega_{1P_n} \wedge \Omega_{2P_n} > 0$, $\Omega_{1P_n} + \Omega_{2P_n} \to 0$ and $\Omega_{4P_n}$ does not tend to zero, we cannot apply Theorem \ref{thm:unif_univ} and thus have no size guarantee in this case.
	\end{enumerate}

We compute rejection rates under the null, by testing for $\theta_0=0$, and under the alternative, by testing for $\theta_0=\theta\ne 0$, with $\theta=0.5$ in DGP1 and $\theta=0.15$ in DGP2 to DGP4. This choice of $\theta$ ensures that power is nontrivial with our sample sizes. We compare our test (``DDG'' in the table) with usual inference (``Usual'' in the table). Recall that $\se_u=\max(0,\widehat{V}_u)^{1/2}$, so that we automatically reject the null hypothesis with usual inference when $\widehat{V}_u \le 0$. We also consider the bootstrap with selection (BS-S) developed by \cite{menzel2021bootstrap}. This bootstrap requires a tuning parameter $\kappa_0$: we consider both $\kappa_0=0.05$, as in the programs accompanying \cite{menzel2021bootstrap}, and a much larger value, $\kappa_0=1.25$.\footnote{In fact, there are two parameters appearing in Menzel's bootstrap with selection, namely $\kappa_a$ and $\kappa_g$ \citep[see Section 3 in][]{menzel2021bootstrap}. But in his simulations, he makes both depend on a single parameter $\kappa_0$, by setting $\kappa_a = \kappa_0 \log(C_1)/C_1$ and $\kappa_g =  \kappa_0 \log(C_2)/C_2$. We do not report here the results of his conservative bootstrap (BS-C), which is very conservative in our simulations.}

\medskip
The results are displayed in Table \ref{tab:univ_mean_all}. As predicted by theory, DDG and usual inference are very close in DGP1, for which the estimator is asymptotically Gaussian and usual inference is valid. For this DGP, the results of the four methods are very similar. In DGP2, on the other hand, the usual variance estimator is negative in around 30\% of the samples. Accordingly, the test is highly distorted. Our test is conservative, but less than BS-S with $\kappa=0.05$; its  power is similar to that of BS-S with $\kappa=1.25$. In DGP3, our test is again conservative but has higher power than BS-S with $\kappa=0.05$. The bootstrap with $\kappa=1.25$ is the most powerful but slightly overrejects. Usual inference is still distorted, though less so than in DGP2. Finally, in the last DGP, for which we do not have any theoretical guarantee, our test turns out to have a level close to the nominal one. Again, it has slightly larger power than BS-S with $\kappa=0.05$. BS-S with $\kappa=1.25$ slightly overrejects, and usual inference is quite distorted.

\medskip
The bottom line is that our method compares well in terms of level and power with the bootstrap and has the advantage of not requiring the choice of a tuning parameter, which may be difficult to choose appropriately and does affect rejection rates.

\begin{table}[H]
\begin{center}
{\small
\begin{tabular}{cc|ccccc|cccc}
		& & \multicolumn{5}{c|}{Level} & \multicolumn{4}{c}{Power} \\
		DGP & $n$ & DDG & Usual & $P(\se_u=0)$ & $\underset{(\kappa_0=0.05)}{\text{BS-S}}$ & $\underset{(\kappa_0=1.25)}{\text{BS-S}}$ & DDG & Usual &  $\underset{(\kappa_0=0.05)}{\text{BS-S}}$ & $\underset{(\kappa_0=1.25)}{\text{BS-S}}$ \\
        \toprule
1 & 10 & 0.121 & 0.122 & 0.000 & 0.104 & 0.118 & 0.342 & 0.345 & 0.334 & 0.343\\
1 & 20 & 0.082 & 0.082 & 0.000 & 0.079 & 0.080 & 0.435 & 0.435 & 0.428 & 0.428\\
1 & 40 & 0.070 & 0.070 & 0.000 & 0.066 & 0.066 & 0.618 & 0.618 & 0.612 & 0.612\\ \midrule
2 & 10 & 0.022 & 0.329 & 0.280 & 0.007 & 0.054 & 0.331 & 0.582 & 0.179 & 0.282\\
2 & 20 & 0.010 & 0.332 & 0.288 & 0.003 & 0.044 & 0.730 & 0.809 & 0.669 & 0.761\\
2 & 40 & 0.005 & 0.339 & 0.303 & 0.001 & 0.045 & 0.961 & 0.964 & 0.963 & 0.984\\  \midrule
3 & 10 & 0.022 & 0.244 & 0.194 & 0.014 & 0.085 & 0.273 & 0.473 & 0.168 & 0.281\\
3 & 20 & 0.012 & 0.230 & 0.193 & 0.006 & 0.072 & 0.627 & 0.714 & 0.578 & 0.689\\
3 & 40 & 0.009 & 0.228 & 0.197 & 0.004 & 0.081 & 0.916 & 0.925 & 0.909 & 0.944\\  \midrule
4 & 10 & 0.071 & 0.231 & 0.128 & 0.033 & 0.104 & 0.336 & 0.435 & 0.270 & 0.370\\
4 & 20 & 0.054 & 0.216 & 0.130 & 0.018 & 0.089 & 0.551 & 0.597 & 0.532 & 0.657\\
4 & 40 & 0.046 & 0.201 & 0.132 & 0.013 & 0.090 & 0.825 & 0.832 & 0.821 & 0.891\\
		\bottomrule
		\multicolumn{11}{p{440pt}}{{\footnotesize Notes: $C_1=C_2=n$, nominal level: 5\%. For power, we test $\theta_0=0.5$ in DGP1 and $\theta_0=0.15$ for DGP2-4. The results are obtained with 5,000 samples in each case.}}
	\end{tabular}}	
	\caption{Performances of various tests on a scalar expectation}
	\label{tab:univ_mean_all}
\end{center}
\end{table}

	
\subsection{Linear regressions} 
\label{sub:lin_reg}

Second, we consider inference in linear regressions, a simple instance of the GMM models discussed in Section~\ref{sec:GMM}. Specifically, we consider the following:
	$$Y_{ij} = X_{ij}'\beta_0 + \eps_{ij}, \; E[X_{ij}\eps_{ij}]=0,$$
	where we wish to conduct inference on $\theta_0$, the second coefficient of $\beta_0$ (corresponding to the first non-constant element of $X_{ij}$). We assume $\theta_0=0$ and consider again four DGPs (where, as above, $(U_{ij})_{i,j\ge 0}$ and $(\widetilde{U}_{ij})_{i,j\ge 0}$ are two independent families of i.i.d. standard normal variables): 	
	\begin{enumerate}
		\item $X_{ij}=(1,U_{i0})'$, $\eps_{ij}=\delta_{1n} \widetilde{U}_{i0} + \delta_{2n} \widetilde{U}_{0j} +  \widetilde{U}_{i0}\widetilde{U}_{0j} + \frac{1}{2} \widetilde{U}_{ij}$ and $\delta_{1n}=\delta_{2n}=1$. This DGP is fixed and Assumption \ref{as:GMM_main} holds so our test is asympotically valid and non-conservative by Theorem \ref{thm:GMM};
		\item Same as above but $\delta_{1n}=0$, $\delta_{2n}=1$. Assumption \ref{as:GMM_main} fails so we have no size guarantee;
		\item Same as above but $\delta_{1n}=\delta_{2n}=1/\sqrt{n}$. Assumption \ref{as:GMM_main} also fails. 
		\item $X_{ij}\in\R^3$, $X_{ij}=(1,U_{i0},U_{ij})'$ and $\eps_{ij}= \widetilde{U}_{0j}+0.1 \widetilde{U}_{ij}$. Assumption \ref{as:GMM_main} still fails.
	\end{enumerate}

We compute the rejection rates under the null and under the alternative by testing for $\theta_0=\theta\ne 0$, with $\theta=0.3$ in DGP1, $\theta=0.15$ in DGP2 and 3 and $\theta=0.13$ in DGP4. Apart from our test and the usual one, we consider CGM's fix detailed in Section \ref{sec:set_up}. We also consider Menzel's bootstrap with selection (BS-S). As with univariate sample means, this bootstrap requires a tuning parameter, which we also call $\kappa_0$: we consider both $\kappa_0=10$, as in the programs accompanying \cite{menzel2021bootstrap}, and a smaller value, $\kappa_0=1$.\footnote{As above, the two tuning parameters $\kappa_a$ and $\kappa_g$ are defined in Menzel's programs as $\kappa_a = \kappa_0 \mu_{4e} \log(C_1)/C_1$ and $\kappa_g =  \kappa_0 \mu_{4e}\log(C_2)/C_2$, with $\mu_{4e}=[2\max(1/100,\overline{\widehat{\eps}^4})]^{1/2}$, where $\widehat{\eps}$ denotes the residual of the regression. The choice $\kappa_0=1$ also appears in the programs but is commented.}

\medskip
The results are displayed in Table \ref{tab:reg}. 
Interestingly, in DGP1 for which the usual inference is asymptotically valid, our test leads to substantial improvements when $n=10$, also over CGM. With $\kappa_0=10$, BS-S does not seem to work properly in this DGP, but using $\kappa_0=1$ yields results broadly similar to those of DDG. 

\medskip
Usual inference is highly distorted in DGP2 to DGP4, with in particular a rejection rate of 1 in DGP4. CGM is less distorted but still rejects between 14\% and 56\% in these three DGPs. Also, as indicated above, inference based on CGM's fix is not invariant to linear change in the regressors. For instance, we obtain a very conservative test, with a rejection rate of 0 under the null, when adding 2 to the first regressor. Conversely, multiplying this regressor by a constant approaching 0 makes the rejection rate tend to 1. Though our theoretical results do not apply for DGP2 to 4, our test seems to behave well in these cases, with rejection rates below 5\% under the null for all sample sizes. The two bootstraps differ in DGP2, with $\kappa_0=10$ leading to conservative inference, but behave very similarly for DGP3 and DGP4. They also appear to slightly overreject with DGP3.

\begin{table}[H]
	\begin{center}
		{\footnotesize\begin{tabular}{cc|cccccc|cccc}
			& & \multicolumn{6}{c|}{Level} & \multicolumn{4}{c}{Power}\\			
DGP & $n$ & DDG & Usual & CGM & $P(\se_u=0)$ & $\underset{(\kappa_0=10)}{\text{BS-S}}$ & $\underset{(\kappa_0=1)}{\text{BS-S}}$ & DDG & CGM & $\underset{(\kappa_0=10)}{\text{BS-S}}$ & $\underset{(\kappa_0=1)}{\text{BS-S}}$ \\
\toprule
1 & 10 & 0.161 & 0.295 & 0.266 & 0.080 & 0.319 & 0.194 & 0.341 & 0.441 & 0.500 & 0.328\\
1 & 20 & 0.110 & 0.127 & 0.125 & 0.004 & 0.489 & 0.147 & 0.385 & 0.439 & 0.753 & 0.347\\
1 & 40 & 0.080 & 0.081 & 0.081 & 0.000 & 0.604 & 0.053 & 0.528 & 0.554 & 0.930 & 0.460\\
\midrule
2 & 10 & 0.039 & 0.679 & 0.297 & 0.593 & 0.029 & 0.063 & 0.345 & 0.606 & 0.110 & 0.296\\
2 & 20 & 0.018 & 0.662 & 0.206 & 0.599 & 0.024 & 0.067 & 0.717 & 0.825 & 0.425 & 0.749\\
2 & 40 & 0.012 & 0.638 & 0.191 & 0.582 & 0.021 & 0.065 & 0.963 & 0.970 & 0.971 & 0.988\\
\midrule
3 & 10 & 0.054 & 0.357 & 0.225 & 0.253 & 0.115 & 0.102 & 0.300 & 0.485 & 0.262 & 0.256\\
3 & 20 & 0.021 & 0.284 & 0.162 & 0.224 & 0.104 & 0.102 & 0.651 & 0.735 & 0.652 & 0.648\\
3 & 40 & 0.011 & 0.240 & 0.143 & 0.201 & 0.109 & 0.109 & 0.903 & 0.913 & 0.975 & 0.974\\
\midrule
4 & 10 & 0.030 & 1 & 0.255 & 1 & 0.010 & 0.011 & 0.240 & 0.505 & 0.183 & 0.193\\
4 & 20 & 0.030 & 1 & 0.355 & 1 & 0.026 & 0.026 & 0.854 & 0.961 & 0.847 & 0.847\\
4 & 40 & 0.041 & 1 & 0.553 & 1 & 0.034 & 0.034 & 1 & 1 & 1 & 1\\
\bottomrule			
			\multicolumn{12}{p{420pt}}{{\scriptsize Notes: $C_1=C_2=n$, nominal level: 5\%. For power, we fix $\theta$ to 0.3 for DGP1, 0.15 for DGP2-3 and 0.13 for DGP4. 5,000 samples for each of the 12 cases.}}
		\end{tabular}}
	\end{center}
	\caption{Performances of various tests on the coefficient of a linear regression}
	\label{tab:reg}
\end{table}


	
\section{Conclusion} 
\label{sec:conclusion}
	
We have shown that suitable, elementary changes in the usual inference with two-way clustering may result in pointwise valid tests even in non-Gaussian regimes. With sample means, this holds under the same moment condition as with i.i.d. data. For GMM estimators, a condition on the rates of convergence of the different moments is required. This condition is specific to the two-way clustering setup and ensures that  the estimator is asymptotically close to the average of its influence functions. We also show uniform validity of the tests over suitable classes of DGPs.
	
\medskip
We leave a few questions for future research. The first is whether we can still obtain asymptotically valid inference under weaker restrictions than those we have imposed. The second is whether our proposal extends to multiway clustering with three or more dimensions of clustering. The third is whether simple, analytic inference for dyadic data is possible, including in non-Gaussian regimes. This may not be straightforward: we show in Appendix \ref{sec:dyadic_data} that the fix we use with two-way clustering does not lead to valid pointwise inference in this setup.
	

\bibliography{biblio}

\newpage
\linespread{1.3}\selectfont
\appendix

\section{Do nonpositive variance estimators arise in practice?}
\label{sec:aer_meta}

We investigate whether variance matrices $\widehat{V}_u^\beta$ of linear regression coefficients (see Section~\ref{sec:GMM} for a definition) are often nonpositive in practice. To this end, we revisit papers published in the {\it American Economic Review} between January 2018 and June 2024. We choose this journal because the supporting data are often available online. To select the relevant papers, we looked for the regular expressions that include ``clust'', possibly separated by dashes or spaces, and starting with an upper or lower case. Next, we reviewed manually all the selected articles to identify the following 15 applied papers using multiway clustering and for which the data are available:
\begin{enumerate}[leftmargin=*,topsep=0pt,itemsep=-1ex,partopsep=1ex,parsep=1ex]
\item ``Legal Origins and Female HIV.''
\item ``Importing Political Polarization? The Electoral Consequences of Rising Trade Exposure.''
\item ``Does the Squeaky Wheel Get More Grease? The Direct and Indirect Effects of Citizen Participation on Environmental Governance in China.''
\item ``Overreaction in Macroeconomic Expectations.'' 
\item ``Heroes and Villains: The Effects of Heroism on Autocratic Values and Nazi Collaboration in France.''
\item ``Measuring Geopolitical Risk.''
\item ``Asymmetric Attention.''
\item ``Partisanship and Fiscal Policy in Economic Unions: Evidence from US States.''
\item ``Job Search and Hiring with Limited Information about Workseekers' Skills.''
\item ``Stock Market Wealth and the Real Economy: A Local Labor Market Approach.''
\item ``The Violent Legacy of Conflict: Evidence on Asylum Seekers, Crime, and Public Policy in Switzerland.''
\item ``The Taxing Deed of Globalization.''
\item ``Information Networks and Collective Action: Evidence from the Women's Temperance Crusade.''
\item ``Geographic Dispersion of Economic Shocks: Evidence from the Fracking Revolution: Comment.''
\item ``Propagation and Insurance in Village Networks.''
\end{enumerate}

\medskip
For each of these papers, we finally selected the first regression in the paper where the authors rely on multiway clustering. For 9 of these regressions, the ``usual'' two-way clustering variance estimator has at least one negative eigenvalue.

\section{Dyadic data} 
	\label{sec:dyadic_data}
	
Dyadic data correspond to variables observed at a pair level, namely between two units belonging to the same population. An important economic example is international trade between countries. To model such data, we often use jointly exchangeable arrays. Namely, we modify Assumption \ref{as:DGP} as follows:
	
	\begin{hyp}\label{as:DGP_joint}
		We observe a sample $(Y_{ij})_{1\le i,j \le C, i\neq j}$, which is extracted from the dissociated and jointly exchangeable array $\bm{Y}:=(Y_{ij})_{(i,j)\in \N^{*2},i\neq j}$. Namely, $\bm{Y}$ satisfies:
		\begin{enumerate}
			\item (dissociation) for any $(E,F) \subset \N^{*2}$ such that $E\cap F=\emptyset$, $(Y_{ij})_{(i,j)\in E^2, i\ne j}$ is independent of $(Y_{ij})_{(i,j)\in F^2, i\ne j}$.
			\item (joint exchangeability) for any permutation $\pi$ on $\mathbb{N}^*$, 
			$$(Y_{ij})_{(i,j)\in \N^{*2},i\ne j}\overset{d}{=}(Y_{\pi(i)\pi(j)})_{(i,j)\in \N^{*2},i\ne j}.$$
		\end{enumerate}
		The distribution of $\bm{Y}$ may depend on $C$. 
	\end{hyp}
	

The variance estimator commonly used in this context, following \cite{fafchamps2007formation} \citep[see also][for a related, earlier proposal]{holland1976local}, is $\widehat{V}_u := \widehat{V}_1 - \widehat{V}_{12}$, with
	\begin{align*}
		\widehat{V}_1 := & \frac{1}{2C^2} \sum_{i=1}^C \left(\frac{1}{C-1} \sum_{j\neq i} (Y_{ij}+Y_{ji}) -2\overline{Y}\right)^2, \\	
		\widehat{V}_{12} := & \frac{1}{2(C(C-1))^2} \sum_{i\neq j} \left(Y_{ij}+Y_{ji} -2\overline{Y}\right)^2.
	\end{align*}
	Intuitively, in this case where rows and columns correspond to the same population, $\widehat{V}_1=\widehat{V}_2$. $\widehat{V}_u$ suffers from the same issues as $\widehat{V}$: it may be negative, even asymptotically, and it does not necessarily lead to valid inference. For instance, if $Y_{ij}=U_i U_j$, with $(U_i)_{i\ge 1}$ i.i.d. with mean zero and variance one, $n\widehat{V}_u \convD 2(\chi^2(1)-1)$, so asymptotically, $\widehat{V}_u$ is negative with a probability of around 68\%.
	
	\medskip
	Remark that  $\widehat{V}_1 \geq \widehat{V}_u$. Thus, a natural extension of our proposal for multiway clustering would simply be to use $\widehat{V}_1$ instead of $\widehat{V}_u$. However, this does not always lead to valid inference. In the example above, one can show that
	$$\frac{\overline{Y}-\theta_0}{\sqrt{\widehat{V}_1}} \convD \frac{Z^2-1}{\sqrt{2}|Z|},$$
	where $Z\sim\mathcal{N}(0,1)$. Then, the rejection rate of a test based on $\widehat{V}_1$ and with nominal level 5\% is around 25.5\%, for instance. Intuitively, the reason why it fails is that in Lemma~\ref{lem:weak_conv_means} below, the $(Z_{k_1,0,0})_{k_1\ge 1}$ and the $(Z_{0,k_2,0})_{k_2\ge 1}$ are now the same. Hence, the argument that $L|(Z_{0,k_2,0})_{k_2\ge 1}\sim\mathcal{N}(0,V_1)$ (see the discussion below Theorem \ref{thm:point_univ}) no longer applies.
	
	

\section{Power loss of our test} 
\label{app:power_loss}

We examine here the degree of power loss one can expect in non-Gaussian regimes when using our test rather than an oracle that would select the appropriate test and would therefore not be conservative. Specifically, we consider a univariate setup with $\Omega_1+\Omega_2=0$ and $\Omega_3>0$ (so that we are, indeed, in a non-Gaussian regimes), and compare the average lengths of the confidence interval based on $\se_1$ (CI$_1$, say) and that based on $\se=\max(\se_1,\se_2,\se_u)$ (CI, say). CI corresponds to our method and is asymptotically conservative, whereas in non-Gaussian regimes, CI$_1$ is not conservative.

\medskip
Lemma \ref{lem:weak_conv_means} in Appendix \ref{ssec:key_lemmas} shows that the asymptotic distribution of  $(\se,\se_1)$, once properly normalized, only depends on $\Omega_4$ and the $(\mu_{\k})_{\k\in\K_3}$. These coefficients appear in the decomposition of the function $\tau$ on an orthonormal basis, see Lemma \ref{lem:mu_nu} below for more details. Without loss of generality, we fix $V(Y_{11})(=\Omega_3+\Omega_4)=1$. Then, we draw $\Omega_4$ according to a uniform distribution, and draw $(\mu_{k_1,k_2,0})_{(k_1,k_2)\in\{1,...,10\}^2}$ uniformly on the sphere of radius $\Omega^{1/2}_3=(1-\Omega_4)^{1/2}$. The other coefficients $\mu_{\k}$, with $\k=(k_1,k_2,k_3)$ satisfying $\max(k_1,k_2,k_3)>10$, are set to 0. For each draw of $\Omega_4$ and the $(\mu_{\k})_{\k\in\K_3}$, we then draw $(\se^a,\se^a_1)$ along the asymptotic distribution of $(\se,\se_1)$, which can be obtained using Lemma \ref{lem:weak_conv_means}. Then, we can approximate by simulations $R:=E[\se^a]/E[\se^a_1]$. Since for any nominal coverage, CI and CI$_1$ use the same quantile of a normal distribution, $R$ corresponds to the ratio of the average lengths of CI and CI$_1$ using the asymptotic distribution of $(\se,\se_1)$ as an approximation of their true distribution.

\begin{figure}[H]
	\centering
	\includegraphics[trim=10mm 80mm 10mm 80mm, scale=0.6, clip=true]{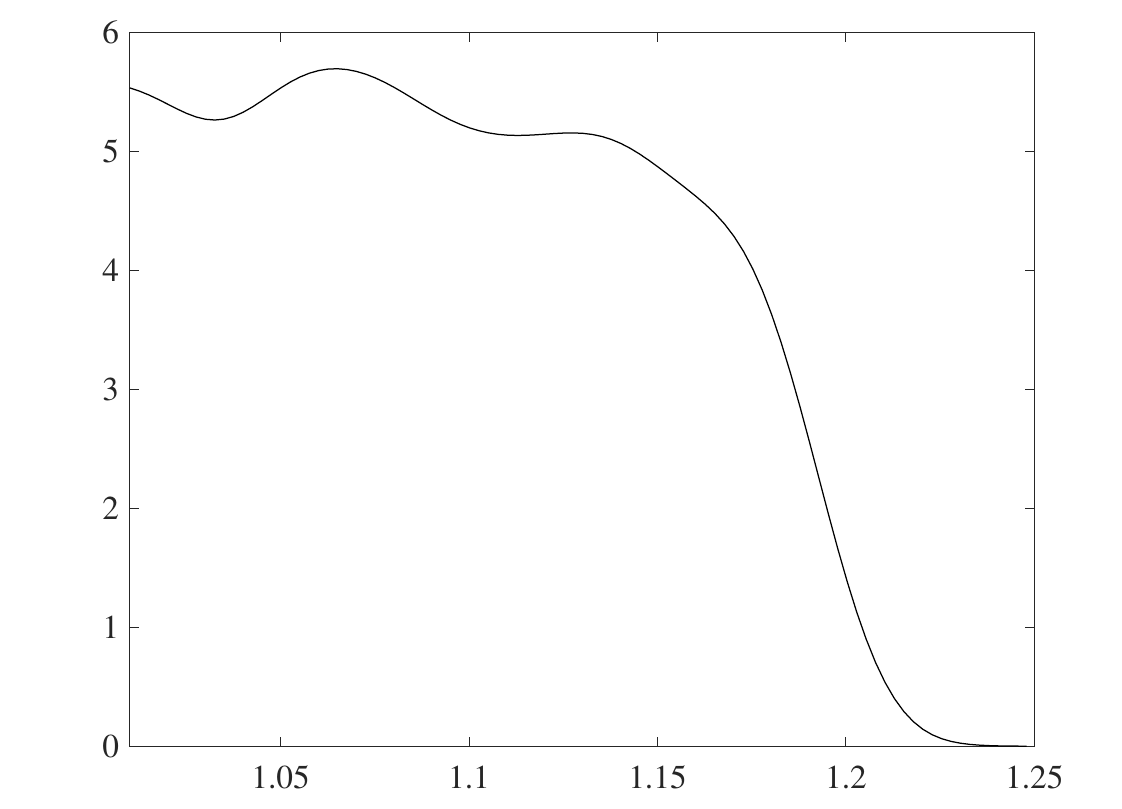}
	\caption{Density of the ratio of the average length of asymptotically conservative vs asymptotically exact CIs.}
	\label{fig:dist_ratio_length}
\end{figure}

\medskip
This way, we can approximate $R$ for each draw of $\Omega_4$ and the $(\mu_{\k})_{\k\in\K_3}$. Figure \ref{fig:dist_ratio_length} plots the density of $R$ across the draws of $\Omega_4$ and the $(\mu_{\k})_{\k\in\K_3}$. 
The distribution of $R$ appears to be roughly uniform between 1 and 1.15, and then decreases until 1.25. On average across the draws of $\Omega_4$ and the $(\mu_{\k})_{\k\in\K_3}$, we obtain an increase of 9\% in the average length of CI compared to the oracle CI$_1$.

\section{Inference on multivariate expectations} 
\label{sec:multiv_extension}

\subsection{Definition of the tests and confidence regions}

In this section, $\theta_0:=E[Y_{ij}]\in \R^d$ and we want to either test that $\theta_0=\theta$ or construct a confidence region on $\theta_0$. For any matrix $A$, let $A^{\otimes 2}$ stand for $AA'$. We now define $\widehat{V}_k$, $k\in\{1,2,12\}$ as follows:
\begin{align*}
	\widehat{V}_1 & := \frac{1}{C_1^2}\sum_{i=1}^{C_1}\left(\frac{1}{C_2}\sum_{j=1}^{C_2}Y_{ij} - \overline{Y} \right)^{\otimes 2}, \\
	\widehat{V}_2 & := \frac{1}{C_2^2}\sum_{j=1}^{C_2}\left(\frac{1}{C_1}\sum_{i=1}^{C_1} Y_{ij} - \overline{Y} \right)^{\otimes 2}, \\
	\widehat{V}_{12} & := \frac{1}{(C_1C_2)^2}\sum_{i=1}^{C_1}\sum_{j=1}^{C_2} \left(Y_{ij} - \overline{Y}\right)^{\otimes 2}.
\end{align*}
As above, $\widehat{V}_u:=\widehat{V}_1+\widehat{V}_2-\widehat{V}_{12}$. The ``usual'' test is $\phi_{u,\alpha} := \ind{\widetilde{F}_u>q_{1-\alpha/2}(d)}$, where $q_{1-\alpha}(d)$ is the quantile of order $1-\alpha$ of a $\chi^2(d)$ and
$$\widetilde{F}_u := \ind{F_u<0} \times \infty + \ind{F_u\ge 0} \times F_u, \quad \text{with } F_u= (\overline{Y}-\theta)' \widehat{V}^{-1}_u (\overline{Y}-\theta).$$
Here we take the convention that $0\times \infty=0$. Thus, when $F_u<0$, $\widetilde{F}_u=\infty$ and we reject the test, as we do in the univariate case when $\widehat{V}_u<0$.\footnote{Neither MH nor CGM consider such a test, but we label it ``usual'' as it seems to be the natural multivariate counterpart of the univariate test they consider.} 

\medskip
We consider the following modification of this $F$-test. Let us first assume that $\widehat{V}_1$, $\widehat{V}_2$ and $\widehat{V}_u$ are invertible; the case where one of them is singular is considered below. For $k\in\{1,2\}$, let
$$F_k := (\overline{Y}-\theta)' \widehat{V}_k^{-1} (\overline{Y}-\theta),$$
while $F_u$ is still defined as above. Note that by construction, $F_k\ge 0$ for $k\in\{1,2\}$. Then, our test is $\phi_{\alpha} := \ind{F>q_{1-\alpha}(d)}$ where
\begin{equation}
	F := \min\left(\widetilde{F}_u, F_1, F_2\right).	
	\label{eq:def_F}
\end{equation}
As above, we thus take a conservative approach by picking the minimum between three $F$-statistics, the usual one (set to $\infty$ if negative) and two others obtained by focusing on one dimension of clustering only.

\medskip
Now, if $\widehat{V}_1$, $\widehat{V}_2$ or $\widehat{V}_u$ is singular, let us define
$$F^\lambda_k := (\overline{Y}-\theta)' \left(\lambda \Id +\widehat{V}_k\right)^{-1} (\overline{Y}-\theta), \quad k\in\{1,2,u\},$$
where $\Id$ denotes the identity matrix. As above, let $\widetilde{F}^\lambda_u := \ind{F^\lambda_u<0} \times \infty + \ind{F^\lambda_u\ge 0} \times F^\lambda_u$.  Then, our test is 
\begin{equation}
\phi_{\alpha} := \ind{F>q_{1-\alpha}(d)}, \; \text{where} \; F := \lim_{\lambda \downarrow 0} \min\{\widetilde{F}^\lambda_u, F^\lambda_1, F^\lambda_2\}.
\label{eq:def_Ftest}
\end{equation}
A simple alternative would be to define $F$ as above, replacing inverses by Moore-Penrose inverses. However, this could lead to conservative inference in some cases where our approach leads to asymptotically non-conservative inference.\footnote{The reason behind is that for any symmetric positive semidefinite matrices $A$ and $B$, $A\gg B$ (meaning that $A-B$ is symmetric positive semidefinite) does not imply that their Moore-Penrose inverses $A^+$ and $B^+$ satisfy  $B^+\gg A^+$. On the other hand, for all $\lambda>0$, we do have $(\lambda \Id + B)^{-1}\gg (\lambda \Id + A)^{-1}$.}

\medskip
We also consider tests based on a Bonferroni correction of univariate tests. Albeit conservative in general, these tests turn out to rely on milder restrictions than the tests above. They also lead to straightforward rectangular confidence regions. Let $\overline{Y}=(\overline{Y^{(1)}},...,\overline{Y^{(d)}})'$, $\theta=(\theta^{(1)},...,\theta^{(d)})'$ and let $\se^\ell_k$ denote the square root of the $\ell$-th diagonal term of $\widehat{V}_k$ for $k\in\{1,2,u\}$. As before, we take the convention that $\se^\ell_u=0$ if the $\ell$-th diagonal term of $\widehat{V}_u$ is negative. Let $\se^\ell:=\max\{\se^\ell_1,\se^\ell_2,\se^\ell_u\}$ and $t^\ell:=(\overline{Y^{(\ell)}}- \theta^{(\ell)})/\se^\ell$. Then, the Bonferroni-based test of $H_0$ is
\begin{equation}
\phi^b_\alpha :=  \max_{\ell=1,...,d} \ind{|t^\ell| > z_{1-\frac{\alpha}{2d}}}.
\label{eq:def_bonf_test}
\end{equation}

\paragraph{Confidence regions.} The usual method to construct confidence regions is to invert tests. Here, we could build a confidence region on $\theta_0$ by inverting the test $F$, which we index by $\theta$ for clarity here:
$$\CR{1} := \left\{\theta: F(\theta)\le q_{1-\alpha}(d)\right\}.$$
Because $\widehat{V}_1, \widehat{V}_2$ and $\widehat{V}_u$ do not depend on $\theta$, it is easy to see that $\CR{1}$ is a star-shaped set: for any $\theta\in\CR{1}$, the segment between $\theta$ and $\overline{Y}$ is also in $\CR{1}$. Still, $\CR{1}$ is not convex in general, because $\theta\mapsto F(\theta)$ itself may not be convex, so it can be costly to compute. One simple solution is to mimic the univariate case, where we recall that the confidence interval is $[\overline{Y} \pm \, z_{1-\alpha/2} \, \se]$, with $\se:=(\max\{\widehat{V}_1,\widehat{V}_2,\widehat{V}_u\})^{1/2}$. The maximum operator does not trivially generalize to matrices, but the following works. For any diagonal matrix $\Delta$ with diagonal elements $(\delta_\ell)_{1\le \ell\le d}$, let $|\Delta|$ be diagonal with elements $(|\delta_\ell|)_{1\le \ell\le d}$. Then, for any symmetric $A$ with eigendecomposition $A=P'\Delta P$, let $|A|:=P'|\Delta|P$. Recall that for two scalars $a$ and $b$, we have $\max(a,b)=(a+b+|a-b|)/2$. Similarly, for any symmmetric matrices $A$ and $B$, let
$$\max(A, B):=\frac{1}{2}\left[A+B + |A-B|\right].$$
Note that $\max(A, B)\gg A$ and $\max(A, B)\gg B$. Then, we consider
$$\widehat{V} := \max\left(\widehat{V}_u, \max(\widehat{V}_1, \widehat{V}_2)\right).$$
Remark that $\widehat{V}$ is positive semi-definite. Then, we let
$$\CR{2} := \left\{m: (\overline{Y}-\theta)'\widehat{V}^{-1}(\overline{Y}-\theta) \le q_{1-\alpha}(d)\right\}.$$
This confidence interval takes the usual form of an ellipsoid. Note that we could also consider tests of $\theta_0=\theta$ using the $F$-statistic $(\overline{Y}-\theta)'\widehat{V}^{-1}(\overline{Y}-\theta)$. However, this would lead to more conservative inference than with the $F$ defined by \eqref{eq:def_F}.

\subsection{Pointwise results}
\label{ssub:pointwise_multiv}

Our result relies on the following condition:

\begin{hyp}
	\label{as:pointw_multiv}
	We either have (i) $\textrm{range}(\Omega_3)\subseteq \textrm{range}(\Omega_1+\Omega_2)$ or (ii) $\Omega_{j}=0$ for some $k\in\{1,2\}$ and $\Omega_1+\Omega_2+\Omega_4$ is invertible.
\end{hyp}

As we show in the proofs, Condition (i) is actually equivalent to $\overline{Y}-\theta_0$ being asymptotically Gaussian, a situation referred to hereafter as the ``Gaussian regime''. Condition (i) holds in particular if $\Omega_1+\Omega_2$ is invertible. Then, as in the univariate case with $\Omega_1+\Omega_2>0$, it turns out that $c'(\overline{Y}-\theta_0)$ has a ``slow'' convergence rate of at most $\max(C_1,C_2)^{1/2}$ for all $c\in\R^d$, $c\ne 0$. Condition (i) also holds if $\Omega_3=0$ or equivalently $\gamma_{ij}=0$ for all $(i,j)$. This is the case for instance in the DGP $Y_{ij}=U_{i0} + U_{0j} + U_{ij}$, where the $(U_{ij})_{i,j\ge 0}$ are i.i.d. random vectors. Finally, Condition (i) also includes cases where different components of $\overline{Y}$ converge at different rates, as for instance with the DGP $Y_{ij}=\theta_0 + (U_{i0}+U_{i0}U_{0j}, U_{ij})'$. In this DGP, the first component converges at the $C_1^{1/2}$ rate, whereas the second component converges at the $(C_1C_2)^{1/2}$ rate. Still, Condition (i) holds as $\textrm{range}(\Omega_3)= \textrm{range}(\Omega_1+\Omega_2)=$span$((1,0)')$.

\medskip
Condition (ii) holds in several non-Gaussian regimes, including cases where the different components of $\overline{Y}-\theta_0$ do not converge at the same rate, as Example \ref{ex:multi2} below illustrates. Example \ref{ex:counter_multi} gives a seemingly close DGP for which Assumption \ref{as:pointw_multiv} fails. 

\begin{example}
	\label{ex:multi2}
	Assume that $Y_{ij}= \theta_0 + (U_{i0}, U_{i0} U_{0j} + U_{ij})'$ where the $(U_{ij})_{i,j\ge 0}$ are as in \eqref{eq:AHK}, with variance one. Then, 
	$$\Omega_1=\begin{pmatrix}
		1 & 0 \\
		0 & 0
	\end{pmatrix}, \; \Omega_2=\begin{pmatrix}
		0 & 0 \\
		0 & 0
	\end{pmatrix} \; \text{and } \Omega_4=\begin{pmatrix}
		0 & 0 \\
		0 & 1
	\end{pmatrix}.$$
	As a result, $\Omega_1+\Omega_2+\Omega_4$ is invertible and Assumption \ref{as:pointw_multiv} holds. 
\end{example}

\begin{example}
	\label{ex:counter_multi}
	If $Y_{ij}= \theta_0 + (U_{i0}  + U_{ij}, U_{i0} U_{0j})'$ where the $(U_{ij})_{i,j\ge 0}$ are as in Example \ref{ex:multi2}. Then,
	$$\Omega_1=\Omega_4=\begin{pmatrix}
		1 & 0 \\
		0 & 0
	\end{pmatrix} \text{and } \Omega_2=\begin{pmatrix}
		0 & 0 \\
		0 & 0
	\end{pmatrix}.$$ This implies that $\Omega_1+\Omega_2+\Omega_4$ is singular. Assumption \ref{as:pointw_multiv} thus fails.
\end{example}

Example~\ref{ex:counter_multi} may seem worrying, as it puts forward a very simple DGP that violates Assumption~\ref{as:pointw_multiv}-(ii). But importantly, Assumption~\ref{as:pointw_multiv}-(ii) still holds whenever cell-level shocks are present in every entry of $Y_{ij}$ and are not collinear, a reasonable situation in practice. 

\begin{theorem}
	Suppose that $P_{\bm{Y}}$ does not depend on $n$, Assumptions \ref{as:DGP}-\ref{as:asym} hold, $\phi_\alpha$ and $\phi^b_\alpha$  be as in \eqref{eq:def_Ftest} and \eqref{eq:def_bonf_test}, respectively and $\theta=\theta_0$. Then, for every $\alpha\in(0,1)$,
	\begin{enumerate}
		\item If Assumption \ref{as:pointw_multiv}-(ii) holds,
		\begin{equation}
			\limsup_{n \to \infty} \E[\phi_\alpha]\le \alpha. \label{eq:mul_pointwise_valid}
		\end{equation}
		\item If Assumption \ref{as:pointw_multiv}-(i) holds,
		\begin{equation}
			\lim_{n \to \infty} \E[\phi_\alpha] = \alpha. \label{eq:mul_pointwise_exact}
		\end{equation}
		\item $\limsup_{n \to \infty} \E[\phi^b_\alpha]\le \alpha$ with possibly strict inequality even if Assumption \ref{as:pointw_multiv}-(i) holds.
	\end{enumerate}
	\label{thm:point_multiv}
\end{theorem}

Theorem \ref{thm:point_multiv} shows that as in the univariate case, the $F$-test we propose is asymptotically valid, although possibly conservative in non-Gaussian regimes. Compared to the univariate case, our result on $\phi_\alpha$ relies on Assumption \ref{as:pointw_multiv}. As Example \ref{ex:counter_multi2} below shows, this condition cannot be removed. We can however dispense with this assumption by using the Bonferroni test. This approach is asymptotically more conservative than the $F$-test we propose in Gaussian regimes, since the latter is asymptotically exact in such cases. In non-Gaussian regimes where both tests are asymptotically conservative, on the other hand, it is unclear from a theoretical standpoint whether one approach is more conservative than the other.

\medskip
As above, Theorem \ref{thm:point_multiv} follows from our uniform result below (Theorem \ref{thm:unif_multiv}), but let us still explain why the three points hold. Point 3 basically follows from  Theorem \ref{thm:point_univ} and a union bound. Regarding Point 2, the main facts we prove are that (i) $\overline{Y}$ is asymptotically Gaussian; (ii) $V(\overline{Y})^{1/2}\widehat{V}_u^{-1}V(\overline{Y})^{1/2} \convP \Id$ and (iii) for $k\in\{1,2\}$, $V(\overline{Y})^{1/2}\widehat{V}_k^{-1}V(\overline{Y})^{1/2} \convP M_k$ for some matrix $M_k$ satisfying $M_k\gg \Id$. Together, these three observations imply that as in the univariate case with a Gaussian regime, both the usual test and ours are asymptotically valid, non-conservative and equivalent. To obtain (i), we use again the decomposition $\overline{Y}-\theta_0=\overline{\alpha}+\overline{\beta}+\overline{\gamma} +\overline{\eps}$. Then, $\textrm{range}(\Omega_3)\subseteq \textrm{range}(\Omega_1+\Omega_2)$ implies that the asymptotically non-Gaussian component $\overline{\gamma}$ is negligible compared to $\overline{\alpha}+\overline{\beta}$ (or both are zero if $\Omega_3=\Omega_1+\Omega_2=0$).

\medskip
Point 1 is more difficult to prove. As in the univariate case, the idea is to show that for some $k\in\{1,2\}$, under Assumption \ref{as:pointw_multiv}-(ii),
\begin{equation}
	\widehat{V}_k^{-1/2} (\overline{Y} -\theta_0)\convNor{\Id}.	
	\label{eq:conv_nonG_mult}
\end{equation}
Unlike the univariate case, we could be in a non-Gaussian regime even when $\Omega_1+\Omega_2\ne 0$. In such a situation, \eqref{eq:conv_nonG_mult} may not hold when both $\Omega_1\ne 0$ and $\Omega_2\ne 0$. This is why we impose $\Omega_1= 0$ or $\Omega_2= 0$ in Assumption \ref{as:pointw_multiv}-(ii). Second, $V(\overline{Y})^{-1/2}\widehat{V}_kV(\overline{Y})^{-1/2}\convD V_k$, a random matrix which may be singular with positive probability. If so, \eqref{eq:conv_nonG_mult} may not hold, even if $\widehat{V}_k$ is invertible with probability one, and our test may not be asymptotically valid, as illustrated in the following example.

\begin{example}
	\label{ex:counter_multi2}
	Consider the DGP $Y_{ij}= \theta_0 + (U_{i0} +U_{0j}+ U_{ij}, U_{i0} U_{0j})'$, for which Assumption \ref{as:pointw_multiv}-(ii) fails. Then,
	$$V_1= \begin{pmatrix}
		1 & Z \\
		Z & Z^2
	\end{pmatrix},$$
	where $Z\sim\mathcal{N}(0,1)$. $V_2$ has the same distribution as $V_1$. Thus, rank$\,(V_1)=$ rank$\,(V_2)=1$. In this case, simulations show that the asymptotic level of $\phi_{0.05}$ is around 30\%, rather than 5\%.
\end{example}

However, we prove that  under Assumption \ref{as:pointw_multiv}-(ii), $V_k$ is invertible almost surely. To this end, we rely on the following lemma on Gaussian matrices, which may be of independent interest. 

\begin{lemma}\label{lem:invertible_QF}
	If $G$ is a $p_1\times p_2$ matrix such that vec$(G)$ is Gaussian, then $P(\det(G^{\otimes 2})>0)\in\{0,1\}$. Moreover, if $E(G)^{\otimes 2}$ is invertible, $G^{\otimes 2}$ is invertible a.s.
\end{lemma}

Note that in the lemma, we do not impose anything on the covariance between the different entries of vec$(G)$. In particular, we allow the corresponding variance matrix to be singular.


\subsection{Uniform results} 
\label{sub:uniform_multiv}

We adapt the definition of $\tau_{1P}$ (resp. $\tau_{2P}$) in the following manner:
\begin{align}
	\tau_{1P}: \; [0,1]^3 & \to  \R^d \label{eq:def_tau1_mult} \\
	(u_1,u_2,u_3) & \mapsto  (\lambda^*_{\min}(\Omega_{1P}) + \mathds{1}\{\lambda^*_{\min}(\Omega_{1P}) = \infty\})^{-1/2}E_P[W_{11} \mid U_{10} = u_1]\notag,	
\end{align}
where $\lambda^*_{\min}(A)$ denotes the smallest strictly positive eigenvalue of $A$ (with the convention that $\lambda^*_{\min}(0)=\infty$). We define $\tau_{2P}$ similarly. Also, let us consider the following two conditions:
\begin{align}
	\textrm{range}(\Omega_{3P}) \subseteq \textrm{range}(\Omega_{1P}+\Omega_{2P}) & \text{ and } \lambda_{\max}(\Omega_{3P}) \leq m^{-1}\lambda_{\min}^*(\Omega_{1P}+\Omega_{2P}), \label{eq:cond_multiv_G} \\
	\|\Omega_{1P}\|\wedge \|\Omega_{2P}\|  =0 & \text{ and } \lambda_{\min}\left(\Omega_{1P}+\Omega_{2P}+\Omega_{4P}\right)\ge m,
	\label{eq:cond_multiv_NG}
\end{align}
where, for any symmetric matrix $A$, $\lambda_{\min}(A)$ denotes its smallest eigenvalue. Then, for $H$ a compact subset of $L_2([0,1]^3,\R^d)$ and $m>0$, we define
\begin{align*}
	\mathcal{P}_{m,H}^d := & \big\{P\in\P: \; \lambda_{\min}\left(V_P(W_{11})\right) \ge m, \; \exists \tau_P \in H \text{ satisfying  \eqref{eq:AHK}} \\
	&  \text{ s.t. } \tau_{kP}\in H \,(k=1,2) \text{ and either } \eqref{eq:cond_multiv_G} \text{ or } \eqref{eq:cond_multiv_NG} \text{ hold}\,\big\},
\end{align*}
Similarly, let $\mathcal{P}_{m,H}^{d,G} := \{P \in \mathcal{P}_{m,H}^d: \eqref{eq:cond_multiv_G} \text{ holds}\,\}$. 
Note that the first conditions in $\mathcal{P}_{m,H}^d$ are the multivariate counterparts of the first two lines in $\mathcal{P}_{m,H}^1$. Condition \eqref{eq:cond_multiv_G} includes DGPs with Gaussian regimes. Its first part is the analogue of Assumption~\ref{as:pointw_multiv}-(i), while its second part is the multivariate version of the restriction $\Omega_{3P} \le m^{-1}(\Omega_{1P}+\Omega_{2P})$ that we imposed in $\mathcal{P}_{m,H}^1$. Condition \eqref{eq:cond_multiv_NG}, which includes DGPs with non-Gaussian regimes, may be seen as a uniform version of Assumption~\ref{as:pointw_multiv}-(ii).

\medskip
Finally, we also introduce a class of probability distributions on $(Y_{ij})_{i,j\ge 1}$ for which the test based on Bonferroni correction is valid. Hereafter, for any $k\in\{1,...,d\}$, $m>0$ and $H'$ a compact subset of $L_2([0,1]^3,\R)$, let $\mathcal{P}_{m,H'}^{1(k)}$  denote the same set as $\mathcal{P}_{m,H'}^1$ but replacing $Y_{11}$ therein by its $k$-th component, $Y_{11}^{(k)}$:
\begin{align*}
	\mathcal{P}_{m,H'}^{d,b} := & \big\{P\in\P: P \in \mathcal{P}_{m,H'}^{1(k)} \quad \forall k=1,...,d\big\}.
\end{align*}
An advantage of this approach is that for appropriate choices of the compact subsets $H$ and $H'$, the set $\mathcal{P}_{m,H'}^{d,b}$ includes $\mathcal{P}_{m,H}^d$.\footnote{Formally, let $\pi_k(x_1,...,x_d)=x_k$ ($k=1,...,d$) and for $H$ a compact subset of $L_2([0,1]^3,\R^d)$, let $H_k=\pi_k(H)$. Then, we can show that $\mathcal{P}_{m,H}^d\subseteq \mathcal{P}_{m,\cup_{k=1}^d H_k}^{d,b}$.} In other words, we can obtain uniform results on larger sets of DGPs when considering the Bonferroni test. 

\begin{theorem}
	\label{thm:unif_multiv}
	Fix $m>0$ and $H$ (\textit{resp.} $H'$) a compact subset of $L_2([0,1]^3,\R^d)$ (\textit{resp.} $L_2([0,1]^3,\R)$). If Assumption~\ref{as:asym} holds, $\phi_\alpha$ and $\phi^b_\alpha$  be as in \eqref{eq:def_Ftest} and \eqref{eq:def_bonf_test}, respectively and $\theta=\theta_{0P}$, we have, for any $\alpha \in (0,1)$,
	\begin{equation}
		\limsup_{n \to \infty} \sup_{P \in \mathcal{P}_{m,H}^d} \E_P[\phi_\alpha] \le \alpha.
		\label{eq:valid_unif_multiv}
	\end{equation}
	Moreover, if $\mathcal{P}_{m,H}^{d,G}\ne\emptyset$,
	\begin{equation}
		\limsup_{n \to \infty} \sup_{P \in \mathcal{P}_{m,H}^{d,G}} \E_P[\phi_\alpha] =
		\liminf_{n \to \infty} \inf_{P \in \mathcal{P}_{m,H}^{d,G}} \E_P[\phi_\alpha]= \alpha.
		\label{eq:exactness_unif_multiv}
	\end{equation}
	Finally,
	\begin{equation}
		\limsup_{n \to \infty} \sup_{P \in \mathcal{P}_{m,H'}^{d,b}} \E_P[\phi^b_\alpha] \le \alpha,
		\label{eq:valid_unif_multiv_bon}
	\end{equation}
\end{theorem}		

With the same reasoning as in the univariate case, we obtain that Theorem \ref{thm:unif_multiv} generalizes the pointwise result above.





\subsection{Simulations on multivariate sample means} 
\label{sub:multivariate_sample_means}

We investigate the performance of our inference procedures for multivariate expectations on simulations. We consider the following DGP:
$$Y_{ij}=\left(U_{i0}, \delta_{1n} \widetilde{U}_{i0} + \delta_{2n} \widetilde{U}_{0j} + \widetilde{U}_{i0} \widetilde{U}_{0j}+\frac{1}{2} \widetilde{U}_{ij}\right)$$
and $(U_{ij})_{i,j\ge 0}$ and $(\widetilde{U}_{ij})_{i,j\ge 0}$ are two independent families of i.i.d. standard normal variables. Again, we consider four DGPs, depending on the values of $(\delta_{1n}, \delta_{2n})$:
\begin{enumerate}
	\item $\delta_{1n}=\delta_{2n}=1$. Assumption \ref{as:pointw_multiv}-(i) holds and this DGP is fixed so Theorem \ref{thm:point_multiv} applies  and our test is asymptotically exact;
	\item $\delta_{1n}=\delta_{2n}=0$. The DGP is fixed and Assumption \ref{as:pointw_multiv}-(ii) holds. Theorem \ref{thm:point_multiv} still applies but our test is expected to be asymptotically conservative;
	\item $\delta_{1n}=1/\sqrt{n}$, $\delta_{2n}=0$. $\Omega_{2P_n}$ is null and $\Omega_{1P_n}$ is invertible. The  DGPs all belong to $\mathcal{P}^2_{m,H,\mathcal{Q}}$ with $\mathcal{Q}=\mathcal{P}$ and for some $m>0$, compact set $H$ and $n$ large enough. Our uniformity result (Theorem \ref{thm:unif_univ}) applies, but our test is expected to be asymptotically conservative;
	\item $\delta_{1n}=\delta_{2n}=1/\sqrt{n}$. $\lambda_{\max}(\Omega_{3P_n})=1$, $\lambda_{\min}(\Omega_{1P_n}+\Omega_{2P_n})\to 0$ and $\|\Omega_{1P_n}\|\wedge\|\Omega_{2P_n}\|>0$, so neither \eqref{eq:cond_multiv_G} nor \eqref{eq:cond_multiv_NG} holds. We cannot apply our uniformity result to the corresponding  DGPs and we have no size guarantee in this case.
\end{enumerate}

As above, we compute rejection rates under the null, by testing for $\theta_0=0$, and under the alternative, by testing for $\theta_0=\theta\ne 0$, with  $\theta=(0.3,0.3)'$ in DGP1 and $\theta=(0.125,0.125)'$ in DGP2 to DGP4. Because Menzel only briefly discusses joint hypothesis testing without developing a code for such tests, we focus hereafter on the comparison between our method and the usual one. Note that for the latter, in the absence of any recommendation by CGM, we simply consider the test $\widetilde{F}_u$, namely, we reject when the ``standard'' F-statistic $F_u$ is negative. We also consider multiple t-tests with a
Bonferroni correction (DDG-B below).

\medskip
The results are displayed in Table \ref{tab:multiv_mean}. As in the univariate case, we observe very similar behaviours of our test and the usual one in the first DGP. On the other hand, in DGP2 to DGP4, the usual method exhibits important distortions, with rejection rates betwen 15 and 25\% under the null. DDG does not seem to be overly conservative in DGP2 and DGP3. As expected, DDG-B is slightly more conservative in these two DGPs, but the power loss seems moderate when $n=40$. In DGP4 for which we do not have theoretical guarantees, DDG displays some moderate overrejection, while for $n=40$, DDG-B exhibits a rejection rate under the null that is close to the nominal level.

\begin{table}[H]
	\begin{center}
		\begin{tabular}{cc|cccc|ccc}
			& & \multicolumn{4}{c|}{Level} & \multicolumn{3}{c}{Power} \\
			DGP & $n$ & DDG & DDG-B & Usual & $P(F_u<0)$ &  DDG  & DDG-B & Usual \\ \toprule
			1 & 10 & 0.165 & 0.138 & 0.170 & 0.000 & 0.348 & 0.309 & 0.356\\
			1 & 20 & 0.101 & 0.093 & 0.102 & 0.000 & 0.412 & 0.374 & 0.414\\
			1 & 40 & 0.079 & 0.071 & 0.079 & 0.000 & 0.604 & 0.550 & 0.605\\ \midrule
			2 & 10 & 0.099 & 0.074 & 0.260 & 0.139 & 0.326 & 0.226 & 0.502\\
			2 & 20 & 0.055 & 0.048 & 0.233 & 0.156 & 0.663 & 0.574 & 0.739\\
			2 & 40 & 0.038 & 0.038 & 0.206 & 0.148 & 0.918 & 0.904 & 0.927\\ \midrule
			3 & 10 & 0.098 & 0.071 & 0.221 & 0.099 & 0.277 & 0.190 & 0.419\\
			3 & 20 & 0.055 & 0.047 & 0.182 & 0.103 & 0.551 & 0.493 & 0.637\\
			3 & 40 & 0.038 & 0.035 & 0.156 & 0.096 & 0.850 & 0.835 & 0.866\\ \midrule
			4 & 10 & 0.157 & 0.101 & 0.264 & 0.119 & 0.380 & 0.309 & 0.448\\
			4 & 20 & 0.096 & 0.065 & 0.197 & 0.099 & 0.491 & 0.445 & 0.525\\
			4 & 40 & 0.085 & 0.060 & 0.194 & 0.107 & 0.754 & 0.734 & 0.763\\
			\bottomrule
			\multicolumn{9}{p{400pt}}{{\footnotesize Notes: $C_1=C_2=n$, nominal level: 5\%. For power, $\theta=(0.3,0.3)'$ in DGP1 and $\theta=(0.125,0.125)'$ in DGP2-4. Results obtained over 5,000 samples for each of the 12 cases.}}
		\end{tabular}
	\end{center}
	\caption{Performances of the tests on a multivariate expectation}
	\label{tab:multiv_mean}
\end{table}


\section{Proofs} 
\label{sec:proofs}

We use the following notation in the proofs. We use $\|\cdot\|$ to denote the Euclidean norm for vectors, and the Frobenius norm for matrices. We let $\mathrm{S}^d_+$ (resp. $\mathrm{S}^d_{++}$) denote the set of symmetric positive semidefinite (resp. definite) $d\times d$ matrices. For any $p\in\N$, $p\ge 1$, we let $\N^{p\ast} := \N^p\backslash\{(0,...,0)\}$. Elements of $\mathbb{N}^3$ are denoted by $\k=(k_1,k_2,k_3)$. We also let
$$\ell^2_d := \set{(u_{\k})_{\k\in\N^{3}}: u_{\k} \in \R^d,\; \sum_{\k\in\N^{3}} \|u_{\k}\|^2<\infty}.$$
With a slight abuse of notation, we may write $(u_{\k})_{\k\in\N^{3\ast}} \in \ell^2_d$ for some $(u_{\k})_{\k\in\N^{3\ast}}$. Then, one should understand that we implicitly extend $(u_{\k})_{\k\in\N^{3\ast}}$ by letting $u_{(0,0,0)}=(0,...,0)$.

	\subsection{Two key lemmas}
	\label{ssec:key_lemmas}
	The following lemmas are crucial for the proofs of our main results. Their proofs appear in the Supplemental Appendix.

	\subsubsection{A representation lemma}
	
Our first lemma is a representation result on $W_{ij} :=  Y_{ij}-E[Y_{11}]$. Let $\psi_0(x)=1$ and $\psi_k(x)=\sqrt{2}\cos(k\pi x)$ for $k\ge 1$. The functions $(\psi_k)_{k\in\N}$ form an orthonormal basis for the Hilbert space $L^2[0,1]$, when considering the usual scalar product $\langle f,g\rangle=\int_0^1 f(x)g(x)dx$.

	\begin{lemma}
		\label{lem:mu_nu}
		Suppose that Assumption~\ref{as:DGP} holds. Then, there exists $\tau \in L_2([0,1]^3,\Rb^d)$, $\bm{\mu}^n := (\mu^n_{\k})_{\k\in\N^{3}} \in \ell^2_d$ and mutually independent standard uniform variables $(U_{ij})_{(i,j)\in\N^2}$ 
		such that for all $(i,j)\in \N^{*2}$, almost surely
		\begin{equation}
			W_{ij} = \tau(U_{i0},U_{0j},U_{ij}) = \sum_{\k \in \N^3}\mu^n_{\k}\psi_{k_1}(U_{i0})\psi_{k_2}(U_{0j})\psi_{k_3}(U_{ij}).
			\label{eq:AHK_l2}
		\end{equation}
		Moreover, $V(\overline{Y})$ is invertible, which implies the following representation holds true almost surely (with $V_n := V(\overline{Y})^{-1/2}$)
		\begin{equation}
			V_nW_{ij} = \sum_{\k \in \N^3}\left(C_1^{\ind{k_1+ k_3>0}}C_2^{\ind{k_2+ k_3>0}}\right)^{1/2}\nu^n_{\k}\psi_{k_1}(U_{i0})\psi_{k_2}(U_{0j})\psi_{k_3}(U_{ij})
			\label{eq:AHK_l2_standardized}
		\end{equation}		
		with
		\begin{equation}
			\nu_{\k}^n := V_n\frac{\mu^n_{\k}}{\left(C_1^{\ind{k_1+ k_3>0}}C_2^{\ind{k_2+ k_3>0}}\right)^{1/2}} \quad \forall \k\in\N^{3}.
			\label{eq:def_nu}
		\end{equation}
	\end{lemma}
	
Equation \eqref{eq:AHK_l2} is obtained by the AHK representation \eqref{eq:AHK} with uniform variables, and the decomposition of the $L_2$-integrable function $\tau$ on the basis $(\psi_k)_{k\in\N}$. Lemma~\ref{lem:mu_nu} provide two different parametrizations of the distribution of $\phi_\alpha$, either in terms of  $\bm{\mu}^n$ or $\bm{\nu}^n$. In subsequent proofs, we alternate between these two parametrizations, since each happens to be particularly convenient in different situations.

\subsubsection{A weak convergence lemma}
\label{ssub:weak_conv_seq}	
	
	Let us define the sets $(\K_j)_{j=1,...,4}$, which form a partition of $\N^{3\ast}$:
	\begin{align*}
		\K_1 &:= \left\{ \k \in \mathbb{N}^{3\ast} : k_1 > 0,\;  k_2 = k_3 = 0 \right\}, \\
		\K_2 &:= \left\{ \k \in \mathbb{N}^{3\ast} : k_2 > 0,\; k_1 = k_3 = 0 \right\}, \\
		\K_3 &:= \{ \k \in \mathbb{N}^{3\ast} : k_1>0, \;k_2>0, \; k_3 = 0 \}, \\
		\K_4 &:= \left\{ \k \in \mathbb{N}^{3\ast} : k_3 > 0 \right\}.	
	\end{align*}
	
	\begin{lemma}
		\label{lem:weak_conv_means}
		Suppose that Assumptions~\ref{as:DGP} and~\ref{as:asym} hold and let $\bm{\nu}^n$ satisfy~\eqref{eq:def_nu} in Lemma~\ref{lem:mu_nu}. If $\bm{\nu}^n\convLtwo \bm{\nu}^\infty$ for some $\bm{\nu}^\infty$, we have:
		$$\bigg[V_n \overline{W}, V_n(\widehat{V}_1, \widehat{V}_2,  \widehat{V}_{12})V_n\bigg] \convD (L, V_1, V_2, V_{12}),$$
		where, letting $(Z_{\k})_{\k \in \N^{3*}}$ denote an array of i.i.d. standard normal variables,
		\begin{align*}
			L & := \sum_{\k\in\K_1}\nu_{\k}^{\infty}Z_{\k} + \sum_{\k\in\K_2}\nu_{\k}^{\infty}Z_{\k} 
			+ \sum_{\k \in \K_3}\nu_{\k}^{\infty}Z_{k_1,0,0}Z_{0,k_2,0} + \sum_{\k \in \K_4}\nu_{\k}^{\infty}Z_{\k}, \\
			V_1  & :=   \sum_{\k\in\K_4}(\nu_{\k}^{\infty})^{\otimes 2} + \sum_{k_1>0}\left( \nu_{k_1,0,0}^{\infty} + \sum_{k_2 > 0}\nu_{k_1,k_2,0}^{\infty}Z_{0,k_2,0} \right)^{\otimes 2} \\
			V_2 & :=   \sum_{\k\in\K_4}(\nu_{\k}^{\infty})^{\otimes 2} +  \sum_{k_2>0}\left( \nu_{0,k_2,0}^{\infty} + \sum_{k_1 > 0}\nu_{k_1,k_2,0}^{\infty}Z_{k_1,0,0} \right)^{\otimes 2} \\
			V_{12} &:=   \sum_{\k\in\K_{3}\cup\K_{4}} (\nu_{\k}^{\infty})^{\otimes 2}.
		\end{align*}	
	\end{lemma}

	\subsection{Proof of the main results and key lemmas}
We recall that $V_n=V(\overline{Y})^{-1/2}$. For $i=1,...,4$, we also let  $\K_i(\overline{k}) := \K_i\cap[0,\overline{k}]^3$, $\K_{i_1...i_p} := \cup_{k=1}^p \K_{i_k}$ and define similarly $\K_{i_1...i_p}(\overline{k})$. Hereafter, we suppose wlog that $\underline{C}:=C_1\wedge C_2\ge 2$ for all $n\ge 1$. The norm symbol $||\cdot||$ stands for the Euclidean norm for vectors and the matrix 2-norm for matrices.
	
	\subsubsection{Lemma~\ref{lem:decomp}}
			
	Theorem 3.2., Statement b in \cite{kallenberg1989} ensures that $\mathcal{S}_1\subset \sigma(U_{i0}:i\geq1)$ and $(W_{ij})_{i,j\geq 1}\indep (U_{i0})_{i\geq 1}|\mathcal{S}_1$. It follows that $E(W_{ij}|U_{i0})=E(W_{ij}|(U_{i'0})_{i'\geq 1})=E(W_{ij}|(U_{i'0})_{i'\geq 1},\mathcal{S}_1)=E(W_{ij}|\mathcal{S}_1)$. Similarly, $E(W_{ij}|U_{0j})=E(W_{ij}|\mathcal{S}_2)$. Theorem 3.2., Statement c in \cite{kallenberg1989} ensures that $\mathcal{S}_{12}\subset \sigma((U_{i0},U_{0j}):i\geq1, j\geq 1)$ and $(W_{ij})_{i,j\geq 1}\indep (U_{i0},U_{0j})_{i,j\geq 1}|\mathcal{S}_{12}$. Next, $E(W_{ij}|U_{i0},U_{0j})=E(W_{ij}|(U_{i'0},U_{0j'})_{i',j'\geq 1})=E(W_{ij}|(U_{i'0},U_{0j'})_{i',j'\geq 1},\mathcal{S}_{12})=E(W_{ij}|\mathcal{S}_{12})$.
	
\medskip
Because $\Cov(W_{11},W_{12}|U_{10})=0$, we have	
$$\Omega_1=V(E(W_{11}|U_{10}))=\Cov(E(W_{11}|U_{10}),E(W_{12}|U_{10}))=\Cov(W_{11},W_{12}).$$
Similarly, $\Omega_2=\Cov(W_{11},W_{12})$. Next, $\Cov(\alpha_i,\beta_j)=0$ by independence of $U_{i0}$ and $U_{0j}$ and then \begin{align*}\Cov(\gamma_{ij},\alpha_i)&=\Cov(E(W_{ij}|U_{i0},U_{0j}),\alpha_i)-V(\alpha_i)\\
	&=E\left[E(W_{ij}|U_{i0},U_{0j})E(W_{ij}|U_{i0})'\right]-E(W_{11})^{\otimes 2}-V(\alpha_i)\\
	&=E(E(W_{ij}|U_{i0})^{\otimes 2})-E(W_{11})^{\otimes 2}-V(\alpha_i)=0\end{align*}
	and similarly $\Cov(\gamma_{ij},\beta_j)=0$. And by definition of $\varepsilon_{ij}$, we have:
	\begin{align*}
	\Cov(\varepsilon_{ij},\alpha_i)&=\Cov(W_{ij},\alpha_i)-V(\alpha_i)\\&=E\left[(W_{ij}-\mu_0)\alpha_i'\right]-E(\alpha_i\alpha_i')\\
	&=E\left[E(W_{ij}-\mu_0|U_{i0})\alpha_i'\right]-E(\alpha_i\alpha_i')=0,
	\end{align*}
	and similarly, $\Cov(\epsilon_{ij},\beta_j)=0$. The last covariance term is
	\begin{align*}
	\Cov(\varepsilon_{ij},\gamma_{ij})&=\Cov(W_{ij},\gamma_i)-V(\gamma_{ij})\\&=E\left[(W_{ij}-E(W_{11}))\gamma_{ij}'\right]-V(\gamma_{ij})\\
	&=E\left[E(W_{ij}-E(W_{11})|U_{i0},U_{0j})\gamma_{ij}'\right]-V(\gamma_{ij})\\
	&=V(\gamma_{ij})+\Cov(\alpha_i,\gamma_{ij})+\Cov(\beta_j,\gamma_{ij})-V(\gamma_{ij})=0.
	\end{align*}
	This ensures that $V(W_{11})=\Omega_1+\Omega_2+\Omega_3+\Omega_4$. $\Box$

\subsubsection{Theorem~\ref{thm:point_univ}}

We prove that Theorem~\ref{thm:point_univ} follows from Theorem~\ref{thm:unif_univ}. With $P$ the probability distribution of $\bm{Y}$, let $\tau_P$ satisfy \eqref{eq:AHK}. Then, define $\tau_{1P}$ according to \eqref{eq:def_tau1} and let $\tau_{2P}$ be defined accordingly. Finally, let $H := \{\tau_P,\tau_{1P},\tau_{2P}\}$. Because $H$ is finite, it is compact. Next, if $\Omega_{1P} + \Omega_{2P} = 0$ or $\Omega_{3P}=0$, let $m := V_P(Y_{11})$; otherwise, let $m := \min(V_P(Y_{11}), (\Omega_{1P} + \Omega_{2P})/\Omega_{3P})$. By construction, $P\in \mathcal{P}_{m,H,\RP}^1$, and \eqref{eq:pointwise_valid}, the first result of Theorem \ref{thm:point_univ},  follows from \eqref{eq:valid_unif}. Moreover, if $\Omega_{1P} + \Omega_{2P} > 0$ or $\Omega_{3P}=0$, then $\Omega_{3P} \le m^{-1} (\Omega_{1P}+\Omega_{2P})$, and thus $P\in \mathcal{P}_{m,H,\RP}^{1,G}$. Hence, \eqref{eq:pointwise_exact}, the second result of Theorem \ref{thm:point_univ},  follows from \eqref{eq:exactness_unif}. $\Box$

\subsubsection{Theorem~\ref{thm:unif_univ}} 

Let us fix a sequence $(P_n)_{n\ge 1}$ in $\mathcal{P}_{m,H}^1$. \eqref{eq:valid_unif} follows if we prove
\begin{align}
	\limsup_{n\to\infty} E_{P_n}[\phi_\alpha] & \le \alpha. \label{eq:for_thm_unif_seq}
\end{align}
To this end, let us consider a subsequence $(P_{\varphi(n)})_{n\ge 1}$. By Lemma \ref{lem:mu_nu}, $(P_n)_{n\ge 1}$ (resp. $(P_{\varphi(n)})_{n\ge 1}$) is associated to a sequence $(\bm{\nu}^n)_{n\ge 1}$ (resp. a subsequence $(\bm{\nu}^{\varphi(n)})_{n\ge 1}$). Now, the proof of Theorem~\ref{thm:unif_univ} is divided in four steps. First, we first prove that there exists a further subsequence $(\bm{\nu}^{\psi(n)})_{n\ge 1}$ that converges in $\ell^2_1$. Second, we show that along $\psi(\cdot)$, $(\overline{Y}-\theta_0)/s\convNor{1}$ for some $s\le \se$. The third step proves that \eqref{eq:for_thm_unif_seq} and thus \eqref{eq:valid_unif} hold. Finally, the fourth step shows how to adapt the reasoning to prove \eqref{eq:exactness_unif}.\\
\textbf{Step 1: existence of a further subsequence $(\bm{\nu}^{\psi(n)})_{n\ge 1}$ converging in $\ell^2_1$.}\\ 
For a fixed $n$, let $\mathcal{V}^n_{m,H}$ denote the set of $\bm{\nu}^n$ corresponding to $\mathcal{P}_{m,H}^1$ (we refer to Lemma \ref{lem:compacity_mu_nu} for a more formal definition of $\mathcal{V}_{m,H}^n$) and let $\mathcal{V}_{m,H} := \cup_{n\ge 1} \mathcal{V}_{m,H}^n$. By Lemma~\ref{lem:compacity_mu_nu}, the closure of $\mathcal{V}_{m,H}$ ($\overline{\mathcal{V}_{m,H}}$) is compact in $\ell^2_1$. Thus, $(\bm{\nu}^{\varphi(n)})_{n\ge 1}$, as a sequence in $\overline{\mathcal{V}_{m,H}}$, admits a converging subsequence, $(\bm{\nu}^{\psi(n)})_{n\ge 1}$ say.\\
\textbf{Step 2: along $\psi(\cdot)$, $(\overline{Y}-\theta_0)/s\convNor{1}$ for some $s\le \se$.}~\\ 
We reason here along the subsequence $\psi(\cdot)$. We denote by $\bm{\nu}^\infty$ the limit of $(\bm{\nu}^{\psi(n)})_{n\ge 1}$ and
$$M_n:= (V_{n}(\overline{Y}-\theta_0), V_{n}^2 \widehat{V}_1, V^2_{n} \widehat{V}_2, V^2_{n} \widehat{V}_u).$$
Lemma~\ref{lem:weak_conv_means} and the continuous mapping theorem (CMT) yield
\begin{align}
	M_{\psi(n)} \convD (L, V_1, V_2, V_u),   \label{eq:conv_dist_theta}
\end{align}
where $(L, V_1, V_2)$ is defined in Lemma \ref{lem:weak_conv_means} and $V_u:=V_1+V_2-V_{12}$.\\
Now, let $\Sigma_j(\bm{\nu}) := \sum_{\k \in \K_j}\nu_{\k}^{2}$ for $j=1,\dots,4$. Since $\Sigma_j(\cdot)$ is continuous, $\Sigma_j(\bm{\nu}^{\psi(n)}) \to\Sigma_j^\infty := \Sigma_j(\bm{\nu}^\infty)$ for $j=1,\dots,4$. Then, Lemma~\ref{lem:limit_behaviour_nu} ensures $\min\left(\Sigma_1^\infty,\Sigma_2^\infty,\Sigma_3^\infty\right) = 0$.\\
Suppose first that $\Sigma_3^\infty=0$. By Lemma \ref{lem:weak_conv_means} again, $L\sim\mathcal{N}(0,1)$ and
$$(V_1,V_2,V_u)=(\Sigma_1^\infty+\Sigma_4^\infty, \Sigma_2^\infty+\Sigma_4^\infty,  \Sigma_1^\infty+\Sigma_2^\infty+\Sigma_4^\infty).$$
Moreover, since $\Sigma_1^n+\Sigma_2^n+\Sigma_3^n+\Sigma_4^n=1$ and $\Sigma_3^\infty=0$,  $\Sigma_1^\infty+\Sigma_2^\infty+\Sigma_4^\infty=1$. Hence, by the CMT again,
\begin{equation}
	\frac{\overline{Y}-\theta_0}{\se} \convNor{1}, \quad \frac{\se_u}{\se} \convP 1.	
	\label{eq:equiv_usual}
\end{equation}
Now, assume that $\Sigma_3^\infty>0$ and suppose wlog that $\Sigma_2^\infty=0$ (the reasoning is the same with $\Sigma_1^\infty=0$). In this case $L = \sum_{\boldsymbol{k} \in \K_{1}\cup\K_4}\nu_{\k}^{\infty}Z_{\k} + \sum_{\k \in \K_3}\nu_{\k}^{\infty}Z_{k_1,0,0}Z_{0,k_2,0}$ and
\begin{align*}
	V_1 = & \sum_{\k\in\K_4}(\nu_{\k}^{\infty})^{2} + \sum_{k_1>0}\left( \nu_{k_1,0,0}^{\infty} + \sum_{k_2 > 0}\nu_{k_1,k_2,0}^{\infty}Z_{0,k_2,0} \right)^2.
\end{align*}
Then, $L|(Z_{0,k_2,0})_{k_2>0} \sim \mathcal{N}(0,V_1)$. Moreover, because $\Sigma_3^\infty>0$, there exists $\nu_{k^*_1,k^*_2,0}^{\infty}\ne 0$. Thus,
$$\nu_{k^*_1,0,0}^{\infty} + \sum_{k_2 > 0}\nu_{k^*_1,k_2,0}^{\infty}Z_{0,k_2,0} \sim \mathcal{N}\left(\nu_{k^*_1,0,0}^{\infty}, \sum_{k_2 > 0}\nu_{k^*_1,k_2,0}^{\infty\, 2}\right),$$
with $\sum_{k_2 > 0}\nu_{k^*_1,k_2,0}^{\infty\, 2}>0$. As a result, $P(V_1>0)=1$. From this, the CMT and \eqref{eq:conv_dist_theta}, we obtain
$$\frac{\overline{Y}-\theta_0}{\se_1} \convNor{1}.$$
\textbf{Step 3: \eqref{eq:for_thm_unif_seq} and \eqref{eq:valid_unif} hold.}\\ 
By Step 2 and the definition of $\phi_\alpha$, the sequence $v_n := \max(E_{P_n}[\phi_\alpha], \alpha)$ satisfies $\lim_{n \to \infty} v_{\psi(n)}\geq\alpha$. This is straightforward when $\Sigma_3^\infty=0$. Otherwise, this follows from $E_{P_n}[\phi_{1\alpha}]\to \alpha$ (with $\phi_{1\alpha}$ defined as $\phi_{\alpha}$, but using $\se_1$ instead of $\se$) and $\phi_{\alpha}\le \phi_{1\alpha}$ a.s. 
Because the initial subsequence was arbitrary, we obtain, by Urysohn's principle, $\lim_{n \to \infty} v_n\geq\alpha$. This is the same as \eqref{eq:for_thm_unif_seq}. Equation \eqref{eq:valid_unif} follows.\\
\textbf{Step 4: Equation \eqref{eq:exactness_unif} holds.}\\  
The reasoning is very similar. In this case, $(P_n)_{n\ge 1}$ is a sequence in $\mathcal{P}_{m,H}^{1,G}$. Step 1 still holds in this case. In Step 2, Lemma \ref{lem:limit_behaviour_nu} shows that $\Sigma_3^\infty =0$. Then, as shown in Step 2, $(\overline{Y}-\theta_0)/\se\convNor{1}$, and $v_n := E_{P_n}[\phi_\alpha]$ satisfies $\lim_{n \to \infty} v_{\psi(n)}=\alpha$. By Urysohn's principle again, $\lim_{n \to \infty} v_n=\alpha$.  Equation \eqref{eq:exactness_unif_multiv} follows. $\Box$

\subsubsection{Theorem \ref{thm:point_multiv}} 
\label{ssub:theorem_ref_thm_point_multiv}

We proceed as in the proof of Theorem~\ref{thm:point_univ}. We simply need to check that there exists $m$ such that $m\le \lambda_{\min}(V_P(Y_{11}))$ and either
\begin{equation}
\left[\textrm{range}(\Omega_3)\subseteq \textrm{range}(\Omega_1+\Omega_2) \text{ and } \lambda_{\max}(\Omega_{3P}) \leq m^{-1}\lambda_{\min}^*(\Omega_{1P}+\Omega_{2P})\right]	
	\label{eq:cond1_for_corr_mult}
\end{equation}
or $\big[\|\Omega_{1P}\|\wedge \|\Omega_{2P}\|=0$ and $\lambda_{\min}\left(\Omega_{1P}+\Omega_{2P}+\Omega_{4P}\right)\ge m\big]$. Suppose first that Assumption \ref{as:pointw_multiv}-(i) holds. If $\lambda_{\max}(\Omega_{3P})>0$, let
$$m := \min\left(\lambda_{\min}(V_P(Y_{11})), \frac{\lambda_{\min}^*(\Omega_{1P}+\Omega_{2P})}{\lambda_{\max}(\Omega_{3P})}\right),$$
otherwise, simply let $m := \lambda_{\min}(V_P(Y_{11}))$. In this latter case, \eqref{eq:cond1_for_corr_mult} obviously holds. In the former case, \eqref{eq:cond1_for_corr_mult} holds as well since $m>0$. In both cases, we also have $m\le \lambda_{\min}(V_P(Y_{11}))$.

\medskip

Now, suppose that Assumption \ref{as:pointw_multiv}-(ii) holds. Let
$$m := \min\left(\lambda_{\min}(V_P(Y_{11})), \lambda_{\min}(\Omega_{1P}+\Omega_{2P}+\Omega_{4P})\right).$$
Assumption \ref{as:pointw_multiv}-(ii) ensures that $m>0$. Moreover, $m\le \lambda_{\min}(V_P(Y_{11}))$ and $\big[\|\Omega_{1P}\|\wedge \|\Omega_{2P}\|=0$ and $\lambda_{\min}\left(\Omega_{1P}+\Omega_{2P}+\Omega_{4P}\right)\ge m\big]$. $\Box$


	

\subsubsection{Lemma~\ref{lem:invertible_QF}}

If $p_2<p_1$ then $G^{\otimes 2}$ and $E(G)^{\otimes 2}$ are singular and we have nothing to prove. For $p_2\geq p_1$, let $G_1,...,G_{p_2}$ the columns of $G$ and let $\mu$ and $\Sigma$ the expectation and covariance matrix of $(G_1',...,G_{p_2}')'$. Let $P'\Delta P$ the singular value decomposition of $\Sigma$ with $P$ an orthogonal matrix and $\Delta$ a non negative diagonal matrix. Let $\overline{u}_i\in \mathbb{R}^{p_1p_2}$, for $i=1,...,p_1p_2$, denote the columns of $P' \Delta^{1/2}$. For $Z\sim \mathcal{N}(0,\textrm{I}_{p_1p_2})$ we have $(G_1',...,G_{p_2}')'\stackrel{d}{=}\sum_{i=1}^{p_1p_2}\overline{u}_iZ_i+\mu$ with $\overline{u}'_i\overline{u}_{j}=0$ if $i\neq j$ and $\sum_{i=1}^{p_1p_2}\overline{u}_i\overline{u}'_i=\Sigma$. To prove $P(\det(GG')=0)\in\{0,1\}$, we can assume wlog that $(G_1',...,G_{p_2}')'=\sum_{i=1}^{p_1p_2}\overline{u}_iZ_i+\mu$.

\medskip
Let $\mu_j\in \mathbb{R}^{p_1}$ such that $\mu=(\mu'_1,...,\mu'_{p_2})'$. Let $u_{ji}\in \mathbb{R}^{p_1}$ for $j=1,...,p_2$ such that $\overline{u}_i=(u_{1i}',...,u_{p_2i}')'$ and for notational convenience let $u_{j0}=\mu_j$ and $Z_{0}=1$. Note that $G_j=\sum_{i=1}^{p_1p_2}u_{ji}Z_i+\mu_j=\sum_{i=0}^{p_1p_2}u_{ji}Z_i$ and $GG'=\sum_{j=1}^{p_2}G_jG_j'=\sum_{j=1}^{p_2}\left(\sum_{i=0}^{p_1p_2}u_{ji}Z_i\right)\left(\sum_{i=0}^{p_1p_2}u'_{ji}Z_{i}\right)$. The event $\det(GG')=0$ is equivalent to
\begin{align*}Q(Z_1,...,Z_{p_1p_2})
	&=\det\left(\sum_{j=1}^{p_2}\left(\sum_{i=0}^{p_1p_2}u_{ji}Z_i\right)\left(\sum_{i=0}^{p_1p_2}u'_{ji}Z_{i}\right)\right) \\
	&=\det\left(\sum_{i=0}^{p_1p_2}\sum_{i'=0}^{p_1p_2}Z_iZ_{i'}\sum_{j=1}^{p_2}u_{ji}u'_{ji'}\right)=0.
\end{align*}
$Q$ is a polynomial of $p_1p_2$ independent Gaussian of degree lower or equal to $2p_1p_2$. The set of roots of a non-zero polynomial has zero Lebesgue measure (this can be easily shown by induction on the number of variables, using Fubini Theorem). It follows that $P\left(\det(GG')=0\right)=1$ if $Q(z_1,...,z_{p_1p_2})=0$ for any $(z_1,...,z_{p_1p_2})\in \mathbb{R}^{p_1p_2}$ and $P\left(\det(GG')=0\right)=0$ otherwise.

\medskip

Now assume that $E[G]^{\otimes 2}$ is invertible. If the $u_{ji}$ vectors are all null, $G = E[G]$ and there is nothing to prove. We thus focus on the alternative scenario. The key step consists in proving that $$P(||G^{\otimes 2} - E[G]^{\otimes 2}|| \le \lambda_{\min}(E[G]^{\otimes 2})/2) > 0.$$
Adding and substracting terms and using the triangle inequality and submultiplicativity of the matrix 2-norm, we remark that $||G^{\otimes 2} - E[G]^{\otimes 2}|| \le 2||E[G]||\,||E[G]-G|| + ||E[G]-G||^2.$ Let $u_{ji}^{(\ell)}$ denote the $\ell$-th entry of the vector $u_{ji}$ and $|u|_\infty := \max_{(i,j,\ell) \in [1,\dots,p_1p_2]\times[1,\dots,p_2]\times[1,\dots,p_1]}|u_{ji}^{(\ell)}|$. We have
$$||E[G]-G|| \le p_1p_2|u|_\infty\max_{1\leq i \leq p_1p_2}|Z_i|.$$
Since $(Z_i)_{i=1}^{p_1p_2}$ are independent standard normal variables, we have, for all $\varepsilon>0$, $P(\max_{1\leq i \leq p_1p_2}|Z_i| \leq \varepsilon) = (1-2\Phi(-\varepsilon))^{p_1p_2}>0$. When $\max_{1\leq i \leq p_1p_2}|Z_i| \leq \varepsilon$, remark that $||G^{\otimes 2} - E[G]^{\otimes 2}|| \le 2||E[G]||\varepsilon + \varepsilon^2$. Moreover, $||E[G]||>0$ since $E[G]^{\otimes 2}$ is invertible. Then, choosing $\varepsilon = \lambda_{\min}(E[G]^{\otimes 2})/(8||E[G]||)$, we can check that  $||G^{\otimes 2} - E[G]^{\otimes 2}|| \le \lambda_{\min}(E[G]^{\otimes 2})/2$ and this event happens with strictly positive probability. On this event, thanks to Weyl's inequalities, we have $\lambda_{\min}(G^{\otimes 2}) \ge \lambda_{\min}(E[G]^{\otimes 2})/2>0$. As a result, $P(\det(G^{\otimes 2})\neq 0)>0$, and thus $P(\det(G^{\otimes 2})\neq 0)=1$, meaning that $G^{\otimes 2}$ is invertible almost surely. $\Box$

\subsubsection{Theorem \ref{thm:unif_multiv}} 
\label{ssub:thm_inf_unif_multiv}

To obtain \eqref{eq:valid_unif_multiv} and \eqref{eq:exactness_unif_multiv} (\eqref{eq:valid_unif_multiv_bon} is proved below), the structure of the proof is the same as that of Theorem \ref{thm:unif_univ}, and only Step 2 needs to be modified. We now prove that $\widetilde{F} \convD \chi^2(d)$ for some $F$-statistic $\widetilde{F}$ satisfying $\widetilde{F}\le F$. We let, as before
$$M_n:=(V_{n}(\overline{Y}-\theta_0), V_n \widehat{V}_1 V_n, V_{n} \widehat{V}_2, V_{n}, V_{n} \widehat{V}_u V_{n}).$$
By Lemma \ref{lem:weak_conv_means}, $M_{\psi(n)}\convD (L,V_1,V_2,V_u)$. As above, we analyze separately the cases $\Sigma_3^{\infty}= 0$ and $\Sigma_3^{\infty}\ne 0$.\\
\textbf{Case $\Sigma_3^{\infty}= 0$.}\\
As in the proof of Theorem \ref{thm:unif_univ}, $L\sim\mathcal{N}(0,\Id)$ and
$$(V_1,V_2,V_u)=(\Sigma_1^\infty+\Sigma_4^\infty, \Sigma_2^\infty+\Sigma_4^\infty,  \Sigma_1^\infty+\Sigma_2^\infty+\Sigma_4^\infty).$$
Moreover, $\Sigma_3^{\infty}= 0$ implies that $\Sigma_1^\infty+\Sigma_2^\infty+\Sigma_4^\infty=\Id$. Let us define
\begin{align*}
	g_n: \; \Rb^d\times(\mathrm{S}_+^d)^{2}\times\mathrm{S}^d &\to \; \Rb \\
	(x_1,x_2,x_3,x_4)  & \mapsto \liminf_{\lambda\downarrow 0} \min\big\{x_1'(\lambda V_n^2 + x_2)^{-1}x_1, x_1'(\lambda V_n^2 + x_3)^{-1}x_1, \\
    & \hspace{2.8cm} x_1'(\lambda V_n^2 + x_4)^{-1}x_1 \big\} \cap \Rb^+, \\
	g:\; \Rb^d\times(\mathrm{S}_+^d)^{2}\times\mathrm{S}_{++}^d & \to \; \Rb \\
	(x_1,x_2,x_3,x_4)  & \mapsto \;  x_1' x_4^{-1} x_1.
\end{align*}
We wish to prove that
\begin{equation}
g_{\psi(n)}(M_{\psi(n)}) \convD g(L,V_1,V_2,V_u) \sim\chi^2(d).
    \label{eq:conv_g_multiv}
\end{equation}
To this end, we check the conditions of Theorem 18.11 in \cite{vanderVaart2000}, which is an extended CMT. It suffices to prove that for every $(x_{1n},x_{2n},x_{3n},x_{4n})_{n\ge 1}$ converging to $(x_1, \Sigma_1^\infty+\Sigma_4^\infty, \Sigma_2^\infty+\Sigma_4^\infty, \Id) \in \Rb^d\times (\mathrm{S}^d_+)^3$, we have
\begin{equation}
	\label{eq:extended_continuity_argument_normal_case}
	\lim_{n\to\infty} g_n(x_{1n},x_{2n},x_{3n},x_{4n}) = g\left(x_1,\Sigma_1^\infty+\Sigma_4^\infty,\Sigma_2^\infty+\Sigma_4^\infty,\Id\right).
\end{equation}
For $n$ large enough, $x_{4n}$ is symmetric positive definite; we consider such $n$'s hereafter. Then, $x_1'(\lambda V_n^2 + x_j)^{-1}x_1\ge 0$ for $j\in\{2,3,4\}$. Hence, for any $x_4$ symmetric positive definite,
\begin{align*}
	g_n(x_1,x_2,x_3,x_4)= & \liminf_{\lambda\downarrow 0} \min\left\{x_1'(\lambda V_n^2 + x_2)^{-1}x_1, x_1'(\lambda V_n^2 + x_3)^{-1}x_1, x_1'(\lambda V_n^2 + x_4)^{-1}x_1 \right\},\\
	= & \min\left\{\lim_{\lambda\downarrow 0} x_1'(\lambda V_n^2 + x_2)^{-1}x_1, \lim_{\lambda\downarrow 0} x_1'(\lambda V_n^2 + x_3)^{-1}x_1, \lim_{\lambda\downarrow 0} x_1'(\lambda V_n^2 + x_4)^{-1}x_1 \right\} \\
	= & \min\left\{\lim_{\lambda\downarrow 0} x_1'(\lambda V_n^2 + x_2)^{-1}x_1, \lim_{\lambda\downarrow 0} x_1'(\lambda V_n^2 + x_3)^{-1}x_1, x_1' x_4^{-1}x_1 \right\},
\end{align*}
where the second equality follows since the minimum function is continuous, and the functions $\lambda \mapsto x_1'(\lambda V_n^2 + x_j)^{-1}x_1$, $j\in\{2,3,4\}$, are decreasing. Let us define
\begin{align*}
	u_n & := \lim_{\lambda\downarrow 0} x_{1n}'(\lambda V_n^2 + x_{2n})^{-1}x_{1n}, \\
	v_n & := \lim_{\lambda\downarrow 0} x_{1n}'(\lambda V_n^2 + x_{3n})^{-1}x_{1n}, \\
	w_n & := x_{1n}' x_{4n}^{-1}x_{1n},
\end{align*}
so that $g_n(x_{1n},x_{2n},x_{3n},x_{4n})=\min\{u_n,v_n,w_n\}$. By continuity, $\lim_{n\to\infty}w_n = x_1' x_1$. Thus, to prove the result, it suffices to show that $\liminf u_n\ge x_1' x_1$ and $\liminf v_n\ge x_1' x_1$. We focus on $u_n$ as the reasoning is the same for $v_n$. Since $x_{2n}\to \Sigma_1^\infty+\Sigma_2^\infty = I - \Sigma_4^\infty$, there exists $A_n$ symmetric positive, $A_n\to 0$, such that $\Id + A_n \gg x_{2n}$. Thus,
$$x_{1n}'(\lambda V_n^2 + x_{2n})^{-1}x_{1n}\ge x_{1n}'(\lambda V_n^2 + \Id + A_n)^{-1}x_{1n}.$$
Letting $\lambda\to 0$ on the left- and then on the right-hand side, we obtain
$$u_n \ge \overline{w}_n:=x_{1n}'(\Id + A_n)^{-1}x_{1n}.$$
Moreover, $\lim_{n\to\infty}\overline{w}_n=x_1' x_1$. Equation~\eqref{eq:extended_continuity_argument_normal_case} follows. Then, by Theorem 18.11 in \cite{vanderVaart2000}, \eqref{eq:conv_g_multiv} holds as well.\\
\textbf{Case $\Sigma_3^{\infty}\ne 0$.}\\
As in the univariate case, since $\min(\|\Sigma_1^{\infty}\|,\|\Sigma_2^{\infty}\|,\|\Sigma_3^{\infty}\|)=0$,
we can assume wlog $\Sigma_2^{\infty}=0$. We wish to prove that
\begin{equation}
\widetilde{F}:=\lim_{\lambda\downarrow 0} (\overline{Y}-\theta_0)'(\lambda \Id+\widehat{V}_1)^{-1} (\overline{Y}-\theta_0) \convD \chi^2(d).
\label{eq:conv_g_nongG}
\end{equation}
To this end, we use
\begin{equation}
(V_n(\overline{Y}-\theta_0), V_n \widehat{V}_1 V_n)\convD (L,V_1)
    \label{eq:conv_with_V1}
\end{equation}
and we apply the same extended CMT as above, to the functions
\begin{align*}
	g_n: \; \Rb^d\times \mathrm{S}_+^d &\to \; \Rb \\
	(x_1,x_2)  & \mapsto \lim_{\lambda\downarrow 0} x_1'(\lambda V_n^2 + x_2)^{-1}x_1, \\
	g: \; \Rb^d\times \mathrm{S}_{++}^d & \to \; \Rb \\
	(x_1,x_2)  & \mapsto \;  x_1' x_2^{-1} x_1.
\end{align*}

We first prove that $V_1$ is invertible almost surely. Let
\begin{equation*}
	V_{1,\overline{k}}:= \sum_{\k \in \K_4(\overline{k})}\nu_{\k}^{\infty \otimes 2}+\sum_{0<k_1\leq \overline{k}}\left(\nu_{k_1,0,0}^{\infty}+\sum_{k_2>0}\nu_{k_1,k_2,0}^{\infty}Z_{0,k_2,0}\right)^{\otimes 2}.
\end{equation*}
Let $G_{\overline{k}}$ denote the matrix with first rows equal to $\nu^{\infty}_{\k}$ for $\k \in \K_3(\overline{k})$ and next rows equal to $\nu_{k_1,0,0}^{\infty}+\sum_{k_2>0}\nu_{k_1,k_2,0}^{\infty}Z_{0,k_2,0}$ for $k_1=1,...,\overline{k}$, so that $V_{1,\overline{k}}=G_{\overline{k}}^{\otimes 2}$. As $\sum_{\k \in \N^{3\ast}}||\nu^{\infty}_{\k}||^2<\infty$, all the components of $\E(G_{\overline{k}})^{\otimes 2}$ are arbitrarily close to the components of $\sum_{\k \in \K_{14}}\nu^{\infty \otimes 2}_{\k}=\Sigma_1^\infty+\Sigma_4^\infty$ for sufficiently large $\overline{k}$. By Lemma \ref{lem:limit_behaviour_nu} and $\Sigma_2^{\infty}=0$, $\lambda_{\min}\left[\Sigma_1^\infty + \Sigma_4^\infty \right] > 0$. Hence, for $\overline{k}$ large enough, $\E(G_{\overline{k}})^{\otimes 2}$ is invertible. Then, by Lemma~\ref{lem:invertible_QF}, $V_{1,\overline{k}}$ is invertible almost surely.
Since $V_{1,\overline{k}}$ and $V_1-V_{1,\overline{k}}$ are both symmetric non-negative matrices,
this implies that $V_1$ is invertible with probability 1. This implies that the support of the distribution of $(L,V_1)$ is at most $\Rb^d\times\mathrm{S}_{++}^d$.

\medskip
Now, to apply Theorem 18.11 in \cite{vanderVaart2000}, we prove that for every sequence $(x_1^n,x_2^n)$ in $\Rb^d\times\mathrm{S}_+^d$ converging to $(x_1,x_2) \in \Rb^d\times\mathrm{S}_{++}^d$, we have
\begin{equation}
	\label{eq:extended_continuity_argument}
	\lim_{n\to\infty} g_n(x_1^n,x_2^n) = g(x_1,x_2),
\end{equation}
Since $x_2^n$ converges to $x_2$, $x_2^n$ is strictly positive definite for $n$ large enough. Then, $x_2^n$ is strictly positive definite and $g_n(x_1^n,x_2^n) = (x_1^n)' (x_2^n)^{-1} x_1^n = g(x_1^n,x_2^n)$. Continuity of $g$ on $\Rb^d\times\mathrm{S}_{++}^d$ ensures that \eqref{eq:extended_continuity_argument} holds. Then, by Theorem 18.11 in \cite{vanderVaart2000} and \eqref{eq:conv_with_V1}, we obtain
$$\widetilde{F} \convD L'V_1^{-1} L.$$
Now, as in Theorem \ref{thm:unif_univ}, we have $L|(Z_{0,k_2,0})_{k_2>0} \sim \mathcal{N}(0,V_1)$. Thus, conditional on the $(Z_{0,k_2,0})_{k_2>0}$ and then unconditionally, $L'V_1^{-1} L \sim \chi^2(d)$. Eq. \eqref{eq:conv_g_nongG} follows.\\ 
\textbf{Proof of \eqref{eq:valid_unif_multiv_bon}} 
\label{par:proof_of_eqref}\\
By definition of $\mathcal{P}_{m,H'}^{d,b}$ and Theorem \ref{thm:unif_univ}, we have, for all $k\in\{1,...,d\}$,
$$\limsup_{n\to\infty} \sup_{P\in\mathcal{P}_{m,H'}^{d,b}} E_{P}[\ind{|t^k|>z_{1-\frac{\alpha}{2d}}}]\le \frac{\alpha}{d}.$$
As a result,
\begin{align*}
	\limsup_{n\to\infty} \sup_{P\in\mathcal{P}_{m,H'}^{d,b}} E_P[\phi_{\alpha}^b] & = \limsup_{n\to\infty} \sup_{P\in\mathcal{P}_{m,H'}^{d,b}} E_P\left[\max_{k=1,...,d}\ind{|t^k|>z_{1-\frac{\alpha}{2d}}}\right] \\
	& \le \limsup_{n\to\infty} \sup_{P\in\mathcal{P}_{m,H'}^{d,b}} \sum_{k=1}^d E_P\left[\ind{|t^k|>z_{1-\frac{\alpha}{2d}}}\right] \\
	& \le \sum_{k=1}^d \limsup_{n\to\infty}  \sup_{P\in\mathcal{P}_{m,H'}^{d,b}}  E_P\left[\ind{|t^k|>z_{1-\frac{\alpha}{2d}}}\right] \\
	& \le \alpha. \qquad\qquad\qquad\qquad\qquad\qquad\qquad\qquad\qquad\qquad
\end{align*}
Note that the first inequality is strict if the $(t^k)_{k=1,...,d}$ are independent. $\Box$



\newpage
\begin{center}
{\huge Online appendix}
\end{center}

\bigskip
This supplement gathers remaining proofs and additional lemmas used in these proofs or those in the main text.

\section{Proofs not in the main paper} 
\label{sec:proofs_not_in_the_main_paper}

\subsection{Lemma~\ref{lem:mu_nu}}
	
The Aldous-Hoover-Kallenberg representation ensures $W_{ij}=\tau(U_{i0},U_{0j},U_{ij})$. Next, considering $\mu_{\k}^n := E\left(W_{ij}\psi_{k_1}(U_{i0})\psi_{k_2}(U_{0j})\psi_{k_3}(U_{ij})\right)$, Equation  \eqref{eq:AHK_l2} holds.
Because $E(W_{11})=0$, we have $\mu^n_{0,0,0}=0$. There remains to prove $V(\overline{Y})$ is nonsingular for every $n \ge 1$ under Assumption~\ref{as:DGP} (the second representation in~\eqref{eq:AHK_l2_standardized} is straightforward once invertibility of $V(\overline{Y})$ has been obtained). Using Assumption~\ref{as:DGP} and the corresponding AHK representation, we can write
\begin{align*}
	V(\overline{Y}) & = (C_1C_2)^{-1}V(Y_{11}) + \frac{(C_2-1)}{C_2C_1}E[Y_{11}Y_{12}'] + \frac{(C_1-1)}{C_1C_2}E[Y_{11}Y_{21}'] \\
	& = (C_1C_2)^{-1}V(Y_{11}) + \frac{(C_2-1)}{C_2C_1}V[E[Y_{11}\mid U_{10}]] + \frac{(C_1-1)}{C_1C_2}V[E[Y_{11}\mid U_{01}]].
\end{align*}
Assumption~\ref{as:DGP} further ensures $V(Y_{11})$ is nonsingular. We conclude that for every $n \ge 1$, $V(\overline{Y}) \gg (C_1C_2)^{-1}V(Y_{11})$ so that $V(\overline{Y})$ is itself nonsingular. $\Box$

\subsection{Lemma~\ref{lem:weak_conv_means}}
	
	Let us define
	\begin{align*}
		\widetilde{V}_1 := & \frac{1}{C_1^2}\sum_{i=1}^{C_1}\left(\frac{1}{C_2}\sum_{j=1}^{C_2}V_nW_{ij}\right)^{\otimes 2}, \\
		\widetilde{V}_2 := & \frac{1}{C_2^2}\sum_{j=1}^{C_2}\left(\frac{1}{C_1}\sum_{i=1}^{C_1}V_nW_{ij}\right)^{\otimes 2},
	\end{align*}
	and define $\widetilde{V}_{12}$ similarly. Let us also introduce $\xi_{ij} := V_nW_{ij}$. For a square matrix $A$, $A^{\dagger}$ denotes its Moore-Penrose inverse. We first prove the convergence in distribution of $\left(V_n\overline{W}, \widetilde{V}_1,\widetilde{V}_2,\widetilde{V}_{12}\right)$ to $(L,V_1,V_2, V_{12})$ (first step). Next, we show that $\widetilde{V}_j=V_n\widehat{V}_jV_n+o_p(1)$ for $j=1,2,12$. This ensures the convergence in distribution of $\left(V_n\overline{W}, V_n(\widehat{V}_1,\widehat{V}_2,\widehat{V}_{12})V_n\right)$ to $(L,V_1,V_2,V_{12})$ (second step).
	
	\medskip
	
	\noindent\textbf{First step: Convergence of $\left(V_n\overline{W}, \widetilde{V}_1,\widetilde{V}_2,\widetilde{V}_{12}\right)$ to $(L,V_1,V_2, V_{12})$.} Prior to proving the result, we need to introduce a number of objects.
	
	\medskip
	Let $\mathcal{K}_j(\overline{k}):=\mathcal{K}_j\cap\{\k\in \mathbb{N}^{3\ast}:\max(k_1,k_2,k_3)\leq \overline{k}\}$ and
	 \begin{align*}\xi_{ij}(\overline{k}):=&\sum_{\k\in \K_1(\overline{k})}\sqrt{C_1}\nu^n_{k_1,0,0}\psi_{k_1}(U_{i0})+\sum_{\k\in \K_2(\overline{k})}\sqrt{C_2}\nu^n_{0,k_2,0}\psi_{k_2}(U_{0j})
		\\&+\sum_{\k \in\K_{3}(\overline{k})\cup \K_{4}(\overline{k})} \sqrt{C_1C_2}\nu^n_{\k}\psi_{k_1}(U_{i0}) \psi_{k_2}(U_{0j})\psi_{k_3}(U_{ij}),\\
		\overline{\xi}(\overline{k}):=&\frac{1}{C_1C_2} \sum_{i,j} \xi_{ij}(\overline{k}),
		\\ \widetilde{V}_1(\overline{k}):=&\frac{1}{C^2_1}\sum_{i=1}^{C_1}\left(\frac{1}{C_2}\sum_{j=1}^{C_2}\xi_{ij}(\overline{k})\right)^{\otimes 2},\\ \widetilde{V}_2(\overline{k}):=&\frac{1}{C^2_2}\sum_{j=1}^{C_2}\left(\frac{1}{C_1}\sum_{i=1}^{C_1}\xi_{ij}(\overline{k})\right)^{\otimes 2}, \\
		\widetilde{V}_{12}(\overline{k}):=&\frac{1}{(C_1C_2)^2}\sum_{i=1}^{C_1}\sum_{j=1}^{C_2}\left(\xi_{ij}(\overline{k})\right)^{\otimes 2}.
	\end{align*}
	Let $\Lambda_n$, $\Lambda_n(\overline{k})$ the following quantities:
	\begin{align*}
		\Lambda_n:=&\left(V_n\overline{W},\widetilde{V}_1,\widetilde{V}_2,\widetilde{V}_{12}\right),\\
		\Lambda_n(\overline{k}):=&\left(\overline{\xi }(\overline{k}),\widetilde{V}_1(\overline{k}),\widetilde{V}_2(\overline{k}),\widetilde{V}_{12}(\overline{k})\right).
	\end{align*}
	We also introduce the limit counterparts of $\Lambda_n$ and $\Lambda_n(\overline{k})$, namely $\Lambda_\infty$ and $\Lambda_{\infty}(\overline{k})$:
	\begin{align*}
		\Lambda_\infty := & \left(L,V_1,V_2,V_{12}\right), \\
		\Lambda_\infty(\overline{k}) := & \left(L(\overline{k}),V_1(\overline{k}),V_2(\overline{k}),V_{12}(\overline{k})\right),
	\end{align*}
	where $L(\overline{k})$, $V_1(\overline{k})$, $ V_2(\overline{k})$ and $V_{12}(\overline{k})$ are similar to $L$, $V_1$, $V_2$ and $V_{12}$ with $(\K_j)_{j=1,\dots,4}$ replaced with $(\K_j(\overline{k}))_{j=1,\dots,4}$ and the sums on $k_j>0$ are replaced with sums on $\overline{k}\geq k_j>0$ for $j=1,2,3$.
	
	\medskip
	
	We wish to prove convergence in distribution of $\Lambda_n$ to $\Lambda_\infty$. This is equivalent to proving that for any bounded Lipschitz function $h$ 
	\begin{equation}
		\label{eq:weak_conv_result}
		\lim_n\left| \E[h(\Lambda_n)] - \E[h(\Lambda_\infty)]  \right| =0.
	\end{equation}
	Let $\mathcal{M}_d(\Rb)$ denote the space of real square matrices of dimension $d$. We remark that $\Lambda_n$ and $\Lambda_\infty$ belong to $\Rb^d \times \mathcal{M}_d(\Rb) \times \mathcal{M}_d(\Rb) \times \mathcal{M}_d(\Rb)$. We need to introduce a norm $||.||_\Lambda$ on that space. For any $A := (A_1,A_2,A_3,A_4) \in \Rb^d \times \mathcal{M}_d(\Rb) \times \mathcal{M}_d(\Rb) \times \mathcal{M}_d(\Rb)$, we write
	\begin{equation*}
		||A||_\Lambda := \sum_{\ell = 1}^4||A_\ell||.
	\end{equation*}
	For some $C_h$, the triangle and Lipschitz inequalities ensure
		\begin{align}
		\left| \E[h(\Lambda_n)] - \E[h(\Lambda_\infty)]  \right| 
		\leq & C_h\left\{ \E\left[||\Lambda_n - \Lambda_n(\overline{k})||_\Lambda\right] + \E\left[||\Lambda_\infty - \Lambda_\infty(\overline{k})||_\Lambda\right] \right\} \nonumber \\
		& + \left| \E[h(\Lambda_n(\overline{k}))] - \E[h(\Lambda_\infty(\overline{k}))]  \right|.
		\label{eq:fundamental_bound}
	\end{align}
and next it is sufficient to prove $\lim_{\overline{k}}\limsup_{n}\E\left[||\Lambda_n - \Lambda_n(\overline{k})||_\Lambda\right] =0$, $\lim_{\overline{k}}\E\left[||\Lambda_\infty - \Lambda_\infty(\overline{k})||_\Lambda\right]=0$ and $\lim_n \left| \E[h(\Lambda_n(\overline{k}))] - \E[h(\Lambda_\infty(\overline{k}))]  \right|=0$ for any $\overline{k}$ .
	\medskip
	
	\textbf{Substep 1:  $\lim_{\overline{k}}\limsup_{n}\E\left[||\Lambda_n - \Lambda_n(\overline{k})||_\Lambda\right]=0$.}
	
	We have:
	\begin{align}
		\E\left[||\Lambda_n - \Lambda_n(\overline{k})||_\Lambda\right] = &  \E\left[\left|\left|V_n\overline{W}-\overline{\xi}(\overline{k})\right|\right|\right] + \E\left[\left|\left|\widetilde{V}_1-\widetilde{V}_1(\overline{k})\right|\right|\right] \nonumber  \\
		&  ~~~ + \E\left[\left|\left|\widetilde{V}_2-\widetilde{V}_2(\overline{k})\right|\right|\right] + \E\left[\left|\left|\widetilde{V}_{12}-\widetilde{V}_{12}(\overline{k})\right|\right|\right] .
		\label{eq:norm_equivalence_decomposition}
	\end{align}
	We handle the terms on the right-hand side of~\eqref{eq:norm_equivalence_decomposition} separately. First, we can write: 
	\begin{align*}
		& \E\left[\left|\left|V_n\overline{W}-\overline{\xi}(\overline{k})\right|\right|\right]^2 \nonumber\leq \E\left[\left|\left|V_n\overline{W}-\overline{\xi}(\overline{k})\right|\right|^2\right]\\
		= & \E\left[\left|\left|\sum_{\k\in \cup_{\ell=1}^4\left(\K_{\ell}\backslash \K_{\ell}(\overline{k})\right)}V_n\mu_{\k}\frac{1}{C_1C_2}\sum_{i=1}^{C_1}\sum_{j=1}^{C_2}\psi_{k_1}(U_{i0})\psi_{k_2}(U_{0j})\psi_{k_3}(U_{ij})\right|\right|^2\right] \nonumber\\
		= & \sum_{\k\in \cup_{\ell=1}^4\left(\K_{\ell}\backslash \K_{\ell}(\overline{k})\right)} ||\nu^n_{\k}||^2
		\label{eq:control_y_bar}
	\end{align*}

Using the matrix identity $A^{\otimes 2}-B^{\otimes 2}=(A+B)(A-B)'/2+(A-B)(A+B)'/2$, plus the triangle and Cauchy-Schwarz inequalities, we obtain:
\begin{align*}
	&\E\left[\left|\left|\widetilde{V}_1-\widetilde{V}_1(\overline{k})\right|\right|\right] \\
	&\leq \frac{1}{2C_1}\E\left[\left|\left|\left(\frac{1}{C_2}\sum_{j=1}^{C_2}\left(\xi_{1j}+\xi_{1,j}(\overline{k})\right)\right)\left(\frac{1}{C_2}\sum_{j=1}^{C_2}\left(\xi_{1j}-\xi_{1j}(\overline{k})\right)\right)'\right|\right|\right]\\
	&~~~~+\frac{1}{2C_1}\E\left[\left|\left|\left(\frac{1}{C_2}\sum_{j=1}^{C_2}\left(\xi_{1j}-\xi_{1j}(\overline{k})\right)\right)\left(\frac{1}{C_2}\sum_{j=1}^{C_2}\left(\xi_{1j}+\xi_{1j}(\overline{k})\right)\right)'\right|\right|\right]\\
	&\leq \frac{1}{C_1}\E\left[\left|\left|\frac{1}{C_2}\sum_{j=1}^{C_2}\left(\xi_{1j}+\xi_{1j}(\overline{k})\right)\right|\right|\times \left|\left|\frac{1}{C_2}\sum_{j=1}^{C_2}\left(\xi_{1j}-\xi_{1j}(\overline{k})\right)\right|\right|\right]
	\\
	&\leq \sqrt{\frac{1}{C_1}\E\left[\left|\left|\frac{1}{C_2}\sum_{j=1}^{C_2}\left(\xi_{1j}+\xi_{1j}(\overline{k})\right)\right|\right|^2\right]}\times\sqrt{\frac{1}{C_1}\E \left[\left|\left|\frac{1}{C_2}\sum_{j=1}^{C_2}\left(\xi_{1j}-\xi_{1j}(\overline{k})\right)\right|\right|^2\right]}
\end{align*}
We further have: 
\begin{align*}
	\frac{1}{C_1}\E\left[\left|\left|\frac{1}{C_2}\sum_{j=1}^{C_2}\left(\xi_{1j}-\xi_{1j}(\overline{k})\right)\right|\right|^2\right] 
	= & \sum_{\k \in \cup_{\ell\neq 2}\left(\K_{\ell}\backslash \K_{\ell}(\overline{k})\right)}||\nu^n_{\k}||^2+\frac{1}{C_1}\sum_{\k \in \K_2\backslash\K_2(\overline{k})}||\nu^n_{\k}||^2 \nonumber \\
	\leq & \sum_{\k\in \cup_{\ell=1}^4\left(\K_{\ell}\backslash\K_{\ell}(\overline{k})\right)}||\nu^n_{\k}||^2
\end{align*}
Orthogonality of the $(\psi_k)_{k\geq 0}$ in $L^2([0,1])$ ensures that $\xi_{1j}-\xi_{1j}(\overline{k})$ and $\xi_{1j}(\overline{k})$ are uncorrelated. Next,
\begin{align*}
\frac{1}{C_1}\E\left[\left|\left|\frac{1}{C_2}\sum_{j=1}^{C_2}\left(\xi_{1j}+\xi_{1j}(\overline{k})\right)\right|\right|^2\right]&=\frac{1}{C_1}\E\left[\left|\left|\frac{1}{C_2}\sum_{j=1}^{C_2}\left(\xi_{1j}-\xi_{1j}(\overline{k})\right)\right|\right|^2\right]+\frac{4}{C_1}\E\left[\left|\left|\frac{1}{C_2}\sum_{j=1}^{C_2}\xi_{1j}(\overline{k})\right|\right|^2\right]\\
&\leq 4 \sum_{\k \in \N^{3*}}	||\nu^n_{\k}||^2
\end{align*}
Similar inequalities hold for $\E\left[\left|\left|\widetilde{V}_2-\widetilde{V}_2(\overline{k})\right|\right|\right]$. As for $\E\left[\left|\left|\widetilde{V}_{12}-\widetilde{V}_{12}(\overline{k})\right|\right|\right]$, the following holds true:
\begin{align*}
	\E\left[\left|\left|\widetilde{V}_{12}-\widetilde{V}_{12}(\overline{k})\right|\right|\right] 
	&\leq \frac{1}{C_1C_2}\E\left[\left|\left|\xi_{11}+\xi_{11}(\overline{k})\right|\right|\times \left|\left|\xi_{11}-\xi_{11}(\overline{k})\right|\right|\right]	\\
	&\leq \sqrt{\frac{1}{C_1C_2}\E\left[\left|\left|\xi_{11}+\xi_{11}(\overline{k})\right|\right|^2\right]}\times\sqrt{\frac{1}{C_1C_2}\E \left[\left|\left|\xi_{11}-\xi_{11}(\overline{k})\right|\right|^2\right]},
\end{align*}
with \begin{align*}&\frac{1}{C_1C_2}\E \left[\left|\left|\xi_{11}-\xi_{11}(\overline{k})\right|\right|^2\right]\\
	&=\frac{1}{C_2}\sum_{\k \in \left(\mathcal{K}_1\backslash\mathcal{K}_1(\overline{k})\right)}\left|\left|\nu^n_{\k}\right|\right|^2+\frac{1}{C_1}\sum_{\k \in \left(\mathcal{K}_2\backslash\mathcal{K}_2(\overline{k})\right)}\left|\left|\nu^n_{\k}\right|\right|^2+\sum_{\k \in \cup_{\ell=3}^4\left(\mathcal{K}_{\ell}\backslash\mathcal{K}_{\ell}(\overline{k})\right)}\left|\left|\nu^n_{\k}\right|\right|^2\\
	&\leq \sum_{\k\in \cup_{\ell=1}^4\left(\K_{\ell}\backslash\K_{\ell}(\overline{k})\right)}||\nu^n_{\k}||^2
	\end{align*}
and
\begin{align*}
	\frac{1}{C_1C_2}\E \left[\left|\left|\xi_{11}+\xi_{11}(\overline{k})\right|\right|^2\right]&=\frac{1}{C_1C_2}\E \left[\left|\left|\xi_{11}-\xi_{11}(\overline{k})\right|\right|^2\right]+\frac{4}{C_1C_2}\E \left[\left|\left|\xi_{11}(\overline{k})\right|\right|^2\right]\\
	&\leq 4\sum_{\k \in \N^{3*}}	||\nu^n_{\k}||^2.
	\end{align*}
Coming back to~\eqref{eq:norm_equivalence_decomposition}, we deduce that 
\begin{align}
		\E\left[||\Lambda_n - \Lambda_n(\overline{k})||_\Lambda\right] \leq & \left(1+6 \left(\sum_{\k \in \N^{3*}}	||\nu^n_{\k}||^2\right)^{1/2}\right)\left( \sum_{\k\in \cup_{\ell=1}^4\left(\K_{\ell}\backslash\K_{\ell}(\overline{k})\right)}||\nu^n_{\k}||^2\right)^{1/2}
	\end{align}
Note that for any $A\subseteq \N^{3*}$, we have by the reverse triangle inequality $$\left|\left(\sum_{\k \in A}	||\nu^n_{\k}||^2\right)^{1/2}-\left(\sum_{\k \in A}	||\nu^{\infty}_{\k}||^2\right)^{1/2}\right|\leq \left(\sum_{\k \in A}	||\nu^n_{\k}-\nu^{\infty}_{\k}||^2\right)^{1/2}.$$ 
This and convergence of $\bm{\nu}^n$ to $\bm{\nu}^{\infty}$ in $\ell^2$ ensure $\lim_n \sum_{\k \in A}	||\nu^n_{\k}||^2=\sum_{\k \in A}	||\nu^{\infty}_{\k}||^2$. It follows that
\begin{align*}
		&\limsup_{n}\E\left[||\Lambda_n - \Lambda_n(\overline{k})||_\Lambda\right]\\
		& \leq  \left(1+6 \left(\sum_{\k \in \N^{3*}}	||\nu^{\infty}_{\k}||^2\right)^{1/2}\right)\left( 	\sum_{\k\in \cup_{\ell=1}^4\left(\K_{\ell}\backslash\K_{\ell}(\overline{k})\right)}||\nu^{\infty}_{\k}||^2\right)^{1/2},
\end{align*}
and because $\sum_{\k\in \N^{3\ast}}||\nu^{\infty}_{\k}||^2<\infty$, we have
$\lim_{\overline{k}}\sum_{\k\in \cup_{\ell=1}^4\left(\K_{\ell}\backslash \K_{\ell}(\overline{k})\right)} ||\nu^{\infty}_{\k}||^2=0$.

	\medskip
	
	\textbf{Substep 2:  $\lim_{\overline{k}}\E\left[||\Lambda_\infty - \Lambda_\infty(\overline{k})||_\Lambda\right]=0$.}
	
	Using the fact that $(Z_{\k})_{\k \in \cup_{\ell =1}^4\K_\ell}$ is a sequence of \textit{i.i.d.} $\mathcal{N}(0,1)$ random variables, we have
	\begin{align*}
		\E\left(\left|\left|L(\overline{k})-L\right|\right|^2\right)= & \sum_{\k\in \cup_{\ell=1}^4\left(\K_{\ell}\backslash\K_{\ell}(\overline{k})\right)}||\nu^{\infty}_{\k}||^2,\\
		\E\left(\left|\left|L\right|\right|^2\right)= & \sum_{\k\in \N^{3\ast}}||\nu^{\infty}_{\k}||^2.
	\end{align*}
	Coming to $V_1(\overline{k})-V_1$ (similar calculations are valid for $V_2(\overline{k})-V_2$), we can write
	\begin{align*}
		& \E\left(\left|\left|V_1(\overline{k})-V_1\right|\right|\right) \\
		&=\E\left(\left|\left|\V\left(L(\overline{k})|(Z_{\k})_{\k\in \K_2}\right)-\V\left(L|(Z_{\k})_{\k\in \K_2}\right)\right|\right|\right) \\
		&=\E\left(\left|\left|\V\left(L-L(\overline{k})|(Z_{\k})_{\k\in \K_2}\right)+2\Cov\left(L(\overline{k})-L,L|(Z_{\k})_{\k\in \K_2}\right)\right|\right|\right) \\
		&\leq \E\left(\left|\left|L(\overline{k})-L\right|\right|^2\right)+2\E\left(\left|\left|\Cov\left(L(\overline{k})-L,L|(Z_{0,k,0})_{1\leq \overline{k}}\right)\right|\right|\right) \\
		&\leq \E\left(\left|\left|L(\overline{k})-L\right|\right|^2\right)+2\E\left[ \left|\left|L-L(\overline{k})-\E\left(L-L(\overline{k})|(Z_{0,k,0})_{1\leq \overline{k}}\right)\right|\right|\times \left|\left|L-\E\left(L|(Z_{0,k,0})_{1\leq \overline{k}}\right)\right|\right|\right] \\
		&\leq \E\left(\left|\left|L(\overline{k})-L\right|\right|^2\right)+2\E\left(\left|\left|L(\overline{k})-L\right|\right|^2\right)^{1/2}\E(\left|\left|L\right|\right|^2)^{1/2}.
	\end{align*}
We also have:	\begin{equation*}
	\left|\left|V_{12}(\overline{k})-V_{12}\right|\right| = \sum_{\k\in \cup_{\ell=3}^4\K_{\ell}\backslash\K_{\ell}(\overline{k})}||\nu^{\infty}_{\k}||^2 \leq \sum_{\k\in \cup_{\ell=1}^4\K_{\ell}\backslash\K_{\ell}(\overline{k})}||\nu^{\infty}_{\k}||^2.
	\end{equation*}
We can conclude that:
\begin{align}
	\E\left[||\Lambda_\infty - \Lambda_\infty(\overline{k})||_\Lambda\right]\leq & ~~4 \sum_{\k\in \cup_{\ell=1}^4\K_{\ell}\backslash\K_{\ell}(\overline{k})}||\nu^{\infty}_{\k}||^2\nonumber\\&+4\left(\sum_{\k\in \cup_{\ell=1}^4\K_{\ell}\backslash\K_{\ell}(\overline{k})}||\nu^{\infty}_{\k}||^2\right)^{1/2}\left(\sum_{\k\in \N^{3\ast}}||\nu^{\infty}_{\k}||^2\right)^{1/2},\label{eq:control_limite_troncation}
	\end{align}	
and next $\lim_{\overline{k}}\E\left[||\Lambda_\infty - \Lambda_\infty(\overline{k}_{\varepsilon_1})||_\Lambda\right]=0$.

	\medskip
		
	\textbf{Substep 3: $\lim_{n}\left| \E[h(\Lambda_n(\overline{k}))] - \E[h(\Lambda_\infty(\overline{k}))]  \right|=0$ for any $\overline{k}>0$.}	
	
	Let
	\begin{align*}
				Z^n_{\k} & := \left\{
		\begin{array}{lll}
			& \frac{1}{C_1^{1/2}}\sum_{i=1}^{C_1} \psi_{k_1}(U_{i0})\text{ if }\k\in \K_1, \\
			& \frac{1}{C_2^{1/2}}\sum_{j=1}^{C_2} \psi_{k_2}(U_{0j})\text{ if }\k\in \K_2,\\
			& \frac{1}{(C_1C_2)^{1/2}}\sum_{i=1}^{C_1}\sum_{j=1}^{C_2} \psi_{k_1}(U_{i0})\psi_{k_2}(U_{0j})\psi_{k_3}(U_{ij})\text{ if }\k\in \K_4
		\end{array}
		\right. \\
		V_{1,n}(\overline{k}) & := \sum_{\boldsymbol{k}\in\K_4(\overline{k})}\nu_{\k}^{n\otimes 2} + \sum_{k_1=1}^{\overline{k}}\left( \nu^n_{k_1,0,0} + \sum_{k_2=1}^{\overline{k}}\nu^n_{k_1,k_2,0}Z_{0,k_2,0}^n \right)^{\otimes 2}, \\
		V_{2,n}(\overline{k}) & := \sum_{\boldsymbol{k}\in\K_4(\overline{k})}\nu_{\k}^{n\otimes 2} + \sum_{k_2=1}^{\overline{k}}\left( \nu^n_{0,k_2,0} + \sum_{k_1=1}^{\overline{k}}\nu^n_{k_1,k_2,0}Z_{k_1,0,0}^n \right)^{\otimes 2}, \\
		V_{12,n}(\overline{k}) & := \sum_{\boldsymbol{k}\in\K_{3}(\overline{k})\cup\K_{4}(\overline{k})}\nu_{\k}^{n\otimes 2}.
	\end{align*}
	We also define $\Lambda_n^\infty(\overline{k}) := \left( \overline{\xi}(\overline{k}), V_{1,n}(\overline{k}), V_{2,n}(\overline{k}), V_{12,n}(\overline{k}) \right)$. By the Lipschitz property of $h$ and the triangle and Jensen inequalities:
	\begin{align*}
		\left| \E[h(\Lambda_n(\overline{k}))] - \E[h(\Lambda_\infty(\overline{k}))]  \right| \leq & \left| \E[h(\Lambda_n(\overline{k})) - h(\Lambda_n^\infty(\overline{k}))]  \right| + \left| \E[h(\Lambda_n^\infty(\overline{k}))] - \E[h(\Lambda_\infty(\overline{k}))]  \right| \\
		\leq & C_h \E\left[\left|\left| \Lambda_n(\overline{k}) - \Lambda_n^\infty(\overline{k})\right|\right|_\Lambda\right] + \left| \E[h(\Lambda_n^\infty(\overline{k}))] - \E[h(\Lambda_\infty(\overline{k}))]  \right|.
	\end{align*}
Lemma \ref{lem:weak_conv_finite_dim_Z} ensures	$(Z^n_{\k})_{\k\in \K_{124}(\overline{k})}$ converges in distribution to $(Z_{\k})_{\k\in \K_{124}(\overline{k})}\sim \mathcal{N}(0,\Id)$.\\ Let $g_n\left((Z^n_{\k})_{\k\in \K_{124}(\overline{k})}\right):=\Lambda_n^\infty(\overline{k})$ and $g_{\infty}\left((Z_{\k})_{\k\in \K_{124}(\overline{k})}\right):=\Lambda_{\infty}(\overline{k})$. For any $\k$, $\lim_n \nu^n_{\k}=\nu^{\infty}_{\k}$ implies: $$\lim_n g_n\left((z^n_{k_1,0,0},z^n_{0,k_2,0})_{1\leq k_1,k_2\leq \overline{k}}\right)=g_{\infty}\left((z^{\infty}_{k_1,0,0},z^{\infty}_{0,k_2,0})_{1\leq k_1,k_2\leq \overline{k}}\right)$$ for $z^n_{\k}$ such that $\lim_n z^n_{\k}=z^{\infty}_{\k}$. Next, Theorem 18.11 in \cite{vanderVaart2000} ensures convergence in distribution of $\Lambda_n^{\infty}(\overline{k})$ to $\Lambda_{\infty}(\overline{k})$ and
\begin{equation}
	\lim_n\left| \E[h(\Lambda_n^\infty(\overline{k}))] - \E[h(\Lambda_{\infty}(\overline{k}))]  \right|=0.
	\label{eq:portemanteau_intermediary}
\end{equation}
It remains to control $\E\left[\left|\left| \Lambda_n(\overline{k}) - \Lambda_n^\infty(\overline{k})\right|\right|_\Lambda\right]$. 
To do so, note that:	\begin{align*}
		\E\left[\left|\left| \Lambda_n(\overline{k}) - \Lambda_n^\infty(\overline{k})\right|\right|_\Lambda\right] =& 
		\E\left[\left|\left| \widetilde{V}_1(\overline{k}) - V_{1,n}(\overline{k}) \right|\right|\right] + \E\left[\left|\left| \widetilde{V}_2(\overline{k}) - V_{2,n}(\overline{k}) \right|\right|\right]  \\
		& + \E\left[\left|\left| \widetilde{V}_{12}(\overline{k}) - V_{12,n}(\overline{k}) \right|\right|\right].
	\end{align*}
We detail first how to control $\E\left[\left|\left| \widetilde{V}_1(\overline{k}) - V_{1,n}(\overline{k}) \right|\right|\right]$ ($\E\left[\left|\left| \widetilde{V}_2(\overline{k}) - V_{2,n}(\overline{k}) \right|\right|\right]$ can be dealt with using similar arguments). We then handle $\E\left[\left|\left| \widetilde{V}_{12}(\overline{k}) - V_{12,n}(\overline{k}) \right|\right|\right]$.

\medskip

\textbf{Subsubstep 1: $\lim_{n}\E\left[\left|\left| \widetilde{V}_1(\overline{k}) - V_{1,n}(\overline{k}) \right|\right|\right] + \E\left[\left|\left| \widetilde{V}_2(\overline{k}) - V_{2,n}(\overline{k}) \right|\right|\right]=0$ for any $\overline{k}>0$.} \\
The term $\widetilde{V}_1(\overline{k})$ can be decomposed as follows 
\begin{align}
	\widetilde{V}_1(\overline{k})&= \frac{1}{C_1^2}\sum_{i=1}^{C_1} \left(T^n_{i,1}+T^n_{i,2}+T^n_{i,3}+T^n_{i,4}\right)^{\otimes 2},
	\label{eq:sigma_a}
\end{align}
with
	\begin{align*}
	T^n_{i,1}&:=\sqrt{C_1}\sum_{\k \in \K_1(\overline{k})}\nu^n_{\k}\psi_{k_1}(U_{i0})\\
	T^n_{i,2}&:=\sqrt{C_2}\sum_{\k \in \K_2(\overline{k})}\nu^n_{\k}\frac{1}{C_2}\sum_{j=1}^{C_2}\psi_{k_2}(U_{0j})=\sum_{\k \in \K_2(\overline{k})}\nu^n_{\k}Z^n_{\k}\\
	T^n_{i,3}&:=\sqrt{C_1C_2}\sum_{\k \in \K_3(\overline{k})}\nu^n_{\k}\frac{1}{C_2}\sum_{j=1}^{C_2}\psi_{k_1}(U_{i0})\psi_{k_2}(U_{0j}^n)=\sqrt{C_1}\sum_{\k \in \K_3(\overline{k})}\nu^n_{\k}\psi_{k_1}(U_{i0})Z^n_{0,k_2,0}\\
	T_{i,4}^n&:=\sqrt{C_1C_2}\sum_{\k \in \K_4(\overline{k})}\nu^n_{\k}\frac{1}{C_2}\sum_{j=1}^{C_2}\psi_{k_1}(U_{i0})\psi_{k_2}(U_{0j})\psi_{k_3}(U_{ij}).
\end{align*}
Let $h_{k_1}^n:=\nu^n_{k_1,0,0}+\sum_{k_2=1}^{\overline{k}}\nu_{k_1,k_2,0}^nZ_{0,k_2,0}^n$. We can write
\begin{align*}
	\frac{1}{C_1^2}\sum_{i=1}^{C_1} (T^n_{i,1}+T^n_{i,3})^{\otimes 2}- \sum_{k_1=1}^{\overline{k}}(h_{k_1}^{n})^{\otimes 2}=& \sum_{k_1=1}^{\overline{k}}(h_{k_1}^n)^{\otimes 2}\left(\frac{1}{C_1}\sum_{i=1}^{C_1}  \psi_{k_1}^2(U_{i0})-1\right)\\
	&+\sum_{1\leq k_1\neq k_1'\leq \overline{k}} h_{k_1}^nh_{k'_1}^{n \prime}\left(\frac{1}{C_1}\sum_{i=1}^{C_1}\psi_{k_1}(U_{i0})\psi_{k'_1}(U_{i0})\right)
\end{align*}
Next, by the triangle inequality and because $(h_k^n)_{k\geq 1}$ and $(U_{i0})_{i\geq1}$ are independent, we have:
\begin{align*}
	&E\left[\left|\left|\frac{1}{C_1^2}\sum_{i=1}^{C_1} (T^n_{i,1}+T^n_{i,3})^{\otimes 2}- \sum_{k_1=1}^{\overline{k}}(h_{k_1}^{n})^{\otimes 2}\right|\right|\right]\leq \sum_{k_1=1}^{\overline{k}}E\left[\left|\left|h_{k_1}^n\right|\right|^2\right]E\left[\left|\frac{1}{C_1}\sum_{i=1}^{C_1}  \psi_{k_1}^2(U_{i0})-1\right|\right]\\
	&~~~~~~~~~~~~~~~~~~~~~~~~~~~~~~~~~~~~~~+\sum_{1\leq k_1\neq k_1'\leq \overline{k}} E\left[||h_{k_1}^n|| ||h_{k'_1}^{n}||\right]E\left[\left|\frac{1}{C_1}\sum_{i=1}^{C_1}\psi_{k_1}(U_{i0})\psi_{k'_1}(U_{i0})\right|\right].
	\end{align*}
Because $E[Z^n_{0,k_2,0}]=0$ and $E[(Z^n_{0,k_2,0})^2]=1$, we have $E[||h_{k_1}^n||^2]=||\nu^n_{k_1,0,0}||^2+\sum_{k_2=1}^{\overline{k}}||\nu^n_{k_1,k_2,0}||^2$, $E[||h_{k_1}^n|| ||h_{k'_1}^{n}||]\leq  E[||h_{k_1}^n|| ^2]/2+E[ ||h_{k'_1}^{n}||^2]/2$. Because $\sup_{u\in[0,1]}\left|\psi_k^2(u)-1\right|=1$ and $E[\psi_k^2(U_{i0})]=1$ and by independence of $U_{i0}$ accross $i$, we have $$E\left[\left|\frac{1}{C_1}\sum_{i=1}^{C_1}  \psi_{k_1}^2(U_{i0})-1\right|\right]\leq E\left[\left|\frac{1}{C_1}\sum_{i=1}^{C_1}  \psi_{k_1}^2(U_{i0})-1\right|^2\right]^{1/2}\leq C_1^{-1/2}$$
and similarly, because $\sup_{u\in[0,1]}\left|\psi_k(u)\psi_{k'}(u)\right|=2$ we have $E[|\frac{1}{C_1}\sum_{i=1}^{C_1}\psi_{k_1}(U_{i0})\psi_{k'_1}(U_{i0})|]\leq 2C_1^{-1/2}$. This ensures:
\begin{align}
	E\left[\left|\left|\frac{1}{C_1^2}\sum_{i=1}^{C_1} (T^n_{i,1}+T^n_{i,3})^{\otimes 2}- \sum_{k_1=1}^{\overline{k}}(h_{k_1}^{n})^{\otimes 2}\right|\right|\right]\leq& \overline{k}( 2\overline{k}-1) C_1^{-1/2}\sum_{\k\in \N^{3\ast}}||\nu^n_{\k}||^2.
		\label{eq:control_T_31}
	\end{align}
$T^n_{i,2}$ does not depend on $i$, $E[Z^n_{0,k_2,0}]=0$ and $E[Z^n_{0,k_2,0}Z^n_{0,k'_2,0}]=\mathds{1}\{k_2=k_2'\}$, which ensures:
\begin{align}
	\E\left[\left|\left| \frac{1}{C_1^2}\sum_{i=1}^{C_1} T^n_{i,2}{}^{\otimes 2} \right|\right|\right] & \le \frac{E[||T_{1,2}^n||^2]}{C_1} = \frac{1}{C_1}\sum_{\k \in \K_2(\overline{k})}\left|\left|\nu^n_{\k}\right|\right|^2 \leq C_1^{-1}\sum_{\k \in \N^{3\ast}}\left|\left|\nu^n_{\k}\right|\right|^2.
	\label{eq:control_T_2}	
\end{align}


Let $V_{ijj'}(\k,\k'):=\psi_{k_1}(U_{i0}^n) \psi_{k'_1}(U_{i0}^n)\psi_{k_2}(U_{0j}^n)\psi_{k_2'}(U_{0j'}^n)\psi_{k_3}(U_{ij}^n)\psi_{k'_3}(U_{ij'}^n)$, we can write with the triangle and Jensen inequalities, as well as subadditivity and submultiplicativity of the matrix 2-norm
\begin{align*}
	 &\E\left[\left|\left| \frac{1}{C_1^2}\sum_{i=1}^{C_1} T^n_{i,4}{}^{\otimes 2} - \sum_{\k\in\K_4(\overline{k})}\nu_{\k}^{\otimes 2} \right|\right|\right]\nonumber\\&=\E\left[\left|\left| \sum_{(\k,\k') \in \K_4(\overline{k})^2} \nu_{\k}\nu_{\k'}'\left( \frac{1}{C_1C_2}\sum_{i=1}^{C_1}\sum_{j=1}^{C_2}\sum_{j'=1}^{C_2}V_{ijj'}(\k,\k') - \mathds{1}\{\k=\k'\} \right)  \right|\right|\right] \nonumber \\
	\leq & \E\left[\left|\left| \sum_{(\k,\k') \in \K_4(\overline{k})^2} \nu_{\k}\nu_{\k'}'\left( \frac{1}{C_1C_2}\sum_{i=1}^{C_1}\sum_{j=1}^{C_2}V_{ijj}(\k,\k') - \mathds{1}\{\k=\k'\} \right)  \right|\right|\right] \nonumber \\
	& + \E\left[\left|\left| \sum_{(\k,\k') \in \K_4(\overline{k})^2} \nu_{\k}\nu_{\k'}'\frac{1}{C_1C_2}\sum_{i=1}^{C_1}\sum_{1\leq j \neq j' \leq C_2}V_{ijj'}(\k,\k') \right|\right|\right] \nonumber \\
	\leq & \sum_{(\k,\k') \in \K_4(\overline{k})^2}||\nu_{\k}||\,||\nu_{\k'}||\left\{ \E\left[\left(\frac{1}{C_1C_2}\sum_{i=1}^{C_1}\sum_{j=1}^{C_2}V_{ijj}(\k,\k') - \mathds{1}\{\k=\k'\}\right)^2\right]^{1/2} \right. \nonumber\\
	& \left. ~~~~~~~~~~~~~~~~~~~~~~~~~~~~~~~~ + \E\left[\left(\frac{1}{C_1C_2}\sum_{i=1}^{C_1}\sum_{1\leq j \neq j' \leq C_2}V_{ijj'}(\k,\k')\right)^2\right]^{1/2} \right\}.
\end{align*}
We remark that $\E\left[V_{ijj}(\k,\k')\right] = \mathds{1}\{\k=\k'\}$ and $||\psi_k||_{\infty}\leq \sqrt{2}$. This and independence of $(U_{i0},U_{0j},U_{ij})$ across $(i,j)$ ensure $\left|\Cov(V_{ijj}(\k,\k'),V_{i'j'j'}(\k,\k'))\right|\leq (2^3-1)^2\mathds{1}\{\k=\k'\}\mathds{1}\{i=i'\text{ or } j=j'\}+2^6\mathds{1}\{\k \neq \k'\}\mathds{1}\{i=i'\text{ or } j=j'\}$. 
\begin{align*}
	\E\left[\left(\frac{1}{C_1C_2}\sum_{i=1}^{C_1}\sum_{j=1}^{C_2}V_{ijj}(\k,\k') - \mathds{1}\{\k=\k'\}\right)^2\right]^{1/2}&\leq 8\left(\frac{1}{C_1C_2}+\frac{1}{C_1}+\frac{1}{C_2}\right)^{1/2}\leq \frac{8\sqrt{2}}{\Cinf^{1/2}}
	\end{align*}
Moreover $\E(V_{ijj'}(\k,\k'))=0$ if $j\neq j'$, and for $i\neq i'$ and $j,j',j'',j'''$ four distinct elements,
$\Cov\left(V_{ijj'}(\k,\k'),V_{ij''j'''}(\k,\k')\right)$, $\Cov(V_{ijj'}(\k,\k'),V_{i'jj''}(\k,\k'))$, $\Cov(V_{ijj'}(\k,\k'),V_{i'jj''}(\k,\k'))$, \\
$\Cov(V_{ij'j}(\k,\k'),V_{ij''j}(\k,\k'))$, $\Cov(V_{ijj'}(\k,\k'),V_{ijj''}(\k,\k'))$ and $\Cov(V_{ijj'}(\k,\k'),V_{i'jj'}(\k,\k'))$ are null. Next, $\left|\Cov\left(V_{ijj'}(\k,\k'),V_{i'j''j'''}(\k,\k')\right)\right|\leq 2^6 \mathds{1}\{i=i'\}\mathds{1}\{\{j,j'\}=\{j'',j'''\}\}$ 
and:
\begin{align*}
	\E\left[\left(\frac{1}{C_1C_2}\sum_{i=1}^{C_1}\sum_{1\leq j \neq j' \leq C_2}V_{ijj'}(\k,\k')\right)^2\right]^{1/2}&\leq 2^3\left(\frac{2C_1C_2(C_2-1)}{C_1^2C_2^2}\right)^{1/2}\leq \frac{8\sqrt{2}}{C_1^{1/2}}
\end{align*}
Since $||\nu_{\k}||\,||\nu_{\k'}||\leq ||\nu_{\k}||^2/2+||\nu_{\k'}||^2/2$ we conclude:
\begin{align}
	\E\left[\left|\left| \frac{1}{C_1^2}\sum_{i=1}^{C_1} T^n_{i,4}{}^{\otimes 2} - \sum_{\k\in\K_4(\overline{k})}\nu_{\k}^{\otimes 2} \right|\right|\right]&\leq \frac{16\overline{k}(\overline{k}+1)^2\sqrt{2}}{\Cinf^{1/2}}\sum_{\k\in\N^{3\ast}}||\nu^n_{\k}||^2
	\label{eq:control_T_4}
\end{align}
We remark that $E(T^{n\prime}_{i,\ell}\ T^{n}_{i',\ell'})=E\left(||T^n_{1,\ell}||^2\right)\mathds{1}\{i=i',\ell=\ell'\}=C_1\left(\sum_{\k\in \K_{\ell}(\overline{k})}||\nu^n_{\k}||^2\right)\mathds{1}\{i=i',\ell=\ell'\}$ for $(\ell,\ell') \in \{1,3,4\}^2$ and $T^n_{i,2}=T^n_{1,2}$. Based on these observations, we get:
\begin{align}
	E\left[\left|\left|\frac{1}{C_1^2}\sum_{i=1}^{C_1}\left(T^n_{i,1}+T^n_{i,3}+T^n_{i,4}\right)T^{n \prime}_{i,2}\right|\right|\right]&\leq E\left[\left|\left|\frac{1}{C_1^2}\sum_{i=1}^{C_1}\left(T^n_{i,1}+T^n_{i,3}+T^n_{i,4}\right)\right|\right| \left|\left|T^{n \prime}_{1,2}\right|\right|\right]\nonumber\\ &\leq C_1^{-1}E\left[\left|\left|\frac{1}{C_1}\sum_{i=1}^{C_1}\left(T^n_{i,1}+T^n_{i,3}+T^n_{i,4}\right)\right|\right|^2\right]^{1/2} \left(\sum_{\k \in \K_2(\overline{k})}\left|\left|\nu^n_{\k}\right|\right|^2\right)^{1/2}\nonumber\\
	&\leq C_1^{-1}\sum_{\k\in\N^{3\ast}}||\nu^n_{\k}||^2.
	\label{eq:control_T_134_2}
\end{align}

Let $F_{ij}(\k,\k'):=\psi_{k_1}(U_{i0}) \psi_{k'_1}(U_{i0})\psi_{k_2'}(U_{0j})\psi_{k_3'}(U_{ij})$. If $\k'\in\K_4$,  we have $\E[F_{ij}(\k,\k')]=0$ and $|E(F_{ij}(\k,\k')F_{i'j'}(\k,\k'))|=E(S^2_{11}(\k,\k'))\mathds{1}\{i=i',j=j'\}\leq 16\mathds{1}\{i=i',j=j'\}$. This implies:
\begin{align}
	 \E\left[\left|\left| \frac{1}{C_1^2}\sum_{i=1}^{C_1} T^n_{i,1} T^{n\prime}_{i,4} \right|\right|\right]\leq&\sum_{\k,\k'\in \K_1(\overline{k})\times\K_4(\overline{k})}||\nu^n_{\k}\nu^{n \prime}_{\k'}|| \E\left[\left|\frac{1}{C_1C_2^{1/2}}\sum_{i=1}^{C_1}\sum_{j=1}^{C_2}F_{ij}(\k,\k')\right|\right]\notag\\
	\leq & \sum_{\k,\k'\in \K_1(\overline{k})\times\K_4(\overline{k})}||\nu_{\k}||\,||\nu_{\k'}|| \E\left[ \left(\frac{1}{C_1C_2^{1/2}}\sum_{i=1}^{C_1}\sum_{j=1}^{C_2} F_{ij}(\k,\k')\right)^2 \right]^{1/2}\notag\\
	\leq& \frac{1}{2}\sum_{\k,\k'\in \K_1(\overline{k})\times\K_4(\overline{k})}\left(||\nu_{\k}||^2+||\nu_{\k'}||^2\right) \frac{1}{C_1^{1/2}}\sup_{\k,\k'}\E\left[ F^2_{11}(\k,\k') \right]^{1/2}\notag\\
	\leq &\frac{2\overline{k}\left(1+(\overline{k}+1)^2\right)}{C_1^{1/2}}\sum_{\k\in \N^{3\ast}}||\nu^n_{\k}||^2.
		\label{eq:control_T_1_4}
	\end{align}
Let $Q_{ijj'}(\k,\k'):=\psi_{k_1}(U_{i0}) \psi_{k'_1}(U_{i0})\psi_{k_2}(U_{0j})\psi_{k_2'}(U_{0j'})\psi_{k_3'}(U_{ij'})$. If $\k\in \K_4$, we have $E(Q_{ijj'}(\k,\k'))=0$ and $|E(Q_{ijj'}(\k,\k')Q_{i'j''j'''}(\k,\k'))|\leq 2^5\mathds{1}\{i=i',j=j'', j'=j'''\}$.
\begin{align}
	\E\left[\left|\left| \frac{1}{C_1^2}\sum_{i=1}^{C_1} T^n_{i,3} T^{n\prime}_{i,4} \right|\right|\right]\leq&\sum_{\k,\k'\in \K_3(\overline{k})\times\K_4(\overline{k})}||\nu^n_{\k}\nu^{n \prime}_{\k'}|| \E\left[\left|\frac{1}{C_1C_2}\sum_{i=1}^{C_1}\sum_{j=1}^{C_2}\sum_{j'=1}^{C_2}Q_{ijj'}(\k,\k')\right|\right]\notag\\
	\leq& \frac{1}{2}\sum_{\k,\k'\in \K_3(\overline{k})\times\K_4(\overline{k})}\left(||\nu_{\k}||^2+||\nu_{\k'}||^2\right) \left(\frac{2^5}{C_1}\right)^{1/2}\notag\\
	\leq &\frac{2^{3/2}\overline{k}\left(\overline{k}+(\overline{k}+1)^2\right)}{C_1^{1/2}}\sum_{\k\in \N^{3\ast}}||\nu^n_{\k}||^2
		\label{eq:control_T_3_4}
	\end{align}

We conclude from \eqref{eq:control_T_31}, \eqref{eq:control_T_2}, \eqref{eq:control_T_4}, \eqref{eq:control_T_134_2}, \eqref{eq:control_T_1_4} and \eqref{eq:control_T_3_4} there exists some universal constant $K_1$ such that:
\begin{align}
	E\left[\left|\left|\widetilde{V}_1(\overline{k}) - V_{1,n}(\overline{k})\right|\right|\right]+E\left[\left|\left|\widetilde{V}_2(\overline{k}) - V_{2,n}(\overline{k})\right|\right|\right]&\leq \frac{K_1\overline{k}^3}{\Cinf^{1/2}}\sum_{\k \in \N^{3\ast}}||\nu^n_{\k}||^2.\label{eq:control_V1}
\end{align}

\medskip

\textbf{Subsubstep 2: $\lim_{n}\E\left[\left|\left| \widetilde{V}_{12}(\overline{k}) - V_{12,n}(\overline{k}) \right|\right|\right]=0$ for any $\overline{k}>0$.} \\
We have \begin{align}
	\widetilde{V}_{12}(\overline{k})&= \frac{1}{C_1^2C_2^2}\sum_{i=1}^{C_1}\sum_{j=1}^{C_2} \left(S^n_{ij,1}+S^n_{ij,2}+S^n_{ij,3}+S^n_{ij,4}\right)^{\otimes 2},
	\label{eq:sigma_gamma_eps}
\end{align}
with
\begin{align*}
	S^n_{ij,1}&:=\sqrt{C_1}\sum_{\k \in \K_1(\overline{k})}\nu^n_{\k}\psi_{k_1}(U_{i0})\\
	S^n_{ij,2}&:=\sqrt{C_2}\sum_{\k \in \K_2(\overline{k})}\nu^n_{\k}\psi_{k_2}(U_{0j})\\
	S^n_{ij,3}&:=\sqrt{C_1C_2}\sum_{\k \in \K_3(\overline{k})}\nu^n_{\k}\psi_{k_1}(U_{i0})\psi_{k_2}(U_{0j}^n)\\
	S^n_{ij,4}&:=\sqrt{C_1C_2}\sum_{\k \in \K_4(\overline{k})}\nu^n_{\k}\psi_{k_1}(U_{i0})\psi_{k_2}(U_{0j}^n)\psi_{k_3}(U_{ij})	.
\end{align*}
For \begin{align*}P_{ij}(\k,\k'):=&C_1^{1-(\mathds{1}\{\k\in \K_2\}+\mathds{1}\{\k'\in \K_2\})/2}C_2^{1-(\mathds{1}\{\k\in \K_1\}+\mathds{1}\{\k'\in \K_1\})/2}\\
	&\left(\psi_{k_1}(U_{i0})\psi_{k'_1}(U_{i0})\psi_{k_2}(U_{0j})\psi_{k'_2}(U_{0j})\psi_{k_3}(U_{ij})\psi_{k'_3}(U_{ij})-\mathds{1}\{\k=\k'\}\right),\end{align*}
we have for $(\ell,\ell')\in\{1,2,3,4\}^2$:
\begin{align}
	&E\left[\left|\left|\frac{1}{C_1^2C_2^2}\sum_{i=1}^{C_1}\sum_{j=1}^{C_2}S^n_{ij,\ell}S^{n\prime}_{ij,\ell'}-\mathds{1}\{\ell=\ell'\}C_1^{-\mathds{1}\{\ell = 2\}}C_2^{-\mathds{1}\{\ell = 1\}}\sum_{\k\in\K_{\ell}(\overline{k})}\nu^{n\otimes 2}_{k}\right|\right|\right]\notag\\
	&\leq \sum_{\k\in \K_{\ell}(\overline{k})}\sum_{\k'\in \K_{\ell'}(\overline{k})}||\nu^n_{\k}||\  ||\nu^n_{\k'}||E\left[\left|\frac{1}{C_1^2C_2^2}\sum_{i=1}^{C_1}\sum_{j=1}^{C_2}P_{ij}(\k,\k')\right|\right]\notag\\
	&\leq \sum_{\k\in \K_{\ell}(\overline{k})}\sum_{\k'\in \K_{\ell'}(\overline{k})}\left(\frac{||\nu^n_{\k}||^2}{2}+\frac{ ||\nu^n_{\k'}||^2}{2}\right)E\left[\left|\frac{1}{C_1^2C_2^2}\sum_{i=1}^{C_1}\sum_{j=1}^{C_2}P_{ij}(\k,\k')\right|\right]\notag\\
	&\leq \overline{k}(\overline{k}+1)^2\left(\sum_{\k\in \N^{3\ast}}||\nu^n_{\k}||^2\right) \left(\sup_{(\k,\k')\in \K_{\ell}(\overline{k})\times \K_{\ell'}(\overline{k})}E\left[\left|\frac{1}{C_1^2C_2^2}\sum_{i=1}^{C_1}\sum_{j=1}^{C_2}P_{ij}(\k,\k')\right|\right]\right).\label{eq:control_S}
\end{align}
If $\{\ell,\ell'\}=\{1,2\}$ we have by independence of $(U_{i0},U_{0j})_{i,j}$, the Cauchy-Schwarz inequality and $\sup_{u\in[0,1]}|\psi_k(u)|\leq \sqrt{2}$:
\begin{align}
	&\sup_{(\k,\k')\in \K_{\ell}(\overline{k})\times \K_{\ell'}(\overline{k})}E\left[\left|\frac{1}{C_1^2C_2^2}\sum_{i=1}^{C_1}\sum_{j=1}^{C_2}P_{ij}(\k,\k')\right|\right]\notag\\
	&= \frac{1}{C_1^{1/2}C_2^{1/2}}\sup_{1\leq k_1,k_2\leq \overline{k}}E\left[\left|\frac{1}{C_1}\sum_{i=1}^{C_1}\psi_{k_1}(U_{i0})\right|\times\left|\frac{1}{C_2}\sum_{j=1}^{C_2}\psi_{k_2}(U_{0j})\right|\right]\notag\\
	&\leq \frac{1}{C_1C_2}\sup_{1\leq k_1,k_2\leq \overline{k}}E(\psi_{k_1}^2(U_{1,0}))^{1/2}E(\psi_{k_2}^2(U_{0,1}))^{1/2}\notag\\
	&\leq \frac{2}{C_1C_2}.\label{eq:control_S_1_2}
	\end{align}
If $\{\ell,\ell'\}=\{1,3\}$ we have:
\begin{align}
	&\sup_{(\k,\k')\in \K_{\ell}(\overline{k})\times \K_{\ell'}(\overline{k})}E\left[\left|\frac{1}{C_1^2C_2^2}\sum_{i=1}^{C_1}\sum_{j=1}^{C_2}P_{ij}(\k,\k')\right|\right]\notag\\	&= \frac{1}{C_2^{1/2}}\sup_{(\k,\k')\in \K_1(\overline{k})\times \K_3(\overline{k})}E\left[\left|\frac{1}{C_1}\sum_{i=1}^{C_1}\psi_{k_1}(U_{i0})\psi_{k'_1}(U_{i0})\right|\times\left|\frac{1}{C_2}\sum_{j=1}^{C_2}\psi_{k'_2}(U_{0j})\right|\right]\notag\\
&\leq \frac{2}{C_2^{1/2}}\sup_{1\leq k_2\leq \overline{k}}E\left[\left|\frac{1}{C_2}\sum_{j=1}^{C_2}\psi_{k_2}(U_{0j})\right|\right]\leq \frac{2}{C_2}\sup_{1\leq k_2\leq \overline{k}}E(\psi_{k_2}^2(U_{0,1}))^{1/2}\leq \frac{2\sqrt{2}}{C_2}\label{eq:control_S_1_3}
\end{align}
Symmetrically, if $\{\ell,\ell'\}=\{2,3\}$, $\sup_{(\k,\k')\in \K_{\ell}(\overline{k})\times \K_{\ell'}(\overline{k})}E\left[\left|\frac{1}{C_1^2C_2^2}\sum_{i=1}^{C_1}\sum_{j=1}^{C_2}P_{ij}(\k,\k')\right|\right]\leq \frac{2\sqrt{2}}{C_1}$.\\ If $\{\ell,\ell'\}=\{1,4\}$, we obtain:
\begin{align}
		&\sup_{(\k,\k')\in \K_{\ell}(\overline{k})\times \K_{\ell'}(\overline{k})}E\left[\left|\frac{1}{C_1^2C_2^2}\sum_{i=1}^{C_1}\sum_{j=1}^{C_2}P_{ij}(\k,\k')\right|\right]\notag\\	&= \frac{1}{C_2^{1/2}}\sup_{(\k,\k')\in \K_{\ell}(\overline{k})\times \K_{\ell'}(\overline{k})}E\left[\left|\frac{1}{C_1C_2}\sum_{i=1}^{C_1}\sum_{j=1}^{C_2}\psi_{k_1}(U_{i0})\psi_{k'_1}(U_{i0})\psi_{k'_2}(U_{0j})\psi_{k'_3}(U_{ij})\right|\right]\notag\\
	&\leq \frac{1}{C_1^{1/2}C_2}\sup_{(\k,\k')\in \K_{\ell}(\overline{k})\times \K_{\ell'}(\overline{k})}E\left[\psi_{k_1}^2(U_{0,1})\psi_{k'_1}^2(U_{0,1})\psi_{k'_2}^2(U_{1,0})\psi_{k'_3}^2(U_{11})\right]^{1/2}\notag\\
	&\leq \frac{4}{C_1^{1/2}C_2}\label{eq:control_S_1_4},
\end{align}
and if $\{\ell,\ell'\}=\{2,4\}$, $\sup_{(\k,\k')\in \K_{\ell}(\overline{k})\times \K_{\ell'}(\overline{k})}E\left[\left|\frac{1}{C_1^2C_2^2}\sum_{i=1}^{C_1}\sum_{j=1}^{C_2}P_{ij}(\k,\k')\right|\right]\leq \frac{4}{C_1C_2^{1/2}}$.\\
If $\{\ell,\ell'\}=\{3,4\}$,
\begin{align}
	&\sup_{(\k,\k')\in \K_{\ell}(\overline{k})\times \K_{\ell'}(\overline{k})}E\left[\left|\frac{1}{C_1^2C_2^2}\sum_{i=1}^{C_1}\sum_{j=1}^{C_2}P_{ij}(\k,\k')\right|\right]\notag\\
	&= \sup_{(\k,\k')\in \K_{\ell}(\overline{k})\times \K_{\ell'}(\overline{k})}E\left[\left|\frac{1}{C_1C_2}\sum_{i=1}^{C_1}\sum_{j=1}^{C_2}\psi_{k_1}(U_{i0})\psi_{k_2}(U_{0j})\psi_{k'_1}(U_{i0})\psi_{k'_2}(U_{0j})\psi_{k'_3}(U_{ij})\right|\right]\notag\\
	&\leq \frac{1}{C_1^{1/2}C_2^{1/2}}\sup_{(\k,\k')\in \K_{\ell}(\overline{k})\times \K_{\ell'}(\overline{k})}E\left[\psi_{k_1}^2(U_{0,1})\psi_{k_2}^2(U_{1,0})\psi_{k'_1}^2(U_{0,1})\psi_{k'_2}^2(U_{1,0})\psi_{k'_3}^2(U_{11})\right]^{1/2}\notag\\
	&\leq \frac{4\sqrt{2}}{C_1^{1/2}C_2^{1/2}}.\label{eq:control_S_3_4}
\end{align}
As $\sup_{k,k',u}|\psi_k(u)\psi_{k'}(u)-\mathds{1}\{k=k'\}|=2$, we get:
\begin{align}
		&\sup_{(\k,\k')\in \K_{1}(\overline{k})^2}E\left[\left|\frac{1}{C_1^2C_2^2}\sum_{i=1}^{C_1}\sum_{j=1}^{C_2}P_{ij}(\k,\k')\right|\right]\notag\\
		&= \frac{1}{C_2}\sup_{1\leq k_1,k_1'\leq \overline{k}}E\left[\left|\frac{1}{C_1}\sum_{i=1}^{C_1}  \psi_{k_1}(U_{i0})\psi_{k'_1}(U_{i0})-\mathds{1}\{k_1=k'_1\}\right|\right]\notag\\
		&\leq \frac{1}{C_2C_1^{1/2}}\sup_{1\leq k_1,k_1'\leq \overline{k}}E\left[\left|  \psi_{k_1}(U_{i0})\psi_{k'_1}(U_{i0})-\mathds{1}\{k_1=k'_1\}\right|^2\right]^{1/2}\leq \frac{2}{C_2C_1^{1/2}}\label{eq:control_S_1},
\end{align}
and similar arguments ensure $\sup_{(\k,\k')\in \K_{2}(\overline{k})^2}E\left[\left|\frac{1}{C_1^2C_2^2}\sum_{i=1}^{C_1}\sum_{j=1}^{C_2}P_{ij}(\k,\k')\right|\right]\leq \frac{2}{C_1C_2^{1/2}}$.
We note that $\sup_{k_1,k'_1,k_2,k'_2,u,v}\left|\psi_{k_1}(u)\psi_{k_2}(v)\psi_{k'_1}(u)\psi_{k'_2}(v)-\mathds{1}\{k_1=k'_1,k_2=k'_2\}\right|=4$, which implies:
\begin{align}
 	&\sup_{(\k,\k')\in \K_{3}(\overline{k})^2}E\left[\left|\frac{1}{C_1^2C_2^2}\sum_{i=1}^{C_1}\sum_{j=1}^{C_2}P_{ij}(\k,\k')\right|\right]\notag \\
 	&=\sup_{(\k,\k')\in \K_{3}(\overline{k})^2}E\left[\left|\frac{1}{C_1C_2}\sum_{i,j}\psi_{k_1}(U_{i0})\psi_{k'_1}(U_{i0})\psi_{k_2}(U_{0j})\psi_{k'_2}(U_{0j})-\mathds{1}\{\k=\k'\}\right|\right]\notag\\
 	&\leq \frac{1}{C_1^{1/2}C_2^{1/2}}\sup_{(\k,\k')\in \K_{3}(\overline{k})^2}E\left[\left(\psi_{k_1}(U_{1,0})\psi_{k'_1}(U_{1,0})\psi_{k_2}(U_{0,1})\psi_{k'_2}(U_{0,1})-\mathds{1}\{\k=\k'\}\right)^2\right]^{1/2}\notag\\
 &\leq \frac{4}{C_1^{1/2}C_2^{1/2}},\label{eq:control_S_3}
 	\end{align}
Since $\sup_{k_1,k'_1,k_2,k'_2,k_3,k'_3,u,v,w}\left|\psi_{k_1}(u)\psi_{k_2}(v)\psi_{k_3}(w)\psi_{k'_1}(u)\psi_{k'_2}(v)\psi_{k'_3}(w)-\prod_{j=1}^3\mathds{1}\{k_j=k'_j\}\right|=8$, we have:
\begin{align}
	\sup_{(\k,\k')\in \K_{4}(\overline{k})^2}E\left[\left|\frac{1}{C_1^2C_2^2}\sum_{i=1}^{C_1}\sum_{j=1}^{C_2}P_{ij}(\k,\k')\right|\right]&\leq \frac{8}{C_1^{1/2}C_2^{1/2}}.\label{eq:control_S_4}
\end{align}
It follows from \eqref{eq:control_S}, \eqref{eq:control_S_1_2}, \eqref{eq:control_S_1_3}, \eqref{eq:control_S_1_4}, \eqref{eq:control_S_3_4}, \eqref{eq:control_S_1}, \eqref{eq:control_S_3} and \eqref{eq:control_S_4} that for some universal constant $K'_1$:
\begin{align}E\left[\left|\left|\widetilde{V}_{12}(\overline{k}) - V_{12,n}(\overline{k})\right|\right|\right]&\leq \frac{K'_1\overline{k}^3}{\Cinf}\sum_{\k \in \N^{3\ast}}||\nu^n_{\k}||^2.\label{eq:control_V12}
\end{align}
Next, we deduce from \eqref{eq:control_V1} and \eqref{eq:control_V12} that for some constant $K>0$:
\begin{align}
	\E\left[\left|\left| \Lambda_n(\overline{k}) - \Lambda_n^\infty(\overline{k})\right|\right|_\Lambda\right]&\leq \frac{K \overline{k}^3}{\Cinf^{1/2}}\left(\sum_{\k\in\N^{3\ast}}||\nu^n_{\k}||^2\right).
	\label{eq:expec_control}
\end{align}
Combining~\eqref{eq:portemanteau_intermediary} and~\eqref{eq:expec_control}, we conclude $\lim_{n}\left| \E[h(\Lambda_n(\overline{k}))] - \E[h(\Lambda_\infty(\overline{k}))]  \right|=0$ for any $\overline{k}>0$ and any bounded Lipschitz function $h$.

\medskip

\textbf{Substep 4: Conclusion}\\
Combining the results of Substeps 1, 2 and 3, we conclude that for any bounded Lipschitz function $h$ $$\lim_n	\left| \E[h(\Lambda_n)] - \E[h(\Lambda_\infty)]  \right|=0.$$
As mentioned above, this is equivalent to convergence in distribution of $\Lambda_n$ to $\Lambda_\infty$.
	
	
\medskip

\textbf{Second step: $\widetilde{V}_j=V_n\widehat{V}_jV_n+o_p(1)$ for $j=1,2,12$}\\
Note that by direct computations, 
$V_n\widehat{V}_1V_n=\widetilde{V}_1-\frac{1}{C_1}\left(V_n\overline{W}\right)^{\otimes 2}$, $V_n\widehat{V}_2V_n=\widetilde{V}_2-\frac{1}{C_2}\left(V_n\overline{W}\right)^{\otimes 2}$ and $V_n\widehat{V}_{12}V_n = \widetilde{V}_{12}-\frac{1}{C_1C_2}\left(V_n\overline{W}\right)^{\otimes 2}$. By definition of the matrix 2-norm, for any real vector $u$, we have $||u^{\otimes 2}|| \le \tr(u^{\otimes 2})$. It follows that
\begin{equation*}
	E\left[\left|\left|\left(V_n\overline{W}\right)^{\otimes 2}\right|\right|\right] \le E\left[\tr\left(\left(V_n\overline{W}\right)^{\otimes 2}\right)\right]=\tr \left(V_nE\left[\overline{W}^{\otimes 2}\right]V_n\right)= d.
\end{equation*} 
We conclude that $\left(V_n\overline{W},V_n(\widehat{V}_1,\widehat{V}_2,\widehat{V}_{12})V_n\right)$
converges in distribution to $\Lambda_{\infty}=\left(L,V_1,V_2,V_{12}\right)$. $\Box$

\subsection{Theorem \ref{thm:GMM}}

By compacity of $\Theta$ and twice continuous differentiability of $\beta \mapsto \psi(a,\beta)$ for every $a \in \R^{d_a}$, we note that the map
\begin{equation*}
	\beta \mapsto \left(\frac{1}{C_1C_2}\sum_{i=1}^{C_1}\sum_{j=1}^{C_2}\psi_{ij}(\beta)\right)'\Wh \left(\frac{1}{C_1C_2}\sum_{i=1}^{C_1}\sum_{j=1}^{C_2}\psi_{ij}(\beta)\right)
\end{equation*}
is minimized over $\Theta$ almost surely. This implies well-definition of $\widehat{\beta} \in \Theta$, a fact that will prove useful in the present proof.

\medskip

Let us introduce:
\begin{align*}
	\widehat{V}_1^{\inf} & := \frac{1}{C_1^2}\sum_{i=1}^{C_1}\left(\frac{1}{C_2}\sum_{j=1}^{C_2}\Psi_{ij} - \overline{\Psi} \right)^{\otimes 2}, \\
	\widehat{V}_2^{\inf} & := \frac{1}{C_2^2}\sum_{j=1}^{C_2}\left(\frac{1}{C_1}\sum_{i=1}^{C_1} \Psi_{ij} - \overline{\Psi} \right)^{\otimes 2}, \\
	\widehat{V}_{12}^{\inf} & := \frac{1}{(C_1C_2)^2}\sum_{i=1}^{C_1}\sum_{j=1}^{C_2} \left(\Psi_{ij} - \overline{\Psi}\right)^{\otimes 2}.
\end{align*}

The proof consists in three steps. We first show
\begin{equation}
	\label{eq:lin_num}
	\left|\left|\V(\overline{\Psi})^{-1/2}\left(\widehat{\theta}-\theta_0-\overline{\Psi}\right)\right|\right| = o_P(1),
\end{equation}
and
\begin{equation}
	\label{eq:lin_den}
	\left|\left|\V(\overline{\Psi})^{-1/2}(\widehat{V}_k-\widehat{V}_k^{\inf})\V(\overline{\Psi})^{-1/2}\right|\right| = o_P(1), \quad k \in \{1,2,u\},
\end{equation}
before applying Theorem~\ref{thm:point_univ} with $Y_{ij} = \Psi_{ij}$ to conclude.

\medskip

\textbf{Proof of~\eqref{eq:lin_num}:} Since $\Psi_{ij}=\deriv{\varphi(\beta_0)}{\beta}'(J'\W J)^{-1}J'\W \psi_{ij}(\beta_0)$, we have:
\begin{align*}
	\E\left[||\Psi_{11}||^2\right]\leq&\E\left[||\psi_{11}(\beta_0)||^2\right]\frac{\lambda_{\max}(\W )^2\lambda_{\max}(JJ')}{\lambda_{\min}(\W )^2\lambda_{\min}(J'J)^2}||\deriv{\varphi(\beta_0)}{\beta}||^2, \\ 
	\lambda_{\min}\left(\V(\Psi_{11})\right)\geq& \lambda_{\min}\left(\V\left(\psi_{11}(\beta_0)\right)\right)\frac{\lambda_{\min}(\W )^2\lambda_{\min}(J'J)}{\lambda_{\max}(\W )^2\lambda_{\max}(J'J)^2}||\deriv{\varphi(\beta_0)}{\beta}||^2. 
	\end{align*}
The quantities $\lambda_{\min}\left(\V\left(\psi_{11}(\beta_0)\right)\right)$, $\lambda_{\min}(J'J)$ and $\lambda_{\min}(\W )$ are positive, $\E\left[||\psi_{11}(\beta_0)||^2\right]$,  $\lambda_{\max}(JJ')=\lambda_{\max}(J'J)$ and $\lambda_{\max}(\W )$ are finite, and $||\deriv{\varphi(\beta_0)}{\beta}||$ is positive and finite. From this, we deduce that $\lambda_{\min}\left(\V(\Psi_{11})\right)>0$ and $\E\left[||\Psi_{11}||^2\right]<\infty$. This and Assumption~\ref{as:DGP_GMMs} ensures that Assumption \ref{as:DGP} holds for $\Psi_{ij}$. Moreover, because $\E\left[\psi_{11}(\beta_0)\right]=0$, we obtain $\E\left[\Psi_{11}\right]=0$. We thus get that Lemma~\ref{lem:mu_nu} hold as well for $\Psi_{ij}$, which implies $\lambda_{\min}\left(\V\left(\overline{\Psi}\right)\right)>0$ and $\V\left(\overline{\Psi}\right)^{-1/2}$ is well defined. 

\medskip 

Let $\mathbb{P}_n(\beta):=\frac{1}{C_1C_2}\sum_{i=1}^{C_1}\sum_{j=1}^{C_2}\psi_{ij}(\beta)$ and $\mathbb{P}(\beta):=\E\left(\psi_{11}(\beta)\right)$. As $\left|\left|\psi_{11}(\beta)-\psi_{11}(\tilde{\beta})\right|\right|\leq \left|\left|\sup_{b\in \Theta}\Deriv{\psi_{11}(b)}{\beta}\right|\right|\cdot\left|\left|\beta-\tilde{\beta}\right|\right|$ for any $(\beta,\tilde{\beta})\in \Theta^2$ and $\Theta$ is bounded, the class $\mathcal{F}:=\{(n,(a_\ell)_{\ell \ge 1}) \mapsto \sum_{\ell=1}^n\psi(a_\ell,\beta), \beta\in \Theta\}$ has finite $L^1(P)$-bracketing number $N_{[\,]}(\varepsilon,L^1(P),\mathcal{F})<\infty$ (see example 19.7 in \citep{vanderVaart2000}). Theorem 3.4 in \cite{ddg2021} ensures that $\sup_{\beta \in \Theta}\left|\left|\mathbb{P}_n(\beta)-\mathbb{P}(\beta)\right|\right|$ converges in probability to 0. Moreover for any $\beta\in\Theta$, $\left|\left|\mathbb{P}(\beta)\right|\right|\leq \E\left[\sup_{b\in \Theta}\left|\left|\Deriv{\psi(b)}{\beta}\right|\right|\right]\text{diam}(\Theta)$. Let $M_n(\beta):=-\mathbb{P}_n(\beta)'\Wh \mathbb{P}_n(\beta)$ and $M_{\infty}(\beta):=-\mathbb{P}(\beta)'\W\mathbb{P}(\beta)$. We have:
\begin{align*}
	\sup_{\beta \in \Theta}\left|M_n(\beta)-M_{\infty}(\beta)\right|\leq& \sup_{\beta\in \Theta}\left| (\mathbb{P}_n(\beta)-\mathbb{P}(\beta))'\Wh (\mathbb{P}_n(\beta)+\mathbb{P}(\beta))\right|\\
	&+\sup_{\beta \in \Theta}\left|\mathbb{P}(\beta)(\W -\Wh )\mathbb{P}(\beta)\right|\\
	\leq&\left(\sup_{\beta\in \Theta}\left|\left|\mathbb{P}_n(\beta)-\mathbb{P}(\beta)\right|\right|\right)^2 \left|\left|\Wh \right|\right|\\
	&+2\sup_{\beta\in \Theta}\left|\left|\mathbb{P}_n(\beta)-\mathbb{P}(\beta)\right|\right| \left|\left|\Wh \right|\right|\E\left[\sup_{\beta\in \Theta}\left|\left|\Deriv{\psi_{11}(\beta)}{\beta}\right|\right|\right]\text{diam}(\Theta)\\
	&+ \left|\left|\Wh-\W\right|\right|\left(\E\left[\sup_{\beta\in \Theta}\left|\left|\Deriv{\psi_{11}(\beta)}{\beta}\right|\right|\right]\text{diam}(\Theta)\right)^2
	\end{align*}
We remark that $\left|\left|\Wh \right|\right|\leq\left|\left|\Wh-\W\right|\right|+\left|\left|\W \right|\right|$. This plus the convergence in probability of $\left|\left|\Wh-\W\right|\right|$ and $\sup_{\beta\in \Theta}\left|\left|\mathbb{P}_n(\beta)-\mathbb{P}(\beta)\right|\right|$ to 0 ensure that $\sup_{\beta \in \Theta}\left|M_n(\beta)-M_{\infty}(\beta)\right|$ converges to 0 in probability as well. We also have
\begin{equation*}
	\sup_{\beta:||\beta-\beta_0||\geq \epsilon}M_{\infty}(\beta)\leq -\inf_{\beta:||\beta-\beta_0||\geq \epsilon}\left|\left|E\left[\psi_{11}(\beta)\right]\right|\right|^2\lambda_{\min}(\W )<0=M_{\infty}(\beta_0).
\end{equation*}
Next, Theorem 5.7 in \cite{vanderVaart2000} yields the convergence in probability of $\widehat{\beta}$ to $\beta_0$.

\medskip

Our final goal is to prove that $\left|\left||\V(\overline{\Psi})^{-1/2}\left(\widehat{\theta}-\varphi(\theta_0)-\overline{\Psi}\right)\right|\right| = o_P(1)$. Let $\mathcal{V}_\eta := \left\{\beta \in\Theta : ||\beta-\beta_0||\le \eta\right\}$, $\mathbb{D}_n(b):=\frac{1}{C_1C_2}\sum_{i=1}^{C_1}\sum_{j=1}^{C_2}\Deriv{\psi_{ij}(b)}{\beta}$ and $\mathbb{D}(b):=\E\left[\Deriv{\psi_{11}(b)}{\beta}\right]$. To derive our result, we first show that for every fixed $\eta>0$, the event 
\begin{equation*}
	\mathcal{E}_n := \left\{ \widehat{\beta} \in \mathcal{V}_\eta\right\} \cap \left\{ ||\Wh-\W|| \le \eta \right\} \cap \left\{ \sup_{\beta\in\Theta}\left|\left| \mathbb{D}_n(\beta)-\mathbb{D}(\beta) \right|\right| \le \eta \right\}
\end{equation*}
has probability tending to one. Consistency of $\widehat{\beta}$ and $\Wh $, and Theorem 3.4 in \cite{ddg2021} yield this result immediately. Second, we show that for $\eta>0$ small enough, the following holds on $\mathcal{E}_n$: $\widehat{J}\Wh \mathbb{P}_n(\widehat{\beta})=0$ and $\inf_{\beta_1,...,\beta_q \in \mathcal{V}_\eta^q}\lambda_{min}\left(\sum_{k=1}^qe_ke_k'\widehat{J}'\Wh \mathbb{D}_n(\beta_k)\right)>0$. For $\eta$ small enough, $\mathcal{V}_\eta \subset \overset{\circ}{\Theta}$. As a result, we obtain $\deriv{M_n(\widehat{\beta})}{\beta}=0$ or equivalently $\widehat{J}\Wh \mathbb{P}_n(\widehat{\beta})=0$. Recall that $\widehat{J} = \mathbb{D}_n(\widehat{\beta})$ and let $A:= \sup_{\beta\in\Theta}\left|\left| \mathbb{D}_n(\beta)-\mathbb{D}(\beta) \right|\right|$, $B := ||\Wh-\W||$ and $C := E\left[\sup_{b\in \Theta}\left|\left|\Deriv{{}^2\psi_{11}(b)}{\beta\partial \beta}\right|\right|\right] \sup_{\beta \in \mathcal{V}_\eta}\left|\left|\beta-\beta_0\right|\right| = \eta E\left[\sup_{b\in \Theta}\left|\left|\Deriv{{}^2\psi_{11}(b)}{\beta\partial \beta}\right|\right|\right]$. By repeated use of the triangle inequality and submultiplicativity of the matrix 2-norm, we can write for $(e_1,...,e_q)$ the canonical basis of $\R^q$ and for any $\beta_1,...,\beta_q \in \mathcal{V}_\eta^q$
\begin{align}
	\left|\left| \sum_{k=1}^qe_ke_k'\widehat{J}'\Wh \mathbb{D}_n(\beta_k) - J'\W J \right|\right| &=\left|\left| \sum_{k=1}^qe_ke_k'\left[\widehat{J}'\Wh \mathbb{D}_n(\beta_k) - J'\W J\right] \right|\right|\notag\\ &\le q\left(||\W || + B\right) \left[ (A+C)^2 + ||J||(A+C) + ||J||^2B \right]\notag\\& \le \gamma(\eta),
	\label{eq:control_sandwich}
\end{align}
with $\gamma: \Rb_+ \to \Rb_+$ strictly increasing and such that $\gamma(0) = 0$ ($\gamma$ depends on $q$, $||\W ||$, $||J||$ and $E\left[\sup_{b\in \Theta}\left|\left|\Deriv{{}^2\psi_{11}(b)}{\beta\partial \beta}\right|\right|\right]$). For $\eta$ sufficiently small $\gamma(\eta) \le \lambda_{min}(J'\W J)/2$, and then $\inf_{\beta_1,...,\beta_q \in \mathcal{V}_\eta^q}\lambda_{min}\left(\sum_{k=1}^qe_ke_k'\widehat{J}'\Wh \mathbb{D}_n(\beta_k)\right)>0$.

\medskip

We are now in a position to conclude. On $\mathcal{E}_n$, the Taylor Theorem with Lagrange remainder ensures existence of $t_1,...,t_q\in[0,1]^q$ and $\widetilde{\beta}_k:=t_k\widehat{\beta}+(1-t_k)\beta_{0}$ such that $0=e'_k\widehat{J}'\Wh  \mathbb{P}_n(\beta_0)+e'_k\widehat{J}'\Wh \sum_{k=1}^p\mathbb{D}_n(\widetilde{\beta}_k)(\widehat{\beta}-\beta_{0})$. Since $\sum_{k=1}^qe_ke_k'=\Id$ and $\sum_{k=1}^qe_ke_k'\widehat{J}'\Wh \mathbb{D}_n(\widetilde{\beta}_k)$ is non singular on $\mathcal{E}_n$, we can write $(\widehat{\beta}-\beta_0) \mathds{1}_{\{\mathcal{E}_n\}}= -\widetilde{T}\mathbb{P}_n(\beta_0)\mathds{1}_{\{\mathcal{E}_n\}}$ with $\widetilde{T}:=(\sum_{k=1}^qe_ke'_k\widehat{J}'\Wh \mathbb{D}_n(\widetilde{\beta}_k))^{-1}\widehat{J}'\Wh $. Letting $T:=(J'\W J)^{-1}J'\W $, we have $\overline{\Psi}=-\deriv{\varphi(\beta_0)}{\beta}'T\mathbb{P}_n(\beta_0)$ and \\ $\left|\left|\V(\overline{\Psi})^{-1/2}\right|\right|\leq\Big(\lambda_{\min}\left(\V(\mathbb{P}_n(\beta_0))\right) \times \lambda_{\min}(TT')||\deriv{\varphi(\beta_0)}{\beta}||^2\Big)^{-1/2}$. Let $L := \sup_{\beta\in\overset{\circ}{\Theta}} \|\Deriv{^{2}\varphi(\beta)}{\beta\partial \beta'}\|$ (finite by Assumption~\ref{as:GMM_regul}). All our previous obserbations, together with a Taylor-Lagrange expansion of order two of $\varphi(\cdot)$ and repeated use of the triangle and Cauchy-Schwarz inequalities enable us to write: 
\begin{align*}
	& \left|\left|\V(\overline{\Psi})^{-1/2}\left(\widehat{\theta}-\theta_0-\overline{\Psi}\right)\right|\right| \\
	\leq
	&
	\left|\left|\V(\overline{\Psi})^{-1/2}\left(\widehat{\theta}-\theta_0-\overline{\Psi}\right)\mathds{1}_{\{\mathcal{E}_n\}}\right|\right|+\left|\left|\V(\overline{\Psi})^{-1/2}\left(\widehat{\theta}-\theta_0-\overline{\Psi}\right)\mathds{1}_{\{\mathcal{E}_n^c\}}\right|\right| \\
	\leq & \left|\left|\V(\overline{\Psi})^{-1/2}\left(\deriv{\varphi(\beta_0)}{\beta}'(\widehat{\beta}-\beta_0)  -\overline{\Psi}\right)\mathds{1}_{\{\mathcal{E}_n\}}\right|\right| \\
	& + \frac{L}{2}\left|\left|\V(\overline{\Psi})^{-1/2}\right|\right|\,\left|\left|\widehat{\beta}-\beta_0\right|\right|^2\mathds{1}_{\{\mathcal{E}_n\}}| + \left|\left|\V(\overline{\Psi})^{-1/2}\left(\widehat{\theta}-\theta_0-\overline{\Psi}\right)\mathds{1}_{\{\mathcal{E}_n^c\}}\right|\right| \\
	\leq & \left|\left|\V(\overline{\Psi})^{-1/2}\deriv{\varphi(\beta_0)}{\beta}'\left(T-\widetilde{T}\right)\mathbb{P}_n(\beta_0)\mathds{1}_{\{\mathcal{E}_n\}}\right|\right| \\
	& + \frac{L}{2}\left|\left|\V(\overline{\Psi})^{-1/2}\right|\right|\,\left|\left|\widetilde{T}\right|\right|^2\,\left|\left|\mathbb{P}_n(\beta_0)\right|\right|^2\mathds{1}_{\{\mathcal{E}_n\}} + \left|\left|\V(\overline{\Psi})^{-1/2}\left(\widehat{\theta}-\theta_0-\overline{\Psi}\right)\mathds{1}_{\{\mathcal{E}_n^c\}}\right|\right| \\
	\leq & \left( \frac{\lambda_{\max}\left(\V(\mathbb{P}_n(\beta_0)\right)}{\lambda_{\max}\left(\V(\mathbb{P}_n(\beta_0)\right)}\lambda_{\min}(TT')^{-1}\right)^{1/2} \left|\left|T-\widetilde{T}\right|\right|\,\left|\left|\V\left(\mathbb{P}_n(\beta_0)\right)^{-1/2}\mathbb{P}_n(\beta_0)\right|\right|\mathds{1}_{\{\mathcal{E}_n\}} \\
	& + L\left( \frac{\lambda_{\max}\left(\V(\mathbb{P}_n(\beta_0)\right)}{\lambda_{\max}\left(\V(\mathbb{P}_n(\beta_0)\right)}\lambda_{\min}(TT')^{-1}||\deriv{\varphi(\beta_0)}{\beta}||^{-2}\right)^{1/2} \left|\left|T-\widetilde{T}\right|\right|^2\mathds{1}_{\{\mathcal{E}_n\}} \\
	& \times \sqrt{\lambda_{\max}\left(\V(\mathbb{P}_n(\beta_0)\right)}\left|\left|\V\left(\mathbb{P}_n(\beta_0)\right)^{-1/2}\mathbb{P}_n(\beta_0)\right|\right|^2 \\
	& + L\left( \frac{\lambda_{\max}\left(\V(\mathbb{P}_n(\beta_0)\right)}{\lambda_{\max}\left(\V(\mathbb{P}_n(\beta_0)\right)}\lambda_{\min}(TT')^{-1}||\deriv{\varphi(\beta_0)}{\beta}||^{-2}\right)^{1/2}\left|\left|\widetilde{T}\right|\right|^2\mathds{1}_{\{\mathcal{E}_n\}} \\
	& \times \sqrt{\lambda_{\max}\left(\V(\mathbb{P}_n(\beta_0)\right)}\left|\left|\V\left(\mathbb{P}_n(\beta_0)\right)^{-1/2}\mathbb{P}_n(\beta_0)\right|\right|^2 \\
	& +\left|\left|\V(\overline{\Psi})^{-1/2}\left(\widehat{\theta}-\theta_0-\overline{\Psi}\right)\right|\right|\mathds{1}_{\{\mathcal{E}_n^c\}}.
\end{align*}
Theorem 3.4 in \cite{ddg2021}, applied to $\mathbb{D}_n$, and the continuous mapping in probability (applied to $\mathbb{D}$ and to the matrix inverse operator in particular) imply $\left|\left|T-\widetilde{T}\right|\right|\mathds{1}_{\{\mathcal{E}_n\}} = o_P(1)$ and $\left|\left|\widetilde{T}\right|\right|\mathds{1}_{\{\mathcal{E}_n\}} = O_P(1)$.
Combining this with $\frac{\lambda_{\max}\left(\V(\mathbb{P}_n(\beta_0)\right)}{\lambda_{\min}\left(\V(\mathbb{P}_n(\beta_0)\right)}=O(1)$,  $\left|\left|\V\left(\mathbb{P}_n(\beta_0)\right)^{-1/2}\mathbb{P}_n(\beta_0)\right|\right|=O_P(1)$ (implied in particular by Assumption~\ref{as:GMM_main}), $\lambda_{\max}\left(\V(\mathbb{P}_n(\beta_0)\right) = o_P(1)$, $||\deriv{\varphi(\beta_0)}{\beta}||>0$ and $\lim_n P(\mathcal{E}_n^c)=0$, we conclude that~\eqref{eq:lin_num} holds.

\medskip

\textbf{Proof of~\eqref{eq:lin_den}:} Let $\widehat{M}:=\deriv{\varphi(\widehat{\beta})}{\beta}'(\widehat{J}'\Wh \widehat{J})^{+}\widehat{J}'\Wh $, we already know that $\widehat{M}$ converges to $M$ in probability. Let $\mathbb{Q}_{1,n}(\beta):=\frac{1}{C_1C_2^2}\sum_{i=1}^{C_1}\left(\sum_{j=1}^{C_2}\left(\psi_{ij}(\beta)-\mathbb{P}_n(\beta)\right)\right)^{\otimes 2} \\ =\frac{1}{C_1C_2^2}\sum_{i=1}^{C_1}\left(\sum_{j=1}^{C_2}\psi_{ij}(\beta)\right)^{\otimes 2}-\mathbb{P}_n(\beta)^{\otimes 2}$. 
We have:
\begin{align*}
	&\left|\left|\V(\overline{\Psi})^{-1/2}(\widehat{V}_1-\widehat{V}_1^{\inf})\V(\overline{\Psi})^{-1/2}\right|\right|\\
	&=\frac{1}{C_1} \left|\left|\V(\overline{\Psi})^{-1/2}\left(\widehat{M}\mathbb{Q}_{1,n}(\widehat{\beta})\widehat{M}'-M\mathbb{Q}_{1,n}(\beta_0)M'\right)\V(\overline{\Psi})^{-1/2}\right|\right|\\
	&\leq\frac{1}{C_1}\left|\left|\V(\overline{\Psi})^{-1}\right|\right|\,\left|\left|\V\left(\mathbb{P}_n(\beta_0)\right)\right|\right|\,\left|\left|\V(\mathbb{P}_n(\beta_0))^{-1/2}\left(\mathbb{Q}_{1,n}(\widehat{\beta})-\mathbb{Q}_{1,n}(\beta_0)\right)\V(\mathbb{P}_n(\beta_0))^{-1/2}\right|\right|\,\left|\left|\widehat{M}\right|\right|^2\\
	&+\frac{1}{C_1}\left|\left|\V(\overline{\Psi})^{-1}\right|\right|\,\left|\left|\V\left(\mathbb{P}_n(\beta_0)\right)\right|\right|\,\left|\left| \V\left(\mathbb{P}_n(\beta_0)\right)^{-1/2}\mathbb{Q}_{1,n}(\beta_0)\V\left(\mathbb{P}_n(\beta_0)\right)^{-1/2}\right|\right| \times \\
	& \,\,\,\, \left(2\left|\left|M\right|\right| \left|\left|\widehat{M}-M\right|\right|+  \left|\left|\widehat{M}-M\right|\right|^2\right)
	\end{align*}
 Since $\left|\left|\V(\overline{\Psi})^{-1}\right|\right|\,\left|\left|\V\left(\mathbb{P}_n(\beta_0)\right)\right|\right|=O_P(1)$ and $\left|\left|\widehat{M}-M\right|\right|=o_P(1)$, we just have to prove that $$\frac{1}{C_1}\left|\left| \V\left(\mathbb{P}_n(\beta_0)\right)^{-1/2}\mathbb{Q}_{1,n}(\beta_0)\V\left(\mathbb{P}_n(\beta_0)\right)^{-1/2}\right|\right|=O_P(1),$$ and $$\frac{1}{C_1}\left|\left|\V(\mathbb{P}_n(\beta_0))^{-1/2}\left(\mathbb{Q}_{1,n}(\widehat{\beta})-\mathbb{Q}_{1,n}(\beta_0)\right)\V(\mathbb{P}_n(\beta_0))^{-1/2}\right|\right|=o_P(1),$$ to ensure~\eqref{eq:lin_den} is satisfied for $\widehat{V}_1$ and $\widehat{V}_1^{inf}$. \\
  
We have: $\frac{1}{C_1}\V\left(\mathbb{P}_n(\beta_0)\right)^{-1/2}\mathbb{Q}_{1,n}(\beta_0)\V\left(\mathbb{P}_n(\beta_0)\right)^{-1/2}=\frac{1}{C^2_1C^2_2}\sum_{i=1}^{C_1}\left(\sum_{j=1}^{C_2}\V\left(\mathbb{P}_n(\beta_0)\right)^{-1/2}\psi_{ij}(\beta_0)\right)^{\otimes 2}-\frac{1}{C_1}\V\left(\mathbb{P}_n(\beta_0)\right)^{-1/2}\mathbb{P}_n(\beta_0)^{\otimes 2}\V\left(\mathbb{P}_n(\beta_0)\right)^{-1/2}$. Given $\V\left(\mathbb{P}_n(\beta_0)\right)^{-1/2}\mathbb{P}_n(\beta_0)^{\otimes 2}\V\left(\mathbb{P}_n(\beta_0)\right)^{-1/2}$ is symmetric positive, we obtain the following inequality: $$\left|\left|\frac{1}{C_1}\V\left(\mathbb{P}_n(\beta_0)\right)^{-1/2}\mathbb{Q}_{1,n}(\beta_0)\V\left(\mathbb{P}_n(\beta_0)\right)^{-1/2}\right|\right|\leq \frac{1}{C^2_1C^2_2}\sum_{i=1}^{C_1}\left|\left|\sum_{j=1}^{C_2}\V\left(\mathbb{P}_n(\beta_0)\right)^{-1/2}\psi_{ij}(\beta_0)\right|\right|^{ 2}.$$ 
We also have:
\begin{align*}
	&\E\left[\frac{1}{C_1^2}\sum_{i=1}^{C_1}\left|\left| \frac{1}{C_2}\sum_{j=1}^{C_2}\V\left(\mathbb{P}_n(\beta_0)\right)^{-1/2}\psi_{ij}(\beta_0) \right|\right|^2\right]\\
	&=\E\left(\frac{1}{C_1C_2^2}\sum_{j,j'=1}^{C_2}\psi_{1j}(\beta_0)'\V\left(\mathbb{P}_n(\beta_0)\right)^{-1}\psi_{1j'}(\beta_0)\right)\\
	&=\E\left(\frac{1}{C_1C_2^2}\sum_{j,j'=1}^{C_2}\tr\left[\psi_{1j}(\beta_0)'\V\left(\mathbb{P}_n(\beta_0)\right)^{-1}\psi_{1j'}(\beta_0)\right]\right)\\
	&=\E\left(\frac{1}{C_1C_2^2}\sum_{j,j'=1}^{C_2}\tr\left[\V\left(\mathbb{P}_n(\beta_0)\right)^{-1/2}\psi_{1j'}(\beta_0)\psi_{1j}(\beta_0)'\V\left(\mathbb{P}_n(\beta_0)\right)^{-1/2}\right]\right)\\
	&=\tr\left[\V\left(\mathbb{P}_n(\beta_0)\right)^{-1/2}\E\left(\frac{1}{C_1C_2^2}\sum_{j,j'=1}^{C_2}\psi_{1j'}(\beta_0)\psi_{1j}(\beta_0)'\right)\V\left(\mathbb{P}_n(\beta_0)\right)^{-1/2}\right]\\
	&= \tr\left[\V\left(\mathbb{P}_n(\beta_0)\right)^{-1/2}\left[ \frac{1}{C_1C_2}\V\left(\psi_{11}(\beta_0)\right)+\frac{C_2-1}{C_1C_2}\E\left(\psi_{11}(\beta_0)\psi_{12}(\beta_0)'\right)\right]\V\left(\mathbb{P}_n(\beta_0)\right)^{-1/2}\right]\\
	&\leq \tr(\Id)=p,
\end{align*}
where the last inequality follows from $\V(\mathbb{P}_n(\beta_0))=\frac{1}{C_1C_2}\V\left(\psi_{11}(\beta_0)\right)+\frac{C_2-1}{C_1C_2}\E\left(\psi_{11}(\beta_0)\psi_{12}(\beta_0)'\right)+\frac{C_1-1}{C_1C_2}\E\left(\psi_{11}(\beta_0)\psi_{21}(\beta_0)'\right)$ with $\E\left(\psi_{11}(\beta_0)\psi_{12}(\beta_0)'\right)$ and $\E\left(\psi_{11}(\beta_0)\psi_{21}(\beta_0)'\right)$ symmetric positive definite. All in all, we can claim by Markov's inequality that $$\frac{1}{C_1} \V\left(\mathbb{P}_n(\beta_0)\right)^{-1/2}\mathbb{Q}_{1,n}(\beta_0)\V\left(\mathbb{P}_n(\beta_0)\right)^{-1/2}=O_P(1).$$
Let us now consider the following decomposition:
\begin{align*}
	\frac{1}{C_1}\V(\mathbb{P}_n(\beta_0))^{-1/2}\left(\mathbb{Q}_{1,n}(\widehat{\beta})-\mathbb{Q}_{1,n}(\beta_0)\right)\V(\mathbb{P}_n(\beta_0))^{-1/2}&=R_1+R_1'+R_2+R_3+R_3'+R_4,
	\end{align*}
where
\begin{align*}
	R_1 & := \frac{1}{C_1^2}\sum_{i=1}^{C_1}\left(\frac{1}{C_2}\sum_{j=1}^{C_2}\V(\mathbb{P}_n(\beta_0))^{-1/2}\big(\psi_{ij}(\widehat{\beta}) - \psi_{ij}(\beta_0)\big)\right)\left(\frac{1}{C_2}\sum_{j=1}^{C_2}\V(\mathbb{P}_n(\beta_0))^{-1/2}\psi_{ij}(\beta_0)\right)',\\
	R_2 & := \frac{1}{C_1^2}\sum_{i=1}^{C_1}\left(\frac{1}{C_2}\sum_{j=1}^{C_2}\V(\mathbb{P}_n(\beta_0))^{-1/2}\big(\psi_{ij}(\widehat{\beta}) - \psi_{ij}(\beta_0)\big)\right)^{\otimes 2}, \\
	R_3 & := -C_1^{-1}\left(\frac{1}{C_1C_2}\sum_{i=1}^{C_1}\sum_{j=1}^{C_2}\V(\mathbb{P}_n(\beta_0))^{-1/2}\big(\psi_{ij}(\widehat{\beta}) - \psi_{ij}(\beta_0)\big)\right)\left(\V(\mathbb{P}_n(\beta_0))^{-1/2}\mathbb{P}_n(\beta_0)\right)', \\
	R_4 & := -C_1^{-1}\left(\frac{1}{C_1C_2}\sum_{i=1}^{C_1}\sum_{j=1}^{C_2}\V(\mathbb{P}_n(\beta_0))^{-1/2}\big(\psi_{ij}(\widehat{\beta}) - \psi_{ij}(\beta_0)\big)\right)^{\otimes 2}.
\end{align*}
Triangle, Cauchy-Schwarz and Lipschitz inequalities ensure:
\begin{align*}
	\left|\left|R_1\right|\right|&\leq \frac{1}{C_1^2}\sum_{i=1}^{C_1}\left|\left|\frac{1}{C_2}\sum_{j=1}^{C_2}\V(\mathbb{P}_n(\beta_0))^{-1/2}\big(\psi_{ij}(\widehat{\beta}) - \psi_{ij}(\beta_0)\big)\right|\right|\left|\left|\frac{1}{C_2}\sum_{j=1}^{C_2}\V(\mathbb{P}_n(\beta_0))^{-1/2}\psi_{ij}(\beta_0)\right|\right|\\
&\leq \frac{\left|\left|\widehat{\beta}-\beta_0\right|\right|}{C_1^2}\left|\left|\V(\mathbb{P}_n(\beta_0))^{-1/2}\right|\right|\sum_{i=1}^{C_1}\left(\frac{1}{C_2}\sum_{j=1}^{C_2}\sup_{b\in \Theta}\left|\left|\Deriv{\psi_{ij}(b)}{\beta}\right|\right|\right)\left|\left|\frac{1}{C_2}\sum_{j=1}^{C_2}\V(\mathbb{P}_n(\beta_0))^{-1/2}\psi_{ij}(\beta_0)\right|\right|\\
	&\leq \frac{||\hat{\beta}-\beta_0||}{\sqrt{C_1}}\times  \lambda_{\min}^{-1/2}\left(\V(\mathbb{P}_n(\beta_0))\right)\\
	&~~~ \times \sqrt{\frac{1}{C_1C_2}\sum_{i,j}\sup_{b\in \Theta}\left|\left|\Deriv{\psi_{ij}(b)}{\beta}\right|\right|^2} \sqrt{\frac{1}{C_1^2C_2^2}\sum_{i=1}^{C_1}\left|\left| \sum_{j=1}^{C_2}\V\left(\mathbb{P}_n(\beta_0)\right)^{-1/2}\psi_{ij}(\beta_0) \right|\right|^2}.
	\end{align*}

We have previously shown that $\frac{1}{C_1^2C_2^2}\sum_{i=1}^{C_1}\left|\left| \sum_{j=1}^{C_2}\V\left(\mathbb{P}_n(\beta_0)\right)^{-1/2}\psi_{ij}(\beta_0) \right|\right|^2=O_P(1)$. We have also established that $(\widehat{\beta}-\beta_0)\mathds{1}\{\mathcal{E}_n\}=-\left(\widehat{J}'\Wh \mathbb{D}_n(\widetilde{\beta})\right)^{+}\widehat{J}\Wh \mathbb{P}_n(\beta_0)\mathds{1}\{\mathcal{E}_n\}$. As $\lim_n P(\mathcal{E}_n)=1$ and $ \left(\widehat{J}'\Wh \mathbb{D}_n(\widetilde{\beta})\right)^{+}\widehat{J}\Wh $ converges in probability to $\left(J'\W J\right)^{-1}J'\W $, it follows that $\left|\left|\widehat{\beta}-\beta_0\right|\right|=O_P\left(\left|\left|\mathbb{P}_n(\beta_0)\right|\right|\right)=O_p\left(\left|\left|\V\left(\mathbb{P}_n(\beta_0)\right)\right|\right|^{1/2}\right)=O_p\left(\lambda_{\max}^{1/2}\left(\V\left(\mathbb{P}_n(\beta_0)\right)\right)\right)$, and finally $\frac{||\hat{\beta}-\beta_0||}{\lambda_{\max}^{1/2}\left(\V(\mathbb{P}_n(\beta_0))\right)}=O_P(1)$. Combining the previous remarks with $$E\left(\frac{1}{C_1C_2}\sum_{i,j}\sup_{b\in \Theta}\left|\left|\Deriv{\psi_{ij}(b)}{\beta}\right|\right|^2\right)=E\left(\sup_{b\in \Theta}\left|\left|\Deriv{\psi_{11}(b)}{\beta}\right|\right|^2\right)<\infty$$ 
and $\frac{\lambda_{\max}^{1/2}\left(\V(\mathbb{P}_n(\beta_0))\right)}{\lambda_{\min}^{1/2}\left(\V(\mathbb{P}_n(\beta_0))\right)}=O(1)$, we conclude that $\left|\left|R_1\right|\right|=\left|\left|R_1'\right|\right|=O_P(C_1^{-1/2})=o_P(1)$. As for $R_2$, we can write
\begin{align*}
	||R_2|| \le C_1^{-1}\frac{||\widehat{\beta}-\beta_0||^2}{\lambda_{\min}\left(\V(\mathbb{P}_n(\beta_0))\right)}\frac{1}{C_1C_2}\sum_{i=1}^{C_1}\sum_{j=1}^{C_2}\sup_{b \in \Theta}\left|\left|\Deriv{\psi_{ij}(b)}{\beta}\right|\right|^2,
\end{align*}
and the right-hand side is $O_P(C_1^{-1}) = o_P(1)$, following the reasoning for $R_1$. Analogous arguments yield $\left|\left|R_3\right|\right| \vee \left|\left|R_4\right|\right| = O_P(C_1^{-1}) = o_P(1)$. This ensures that~\eqref{eq:lin_den} holds for $\widehat{V}_1$ and $\widehat{V}^{\inf}_1$. Similar reasoning holds for $\widehat{V}_2$ and $\widehat{V}^{\inf}_2$. 

\medskip

Let $\mathbb{Q}_{12,n}(\beta):=\frac{1}{C_1C_2}\sum_{i=1}^{C_1}\sum_{j=1}^{C_2}\left(\psi_{ij}(\beta)-\mathbb{P}_n(\beta)\right)^{\otimes 2}=\frac{1}{C_1C_2}\sum_{i=1}^{C_1}\sum_{j=1}^{C_2}\psi_{ij}(\beta)^{\otimes 2}-\mathbb{P}_n(\beta)^{\otimes 2}$. We can write:
\begin{align*}
	&\left|\left|\V(\overline{\Psi})^{-1/2}(\widehat{V}_{12}-\widehat{V}_{12}^{\inf})\V(\overline{\Psi})^{-1/2}\right|\right|\\
	&\leq\frac{1}{C_1C_2}\left|\left|\V(\overline{\Psi})\right|\right|^{-1}\left|\left|\V\left(\mathbb{P}_n(\beta_0)\right)\right|\right|\left|\left|\V(\mathbb{P}_n(\beta_0))^{-1/2}\left(\mathbb{Q}_{12,n}(\widehat{\beta})-\mathbb{Q}_{12,n}(\beta_0)\right)\V(\mathbb{P}_n(\beta_0))^{-1/2}\right|\right|\left|\left|\widehat{M}\right|\right|^2\\
	&+\frac{1}{C_1C_2}\left|\left|\V(\overline{\Psi})\right|\right|^{-1}\left|\left|\V\left(\mathbb{P}_n(\beta_0)\right)\right|\right|\left|\left| \V\left(\mathbb{P}_n(\beta_0)\right)^{-1/2}\mathbb{Q}_{12,n}(\beta_0)\V\left(\mathbb{P}_n(\beta_0)\right)^{-1/2}\right|\right| \times \\ 
	& \,\,\,\, \left(2\left|\left|M\right|\right| \left|\left|\widehat{M}-M\right|\right|+  \left|\left|\widehat{M}-M\right|\right|^2\right)\\
	&\leq \frac{1}{C_1C_2}\left|\left|\V(\mathbb{P}_n(\beta_0))^{-1/2}\left(\mathbb{Q}_{12,n}(\widehat{\beta})-\mathbb{Q}_{12,n}(\beta_0)\right)\V(\mathbb{P}_n(\beta_0))^{-1/2}\right|\right| O_P(1)\\
	&+\frac{1}{C_1C_2}\left|\left| \V\left(\mathbb{P}_n(\beta_0)\right)^{-1/2}\mathbb{Q}_{12,n}(\beta_0)\V\left(\mathbb{P}_n(\beta_0)\right)^{-1/2}\right|\right|o_P(1)
\end{align*}
Using the fact that $ \V\left(\mathbb{P}_n(\beta_0)\right)^{-1/2}\mathbb{Q}_{12,n}(\beta_0)\V\left(\mathbb{P}_n(\beta_0)\right)^{-1/2}=\frac{1}{C_1C_2}\sum_{i=1}^{C_1}\sum_{j=1}^{C_2}\left(\V\left(\mathbb{P}_n(\beta_0)\right)^{-1/2}\psi_{ij}(\beta_0)\right)^{\otimes 2}-\left(\V\left(\mathbb{P}_n(\beta_0)\right)^{-1/2}\mathbb{P}_n(\beta_0)\right)^{\otimes 2}$
and that $\left(\V\left(\mathbb{P}_n(\beta_0)\right)^{-1/2}\mathbb{P}_n(\beta_0)\right)^{\otimes 2}$ is a symmetric positive matrix, we get 
$$\left|\left|\V\left(\mathbb{P}_n(\beta_0)\right)^{-1/2}\mathbb{Q}_{12,n}(\beta_0)\V\left(\mathbb{P}_n(\beta_0)\right)^{-1/2}\right|\right|\leq \frac{1}{C_1C_2}\sum_{i=1}^{C_1}\sum_{j=1}^{C_2}\left|\left|\V\left(\mathbb{P}_n(\beta_0)\right)^{-1/2}\psi_{ij}(\beta_0)\right|\right|^2.$$
We also have:
\begin{align*}
E\left[	\frac{1}{C_1^2C_2^2}\sum_{i=1}^{C_1}\sum_{j=1}^{C_2}\left|\left|\V\left(\mathbb{P}_n(\beta_0)\right)^{-1/2}\psi_{ij}(\beta_0)\right|\right|^2\right]&=\frac{1}{C_1C_2}E\left(\psi_{11}(\beta_0)'\V\left(\mathbb{P}_n(\beta_0)\right)^{-1}\psi_{11}(\beta_0)\right)\\
&=\tr\left[\V\left(\mathbb{P}_n(\beta_0)\right)^{-1/2}\frac{1}{C_1C_2}\V(\psi_{11}(\beta_0))\V\left(\mathbb{P}_n(\beta_0)\right)^{-1/2}\right]\\
&\leq \tr(\Id)=p
	\end{align*}
where the last inequality follows from $\V(\mathbb{P}_n(\beta_0))=\frac{1}{C_1C_2}\V\left(\psi_{11}(\beta_0)\right)+\frac{C_2-1}{C_1C_2}\E\left(\psi_{11}(\beta_0)\psi_{12}(\beta_0)'\right)+\frac{C_1-1}{C_1C_2}\E\left(\psi_{11}(\beta_0)\psi_{21}(\beta_0)'\right)$ with $\E\left(\psi_{11}(\beta_0)\psi_{12}(\beta_0)'\right)$ and $\E\left(\psi_{11}(\beta_0)\psi_{21}(\beta_0)'\right)$ some symmetric positive matrices.
We conclude that 
\begin{equation}
	\label{eq:control_V_12_1}
	\frac{1}{C_1C_2}\left|\left| \V\left(\mathbb{P}_n(\beta_0)\right)^{-1/2}\mathbb{Q}_{12,n}(\beta_0)\V\left(\mathbb{P}_n(\beta_0)\right)^{-1/2}\right|\right|=O_P(1).
\end{equation}
Moreover:
\begin{align*}
&\frac{1}{C_1C_2}\left|\left|\V(\mathbb{P}_n(\beta_0))^{-1/2}\left(\mathbb{Q}_{12,n}(\widehat{\beta})-\mathbb{Q}_{12,n}(\beta_0)\right)\V(\mathbb{P}_n(\beta_0))^{-1/2}\right|\right|	\\
& \leq 2||\widetilde{R}_1|| + ||\widetilde{R}_2|| + (C_1C_2)^{-1}\left|\left|\left(\V(\mathbb{P}_n(\beta_0))^{-1/2}\mathbb{P}_n(\beta_0)\right)^{\otimes 2}-\left(\V(\mathbb{P}_n(\beta_0))^{-1/2}\mathbb{P}_n(\widehat{\beta})\right)^{\otimes 2}\right|\right|,
\end{align*}
where $\widetilde{R}_1 :=\frac{1}{C_1^2C_2^2}\sum_{i=1}^{C_1}\sum_{j=1}^{C_2}\V(\mathbb{P}_n(\beta_0))^{-1/2}\left(\psi_{ij}(\widehat{\beta})-\psi_{ij}(\beta_0)\right)\left(\V(\mathbb{P}_n(\beta_0))^{-1/2}\psi_{ij}(\beta_0)\right)'$ and $\widetilde{R}_2:=\frac{1}{C_1^2C_2^2}\sum_{i=1}^{C_1}\sum_{j=1}^{C_2}\left(\V(\mathbb{P}_n(\beta_0))^{-1/2}\left(\psi_{ij}(\widehat{\beta})-\psi_{ij}(\beta_0)\right)\right)^{\otimes 2}.$ We can bound $\widetilde{R}_1$ and $\widetilde{R}_2$ as follows:
\begin{align*}
	& ||\widetilde{R}_1|| \\
	\le &  (C_1C_2)^{-1/2}\left|\left|\V(\mathbb{P}_n(\beta_0))^{-1/2}\right|\right|\,\left|\left|\widehat{\beta}-\beta_0\right|\right|\sqrt{\frac{1}{C_1C_2}\sum_{i,j}\left|\left|\sup_{b\in \Theta}\Deriv{\psi_{ij}(b)}{\beta}\right|\right|^2}\sqrt{\frac{1}{C_1^2C_2^2}\sum_{i,j}\left|\left|\V(\mathbb{P}_n(\beta_0))^{-1/2}\psi_{ij}(\beta_0)\right|\right|^2} \\
	= & O_P((C_1C_2)^{-1/2})
\end{align*} 
and
\begin{align*}
	||\widetilde{R}_2|| \le (C_1C_2)^{-1}\left|\left|\V(\mathbb{P}_n(\beta_0))^{-1/2}\right|\right|^2\,\left|\left|\widehat{\beta}-\beta_0\right|\right|^2\frac{1}{C_1C_2}\sum_{i,j}\left|\left|\sup_{b\in \Theta}\Deriv{\psi_{ij}(b)}{\beta}\right|\right|^2 = O_P((C_1C_2)^{-1}).
\end{align*}
The term
\begin{equation*}
	(C_1C_2)^{-1}\left|\left|\left(\V(\mathbb{P}_n(\beta_0))^{-1/2}\mathbb{P}_n(\beta_0)\right)^{\otimes 2}-\left(\V(\mathbb{P}_n(\beta_0))^{-1/2}\mathbb{P}_n(\widehat{\beta})\right)^{\otimes 2}\right|\right|
\end{equation*}
can be controlled in a similar fashion as $R_3+R_3'+R_4$, and we deduce that this term is $O_P((C_1C_2)^{-1})$. Gathering all intermediary results, 
\begin{equation}
	\label{eq:control_V_12_2}
	\frac{1}{C_1C_2}\left|\left|\V(\mathbb{P}_n(\beta_0))^{-1/2}\left(\mathbb{Q}_{12,n}(\widehat{\beta})-\mathbb{Q}_{12,n}(\beta_0)\right)\V(\mathbb{P}_n(\beta_0))^{-1/2}\right|\right| = o_P(1).
\end{equation}
Equations~\eqref{eq:control_V_12_1} and~\eqref{eq:control_V_12_2} ensure~\eqref{eq:lin_den} is satisfied by $\widehat{V}_{12}$ and $\widehat{V}_{12}^{\inf}$. This achieves the proof. 

\medskip

\textbf{Proof of asymptotic validity of our inference method:} Let us call $\phi_\alpha$ the test based on the $t$-stat introduced in Section~\ref{sec:GMM}. 
The array $\bm{A}^\infty$ is dissociated and separately exchangeable with a distribution independent from $n$. This is also the case for $\bm{\Psi}$. Besides, Assumption~\ref{as:asym} has been imposed and the start of the proof of~\eqref{eq:lin_num} ensures $E[\Psi_{11}^2]<\infty$ and $V(\Psi_{11})>0$. As a result, the proof of Theorem~\ref{thm:unif_univ} applies to the array $(\Psi_{ij})_{i,j\ge 1}$. Then, 
$$\left(V(\overline{\Psi})^{-1/2}\overline{\Psi},V(\overline{\Psi})^{-1}(\widehat{V}_1^{\inf},\widehat{V}_2^{\inf},\widehat{V}_{12}^{\inf})\right) \convD (L,V_1,V_2,V_{12}),$$
where $(L,V_1,V_2,V_{12})$ are defined in Lemma~\ref{lem:weak_conv_means}. Then, \eqref{eq:lin_num}, \eqref{eq:lin_den} and the continuous mapping theorem imply that
$$\left(V(\overline{\Psi})^{-1/2}(\widehat{\theta}-\theta_0),V(\overline{\Psi})^{-1}(\widehat{V}_1,\widehat{V}_2,\widehat{V}_{12})\right) \convD (L,V_1,V_2,V_{12}).$$
The result follows with the same exact reasoning as below \eqref{eq:conv_dist_theta} in the proof Theorem~\ref{thm:unif_univ}. $\Box$

\section{Additional lemmas and proofs}

\begin{lemma}
	Let $X$ a closed and bounded subset of a separable Hilbert space $H$ equipped with a scalar product $\left\langle .,.\right\rangle$. Let $(e_k)_{k\in \mathbb{N}}$ an orthonormal basis of $H$. $X$ is compact if and only if $\lim_{N\rightarrow\infty}\sup_{x\in X}\sum_{k>N}\left\langle x,e_k \right\rangle^2=0$.\label{lem:compact_Hilbert}
\end{lemma}

\medskip

\noindent
\textit{Proof of Lemma~\ref{lem:compact_Hilbert}.} If $X$ is compact and  $\lim_{N\rightarrow\infty}\sup_{x\in X}\sum_{k>N}\left\langle x,e_k \right\rangle^2>0$, then for $\varepsilon:=\frac{1}{2}\lim_{N\rightarrow\infty}\sup_{x\in X}\sum_{k>N}\left\langle x,e_k \right\rangle^2$, we have a sequence $x_n\in X$ such that $\sum_{k>N}\left\langle x_N,e_k \right\rangle^2>\varepsilon$. We can extract from $x_n$ a subsequence $x_{\sigma(n)}$ that converges to some $y\in X$. It follows that \begin{align*}\varepsilon&<\sum_{k>\sigma(N)}\left\langle x_{\sigma(N)},e_k \right\rangle^2\\
	&\leq  2\sum_{k>\sigma(N)}\left\langle x_{\sigma(N)}-y,e_k \right\rangle^2+2\sum_{k>\sigma(N)}\left\langle y,e_k \right\rangle^2\\
	&\leq 2 ||x_{\sigma(N)}-y||^2+2\sum_{k>\sigma(N)}\left\langle y,e_k \right\rangle^2\\
	&<\varepsilon/2 \text{ for sufficiently large }N, \text{ a contradiction.}
\end{align*}
Reciprocally assume that $\lim_{N\rightarrow\infty}\sup_{x\in X}\sum_{k>N}\left\langle x,e_k \right\rangle^2=0$. Let $(x_n)_{n\geq 1}$ a sequence in $X$. Since for any $k$, $\left\langle x_n,e_k \right\rangle^2\leq \sup_{x\in X}||x||^2<\infty$, there exists a subsequence $x_{\sigma_1(n)}$ such that $\left\langle x_{\sigma_1(n)},e_1 \right\rangle$ converges. Next there exists a sub-sub sequence such that $\left\langle x_{\sigma_2\circ \sigma_1(n)},e_2 \right\rangle$ converges, and so on. Let $y_n=x_{\sigma_n\circ \cdots \circ \sigma_1(n)}$. This is a subsequence of $(x_n)_{n\geq 1}$ and for any $k$, $\left(\left\langle y_n,e_k\right\rangle\right)_{n}$ converges (and next is Cauchy). Fix an arbitrary $\varepsilon>0$ and $N$ such that $\sup_{x\in X}\sum_{k>N}\left\langle x,e_k \right\rangle^2\leq \varepsilon$. There exists $N'$ such that $\sup_{k\leq N}\sup_{m,m'\geq N'}\left|\left\langle y_m,e_k\right\rangle-\left\langle y_{m'},e_k\right\rangle\right|\leq \sqrt{\varepsilon/N}$. Next, for $m,m'\geq N'$:
\begin{align*}
	\norm{y_m-y_{m'}}^2&=\sum_{k\geq1}\left\langle y_m-y_{m'},e_k\right\rangle^2\\
	&\leq \sum_{k\leq N}\left\langle y_m-y_{m'},e_k\right\rangle^2+2 \sum_{k>N}\left\langle y_m,e_k\right\rangle^2+2 \sum_{k>N}\left\langle y_{m'},e_k\right\rangle^2\\
	&\leq 5\varepsilon.
\end{align*}
This means that $(y_n)_n$ is Cauchy and next converges (because an Hilbert is complete). Because $X$ is closed, $(y_n)_n$ converges in $X$. This ensures that $X$ is compact. $\Box$

\begin{lemma}\label{lem:compacity_mu_nu}
	The set $\overline{\mathcal{V}_{m,H}}$ is compact in $\ell^2_d$.
\end{lemma}

\medskip

\noindent
\textit{Proof of Lemma~\ref{lem:compacity_mu_nu}.} We only prove the result for $d>1$ since the proof is analoguous with $d=1$. We first define $\mathcal{V}_{m,H}^n$ formally. For any  $h \in H$, let $X_h:=\E[h(U_{10},U_{01},U_{11})\mid U_{10}]$ and  define
$$\begin{array}{rrcl} h_1: & [0,1]^3 & \to & \R^d \\
 & (u_1,u_2,u_3) & \mapsto & \left(\lambda_{\min}^*(V(X_h)) + \mathds{1}\{\lambda_{\min}^*(V(X_h))=\infty\}\right)^{-1/2}X_h\end{array}.	$$
We define $h_2$ symmetrically. Then, for every $\bm{\mu} \in \ell^2_d$, let $ \Omega_j(\bm{\mu}) := \sum_{\k \in \K_j} \mu_{\k}^{\otimes 2}$ and let us introduce the sets $H_0 := \left\{ h \in H: h_j \in H, j=1,2 \right\}$ and 
	\begin{align*}
		K_H:=&\left\{\bm{\mu} \in \ell_d^2, \bm{\mu}=\left(\int h(u) \psi_{k_1}(u_1)\psi_{k_2}(u_2)\psi_{k_3}(u_3)d\lambda^{\otimes 3}(u)\right)_{\k\in \mathbb{N}^3}\text{ for } h \in H \right\},\\
		K_m := & \Bigg\{ \bm{\mu} \in \ell_d^2, \lambda_{\min}\left(\sum_{j=1}^4\Omega_j(\bm{\mu})\right) \ge m, \\
		& \quad \text{ either } \big[\textrm{range}(\Omega_3(\bm{\mu}))\subseteq \textrm{range}(\Omega_1(\bm{\mu})+\Omega_2(\bm{\mu})) \\
        & \quad \text{ and } \lambda_{\max}(\Omega_3(\bm{\mu})) \le m^{-1} \lambda_{\min}^*(\Omega_1(\bm{\mu})+\Omega_2(\bm{\mu}))\big], \\
		& \quad \text{ or } \big[\Omega_{1}(\bm{\mu})\wedge \Omega_{2}(\bm{\mu})= 0 \text{ and } \lambda_{\min}\left(\Omega_1(\bm{\mu})+\Omega_{2}(\bm{\mu}) +\Omega_4(\bm{\mu})\right)\ge m\big] \Bigg\},\\
        \mathcal{V}_{m,H}^n:=&\left\{\bm{\nu}^n \in \ell^2\left(\mathbb{N}^3,\mathbb{R}^d\right) : \nu_{\k}^n \text{ satisfies~\eqref{eq:def_nu} }\forall\k\in\N^{3}, \bm{\mu} \in K_H \cap K_m \right\},
	\end{align*}
The set $\overline{\mathcal{V}_{m,H}}$ is closed in $\ell^2_d$. Let us prove that $\overline{\mathcal{V}_{m,H}}$ is bounded and satisfies
\begin{equation*}
	\lim_{\underline{k}\rightarrow \infty} \sup_{\bm{\nu} \in \overline{\mathcal{V}_{m,H}}}\sum_{\k:\max(k_1,k_2,k_3)>\underline{k}}\norm{\nu_{\k}}^2=0.
\end{equation*}

\medskip

Let $(\bm{\nu}^n)_n$ be some arbitrary sequence that satisfies $\bm{\nu}^n \in \mathcal{V}_{m,H}^n$ for every $n$. By definition, this sequence is associated with a sequence $(\bm{\mu}^n)_n$ such that $\bm{\mu}^n \in K_m \cap K_H$ for every $n$.
The key step is to derive an upper bound on $||\nu_{\k}^n||$ that solely depends on $\bm{\mu}^n$, $m$ and $H$ for every $\k \in \N^{3\ast}$. In the rest of the proof, we use the shortcut $\Omega_j^n$ for $\Omega_j(\bm{\mu}^n)$ and resort to the following useful observation: any $\bm{\mu}_n \in K_m \cap K_H$ satisfies
\begin{equation}
	\label{eq:invertibility_V_Y}
	\lambda_{\min}\left(\Omega_1^n+\Omega_2^n+\Omega_3^n+\Omega_4^n\right) \ge m.
\end{equation}

\medskip

\textbf{Upper bound on $||\nu_{\k}^n||$:} By definition of $\mathcal{V}_{m,H}^n$, $\nu_{\k}^n$ satisfies
$$\nu_{\k}^n = \left[\Omega_3^n+\Omega_4^n+C_2\Omega_1^n + C_1\Omega_2^n\right]^{-1/2} \gamma_{\k,n} \mu_{\k}^n,$$
where
$$\gamma_{\k,n}=\left|\begin{array}{cl} C_2^{1/2} & \text{ if } \k\in\K_1 \\ C_1^{1/2} & \text{ if } \k\in\K_2 \\ 1 & \text{ otherwise.} \end{array}\right.$$

\medskip

Let us focus on $\k \in \K_1$ first. Note $A_1 \gg A_2$ when $A_1-A_2$ is symmetric semi-definite positive. Then, using~\eqref{eq:invertibility_V_Y},
$$\Omega_3^n+\Omega_4^n+C_2\Omega_1^n + C_1\Omega_2^n \gg C_2\left[\frac{m}{C_2} \Id + \left(1-\frac{1}{C_2}\right) \Omega_1^n\right].$$
Hence, for all $k\in\K_1$.
$$\|\nu_{\k}^n\|^2 \le (\mu_{\k}^n)'\left[\frac{m}{C_2} \Id + \left(1-\frac{1}{C_2}\right) \Omega_1^n\right]^{-1} \mu_{\k}^n.$$
Now, let $B:=$span$(\mu_{\k}^n: \; \k\in\K_1)$ and $x=\sum_{\k\in\K_1} \alpha_{\k} \mu_{\k}^n \in B$. If $B=\{0\}$, $\|\nu_{\k}^n\|^2=0$. Otherwise, if $\Omega_1^nx=0$, then $x'\Omega_1^nx=0$ and thus $(\mu_k^n)'x=0$ for all $k$. Hence,
$(\sum_{\k\in\K_1} \alpha_{\k}\mu_{\k}^n)'x=0$, implying that $x=0$. Let $f$ be the endomorphism associated to $\Omega_1^n$.  Then $f(B)\subseteq B$ and since Ker$(f)\cap B=\{0\}$, $f_{|B}$ is invertible with smallest eigenvalue equal to $\lambda_{\min}^*(\Omega_1^n)$. Then, for all $x\in B$, because $C_2\ge 2$,
$$x' \left[\frac{m}{C_2} \Id + \left(1-\frac{1}{C_2}\right) \Omega_1^n\right] x \ge \frac{\lambda_{\min}^*(\Omega_1^n)}{2} \|x\|^2.$$
As a result, for all $\k\in\K_1$,
\begin{equation}
	\label{eq:up_bound_nu_k1}
	\|\nu_{\k}^n\|^2 \le (\mu_{\k}^n)' \left[\frac{m}{C_2} \Id + \left(1-\frac{1}{C_2}\right) \Omega_1^n\right]^{-1} \mu_{\k}^n \le 2\lambda_{\min}^*(\Omega_1^n)^{-1} \|\mu_{\k}^n\|^2
\end{equation}
since $\mu_{\k}^n\in B$ for all $\k\in\K_1$. Similarly, for all $\k\in\K_2$,
\begin{equation}
	\label{eq:up_bound_nu_k2}
	\|\nu_{\k}^n\|^2 \le 2\lambda_{\min}^*(\Omega_2^n)^{-1} \|\mu_{\k}^n\|^2.
\end{equation}
For $\k\in\K_{34}$, we simply use $\Omega_3^n+\Omega_4^n+C_2\Omega_1^n + C_1\Omega_2^n \gg \Omega_1^n+\Omega_2^n+\Omega_3^n+\Omega_4^n\gg m \Id$. Hence,
\begin{equation}
	\label{eq:up_bound_nu_k3_k4}
	\|\nu_{\k}^n\|^2 \le \lambda_{\min}\left(\Omega_1^n+\Omega_2^n+\Omega_3^n+\Omega_4^n\right)^{-1} \|\mu_{\k}^n\|^2 \le m^{-1}\|\mu_{\k}^n\|^2.
\end{equation}

\medskip

\textbf{Boundedness and uniform tail control over $\overline{\mathcal{V}_{m,H}}$:} Putting all steps together (and recalling that $\lambda_{\min}^*(\Omega_j^n) = \infty$ when $\mu_k^n = \zero$ for every $\k \in \K_j$), we obtain
\begin{align*}
	\sup_{n\ge 1}\sup_{\bm{\nu} \in \mathcal{V}_{m,H}^n}||\bm{\nu}||^2
	\le & m^{-1}\sup_{\bm{\mu} \in K_m \cap K_H}\bigg\{\sum_{\k \in \K_{34}}||\mu_{\k}||^2 \bigg\} \\
	 & + \sup_{\bm{\mu} \in K_m \cap K_H}\bigg\{\sum_{\k\in\K_{12}}\frac{||\mu_{\k}||^2}{\lambda_{\min}^*(\Omega_j^n) + \mathds{1}\{\lambda_{\min}^*(\Omega_j^n) = \infty\}}\bigg\}.
\end{align*}
We recall that $H$ is compact (hence bounded) in $L_2([0,1]^3,\Rb^d)$ and $\tau$, $\tau_1$ and $\tau_2$ belong to $H$. As a result, there exists $M_H$ such that
\begin{equation*}
	\sup_{\bm{\mu} \in K_m \cap K_H}\bigg\{\sum_{\k \in \K_{34}}||\mu_{\k}||^2 \bigg\} \vee \sup_{\bm{\mu} \in K_m \cap K_H}\bigg\{\sum_{\k\in\K_{12}}\frac{||\mu_{\k}||^2}{\lambda_{\min}^*(\Omega_j^n) + \mathds{1}\{\lambda_{\min}^*(\Omega_j^n) = \infty\}}\bigg\} \le M_H.
\end{equation*}
Thus, $\sup_{n\ge 1}\sup_{\bm{\nu} \in \mathcal{V}_{m,H}^n}||\bm{\nu}||^2 \le (m^{-1}+1)M_H$. Using the fact that
\begin{equation}
	\label{eq:sup_equivalence}
	\sup_{n\ge 1}\sup_{\bm{\nu} \in \mathcal{V}_{m,H}^n}||\bm{\nu}||^2 = \sup_{\bm{\nu} \in \cup_{n\ge 1}\mathcal{V}_{m,H}^n}||\bm{\nu}||^2 = \sup_{\bm{\nu} \in \overline{\mathcal{V}_{m,H}}}||\bm{\nu}||^2,
\end{equation}
we can claim that $\overline{\mathcal{V}_{m,H}}$ is bounded in $\ell^2_d$.

\medskip

There remains to prove
\begin{equation*}
	\lim_{\underline{k}\rightarrow \infty} \sup_{\bm{\nu} \in \overline{\mathcal{V}_{m,H}}}\sum_{\k:\max(k_1,k_2,k_3)>\underline{k}}\norm{\nu_{\k}}^2=0.
\end{equation*}
By construction, $\tau$, $\tau_1$ and $\tau_2$ belong to $H$ which implies
\begin{equation}
	\label{eq:unif_tail_condition_tau}
	\lim_{\underline{k} \to \infty} \sup_{\bm{\mu} \in K_m \cap K_H} \sum_{\k : \max(k_1,k_2,k_3) > \underline{k}} ||\mu_{\k}||^2 = 0,
\end{equation}
and
\begin{equation}
	\label{eq:unif_tail_condition_tauj}
	\lim_{\underline{k} \to \infty} \sup_{\bm{\mu} \in K_m \cap K_H} \sum_{\k \in \K_j : \max(k_1,k_2,k_3) > \underline{k}} \frac{||\mu_{\k}||^2}{\lambda_{\min}^*(\Omega_j^n) + \mathds{1}\{\lambda_{\min}^*(\Omega_j^n) = \infty\}} = 0, \, j = 1,2.
\end{equation}
Combining~\eqref{eq:up_bound_nu_k1},~\eqref{eq:up_bound_nu_k2},~\eqref{eq:up_bound_nu_k3_k4},~\eqref{eq:sup_equivalence},~\eqref{eq:unif_tail_condition_tau} and~\eqref{eq:unif_tail_condition_tauj}, we obtain
\begin{equation*}
	\lim_{\underline{k}\rightarrow \infty} \sup_{n\ge 1}\sup_{\bm{\nu} \in \mathcal{V}_{m,H}^n}\sum_{\k:\max(k_1,k_2,k_3)>\underline{k}}\norm{\nu_{\k}}^2 = \lim_{\underline{k}\rightarrow \infty} \sup_{\bm{\nu} \in \overline{\mathcal{V}_{m,H}}}\sum_{\k:\max(k_1,k_2,k_3)>\underline{k}}\norm{\nu_{\k}}^2 = 0.
\end{equation*}

\medskip

\textbf{Conclusion:} We consider $(e_1,...,e_d)$ the canonical basis of $\mathbb{R}^d$ and $(u_j)_{j\ge 1}$ a sequence of elements in $\ell^2_d$ such that $u_{j\k}:=e_{j-\lfloor\frac{j-1}{d}\rfloor d}\mathds{1}\{\sigma\left(\lfloor\frac{j-1}{d}\rfloor\right)=\k\}$ for $\sigma$ a one-to-one mapping from $\N$ to $\N^{3}$. For any $\nu,\tilde{\nu} \in (\ell^2_d)^2$, we have: $\langle\nu,\tilde{\nu}\rangle_{\ell^2_d}=\sum_{\k\in\N^3}\nu_{\k}'\tilde{\nu}_{\k}=\sum_{\k\in\N^3}\sum_{\ell=1}^d\nu_{\k,\ell}\tilde{\nu}_{\k,\ell}$. We have: $\langle u_j,u_{j'}\rangle_{\ell^2_d}=\mathds{1}\{j=j'\}$ and next $(u_j)_{j\ge 1}$ is an orthonormal basis of $\ell^2_d$. We also note: $\norm{\nu_{\k}}^2= \sum_{j:\sigma\left(\lfloor\frac{j-1}{d}\rfloor\right)=\k}\langle \nu,u_{j}\rangle^2_{\ell^2_d}$. For any $\underline{j}$, let $$\underline{k}(\underline{j}):=\min_{j\geq \underline{j}}\max\left(\sigma\left(\lfloor\frac{j-1}{d}\rfloor\right)_1,\sigma\left(\lfloor\frac{j-1}{d}\rfloor\right)_2,\sigma\left(\lfloor\frac{j-1}{d}\rfloor\right)_3\right)-1.$$ We have $\sum_{j>\underline{j}}\langle \nu,u_{j}\rangle^2_{\ell^2_d}\leq \sum_{\k:\max(k_1,k_2,k_3)>\underline{k}(\underline{j})}\norm{\nu_{\k}}^2$ and $\lim_{\underline{j}\rightarrow \infty}\underline{k}(\underline{j})=\infty$. This ensures:
\begin{align*}
	\lim_{\underline{j}\rightarrow \infty} \sup_{\bm{\nu} \in \overline{\mathcal{V}_{m,H}}}\sum_{j>\underline{j}}\langle \nu,u_{j}\rangle^2_{\ell^2_d}&\leq \lim_{\underline{k}\rightarrow \infty} \sup_{\bm{\nu} \in \overline{\mathcal{V}_{m,H}}}\sum_{\k:\max(k_1,k_2,k_3)>\underline{k}}\norm{\nu_{\k}}^2=0,
\end{align*}
and next Lemma \ref{lem:compact_Hilbert} ensures $\overline{\mathcal{V}_{m,H}}$ is compact. $\Box$

\medskip

In the next lemmas, we use the notation $\Sigma_j^n := \sum_{\k \in \K_j}(\nu_{\k}^n)^{\otimes 2}$ and, for any $\bm{\nu}^n \convLtwo \bm{\nu}^\infty$, $\Sigma_j^\infty := \sum_{\k \in \K_j}(\nu_{\k}^\infty)^{\otimes 2}$.

\begin{lemma}\label{lem:limit_behaviour_nu}
	We have
	\begin{equation}
		\label{eq:min_to_0}
		\lim_{n\to\infty}\sup_{\bm{\nu}^n \in \mathcal{V}_{m,H}^n}\min(\|\Sigma_1^n\|,\|\Sigma_2^n\|,\|\Sigma_3^n\|) = 0,
	\end{equation}
	\begin{equation}
		\label{eq:det_lower_bound}
		\liminf_{n\to\infty}\inf_{\bm{\nu}^n \in \mathcal{V}_{m,H}^n} \det(\Sigma_1^n+\Sigma_2^n+\Sigma_4^n) > 0,
	\end{equation}
	and
	\begin{equation}
		\label{eq:sigma_4_to_0}
		\lim_{n\to\infty}\sup_{\bm{\nu}^n \in \mathcal{V}_{m,H}^{n,G}}\|\Sigma_3^n\| = 0.
	\end{equation}
\end{lemma}

\medskip

\noindent
\textit{Proof of Lemma~\ref{lem:limit_behaviour_nu}.}
In the present lemma, the parameter $\bm{\mu}^n$ associated with $\bm{\nu}^n$ may depend on $n$. In that case, quantities such as $\Omega_j$, $j = 1,\dots,4$, also depend on $n$. We use the notation $\Omega_j^n$ to stress that dependence in the rest of the proof.

\medskip

\textbf{Proof of~\eqref{eq:min_to_0}:} 
First suppose $d=1$. If $\min(\Omega_1^n,\Omega_2^n,\Omega_3^n)=0$, we also have $\min(\|\Sigma_1^n\|,\|\Sigma_2^n\|,$ $\|\Sigma_3^n\|)=0$. Otherwise, using
$\Omega_1^n + \Omega_2^n + \Omega_4^n + \Omega_3^n = V(Y_{11}) \ge m$, $(\Omega_1^n+\Omega_2^n)/\Omega_3^n \ge m$ we get
\begin{align*}
	\Omega_3^n+\Omega_4^n+C_2\Omega_1^n + C_1\Omega_2^n \geq V(Y_{11}) + (\Cinf - 1)(\Omega_1^n+\Omega_2^n) \ge m + m(\Cinf - 1)\Omega_3^n.
\end{align*}
Then, using $\underline{C}\ge 2$ and the definition of $(\nu_{\k}^n)_{\k\in\K_3}$,
\begin{align*}
	\Sigma_3^n \leq \frac{\Omega_3^n}{\Cinf\left(\Cinf^{-1} m + (1-\Cinf^{-1})m\Omega_3^n\right)} \le \frac{1}{\Cinf(1-\Cinf^{-1})m} \le 2(m\Cinf)^{-1}.
\end{align*}
Hence, we  either have $\min(\|\Sigma_1^n\|,\|\Sigma_2^n\|,\|\Sigma_3^n\|)=0$ or $\min(\|\Sigma_1^n\|,\|\Sigma_2^n\|,\|\Sigma_3^n\|) \le 2(m\Cinf)^{-1}$. Equation \eqref{eq:min_to_0} follows. 

\medskip
Now suppose $d>1$. If $\|\Omega_1^n\|\wedge \|\Omega_2^n\|=0$, then $\min(\|\Sigma_1^n\|,\|\Sigma_2^n\|,\|\Sigma_3^n\|)=0$. Otherwise, we have $\textrm{range}(\Omega_3^n)\subseteq \textrm{range}(\Omega_1^n+\Omega_2^n)$ and $\lambda_{\max}(\Omega_3^n) \le m^{-1}\lambda_{\min}^*(\Omega_1^n+\Omega_2^n)$. Since $\lambda_{\min}\left( \sum_{j=1}^4\Omega_j^n \right) \ge m$, we obtain
$$\Omega_3^n+\Omega_4^n+C_2\Omega_1^n + C_1\Omega_2^n \gg m \Id + (\underline{C}-1) (\Omega_1^n +\Omega_2^n).$$
This ensures
\begin{equation}
	\|\Sigma_3^n\| \le \sum_{\k \in \K_3} ||\nu_{\k}^n||^2 \le \frac{1}{\underline{C}} \sum_{\k\in\K_3} (\mu_{\k}^n)' \left[\frac{m}{\underline{C}} \Id + \left(1-\frac{1}{\underline{C}}\right) (\Omega_1^n +\Omega_2^n)\right]^{-1} \mu_{\k}^n.	
	\label{eq:ineg_Sigma4}
\end{equation}
The same reasoning as in the proof of Lemma~\ref{lem:compacity_mu_nu} shows that the endomorphism associated to $\Omega_1^n+\Omega_2^n$ is invertible on its range, which is equal to $B:=$span$(\mu_{\k}^n: \; \k\in\K_{12})$. Also, $\textrm{range}(\Omega_3^n)=$span$(\mu_{\k}^n: \; \k\in\K_3)$, which implies that $\mu_{\k}^n\in B$ for all $\k\in\K_3$. Then, reasoning as above, we obtain
\begin{align*}
	\sum_{\k\in\K_3} (\mu_{\k}^n)' \left[\frac{m}{\underline{C}} \Id + \left(1-\frac{1}{\underline{C}}\right) (\Omega_1^n +\Omega_2^n)\right]^{-1} \mu_{\k}^n & \le \frac{1}{\left(1-\Cinf^{-1}\right)\lambda_{\min}^*\left(\Omega_1^n+\Omega_2^n\right)} \sum_{\k\in\K_3}||\mu_k^n||^2 \\
	& \le \frac{2\tr\left(\Omega_3^n\right)}{\lambda_{\min}^*\left(\Omega_1^n+\Omega_2^n\right)} \le \frac{2\lambda_{\max}\left(\Omega_3^n\right)}{\lambda_{\min}^*\left(\Omega_1^n+\Omega_2^n\right)} \le 2m^{-1}.
\end{align*}
This implies, in view of \eqref{eq:ineg_Sigma4}, that $\|\Sigma_3^n\| \le 2(m\Cinf)^{-1}$. Thus, in the end, we either have $\min(\|\Sigma_1^n\|,\|\Sigma_2^n\|,\|\Sigma_3^n\|)=0$ or $\min(\|\Sigma_1^n\|,\|\Sigma_2^n\|,\|\Sigma_3^n\|) \le 2(m\Cinf)^{-1}$, which implies \eqref{eq:min_to_0}.

\medskip
\textbf{Proof of~\eqref{eq:det_lower_bound}:} When $\textrm{range}(\Omega_3^n)\subseteq \textrm{range}(\Omega_1^n+\Omega_2^n)$ and $\lambda_{\max}(\Omega_3^n) \leq m^{-1}\lambda_{\min}^*(\Omega_1^n+\Omega_2^n)$, we have just shown that $\|\Sigma_3^n\|\le 2(m\Cinf)^{-1}$. Since $\Sigma_1^n+\Sigma_2^n+\Sigma_3^n+\Sigma_4^n=\Id$, we obtain that for all $n$ large enough, $\textrm{range}(\Omega_3^n)\subseteq \textrm{range}(\Omega_1^n+\Omega_2^n)$ and $\lambda_{\max}(\Omega_3^n) \leq m^{-1}\lambda_{\min}^*(\Omega_1^n+\Omega_2^n)$ imply
$\det(\Sigma_1^n+\Sigma_2^n+\Sigma_4^n) > 1/2$. Now, assume instead that $\|\Omega_1^n\|\wedge \|\Omega_2^n\|=0$ and $\lambda_{\min}\left(\Omega_1^n+\Omega_2^n+\Omega_4^n\right)\ge m$. Then, $\Omega_1^n+\Omega_2^n+\Omega_4^n\gg m \Id$ and thus,
$$\Omega_3^n+\Omega_4^n+C_2 \Omega_1^n + C_1 \Omega_2^n \gg \Omega_3^n + m \Id.$$
Hence, for any $x\in\R^d$ such that $x'x=1$,
\begin{align*}
	x'\Sigma_3^n x \le & x'\left(\Omega_3^n + m \Id\right)^{-1/2} \Omega_3^n \left(\Omega_3^n + m \Id\right)^{-1/2}x \\
	= & x'\left(\Omega_3^n + m \Id\right)^{-1/2} (\Omega_3^n + m \Id - m \Id) \left(\Omega_3^n + m \Id\right)^{-1/2}x \\
	= & \|x\|^2 -  x'\left(\Omega_3^n + m \Id\right)^{-1/2} (m \Id) \left(\Omega_3^n + m \Id\right)^{-1/2}x \\
	= & x'(\Sigma_1^n+\Sigma_2^n+\Sigma_3^n+\Sigma_4^n) x - m x'\left(\Omega_3^n + m \Id\right)^{-1} x.
\end{align*}
Hence,
\begin{equation}
	x'(\Sigma_1^n+\Sigma_2^n+\Sigma_4^n) x \ge m x'\left(\Omega_3^n + m \Id\right)^{-1} x.	
	\label{eq:for_condit_det}
\end{equation}
Moreover,
$$\Omega_3^n + m \Id \ll \Omega_3^n + \Omega_1^n+\Omega_2^n+\Omega_4^n = V(Y_{11}).$$
As a result,
$$x'\left(\Omega_3^n + m \Id\right)^{-1} x \ge \frac{1}{\lambda_{\max}(V(Y_{11}))}.$$
Since $H$ is compact, there exists $M_H$ such that $\lambda_{\max}\left(V(Y_{11})\right) \le M_H$. When combined with \eqref{eq:for_condit_det}, this implies
$$\lambda_{\min}(\Sigma_1^n+\Sigma_2^n+\Sigma_4^n) \ge \frac{m}{\lambda_{\max}(V(Y_{11}))} \ge \frac{m}{M_H} > 0.$$
Hence, in the end, for all $n$ large enough, we have $\lambda_{\min}(\Sigma_1^n+\Sigma_2^n+\Sigma_4^n) \ge (m/M_H) \wedge 1/2$. The result follows.

\medskip

\textbf{Proof of~\eqref{eq:sigma_4_to_0}:} the result follows from the proof of~\eqref{eq:min_to_0} and the definition of $\mathcal{V}_{m,H}^{n,G}$. $\Box$
\begin{lemma}
	\label{lem:weak_conv_finite_dim_Z}
	For every $\overline{k}<+\infty$, $Z^n := vec\left((Z_{\k}^n)_{\k\in\K_{124}(\overline{k})}\right) \in \Rb^{\overline{k}\left(2+(\overline{k}+1)^2\right)}$ satisfies
	\begin{equation*}
		Z^n \convD \mathcal{N}\left(\mathbf{0},\Id\right).
	\end{equation*}
\end{lemma}

\noindent
\textit{Proof of Lemma~\ref{lem:weak_conv_finite_dim_Z}.} We first remark that by construction, $Z_1^n := vec\left((Z_{\k}^n)_{\k\in\K_1(\overline{k})}\right)$ and $Z_2^n := vec\left((Z_{\k}^n)_{\k\in\K_2(\overline{k})}\right)$ are two $\overline{k}$-dimensional vectors of sample means that depend on $(U_{i0}^n)_{1\leq i \leq C_1}$ and $(U_{0j}^n)_{1\leq j \leq C_2}$ respectively. To prove asymptotic normality of $Z_1^n$ and $Z_2^n$, we need to verify the Lindeberg-Feller condition due to the triangular array structure at play. The $\psi_k$s form an orthonormal basis of $L_2([0,1])$ and are all uniformly bounded, while $(U_{i0}^n)_{1\leq i \leq C_1}$ and $(U_{0j}^n)_{1\leq j \leq C_2}$ are sequences of \textit{i.i.d.} standard uniform random variables. As a result, the conditions of the Lindeberg-Feller CLT can be easily checked and we conclude that $Z_1^n \convD \mathcal{N}\left(\mathbf{0},\Id\right)$ and $Z_2^n \convD \mathcal{N}\left(\mathbf{0},\Id\right)$. Since $(U_{i0}^n)_{1\leq i \leq C_1}$ and $(U_{0j}^n)_{1\leq j \leq C_2}$ are independent sequences, we can also claim that $Z_{12}^n := ((Z_1^n)',(Z_2^n)')' \in \Rb^{2\overline{k}}$ satisfies
\begin{equation*}
	Z_{12}^n \convD \mathcal{N}\left(\mathbf{0},\Id\right).
\end{equation*}

We now wish to show that for every $t \in \Rb^{\overline{k}\left(2+(\overline{k}+1)^2\right)}$
\begin{equation}
	\label{eq:weak_conv_final_result}
	t'Z^n \convD \mathcal{N}(0,||t||^2),
\end{equation}
which is equivalent to the statement of the theorem by the Cramer-Wold device. We decompose $t$ in two parts, $t_{12} \in \Rb^{2\overline{k}}$ and $t_4 \in \Rb^{\overline{k}(\overline{k}+1)^2}$ and write
\begin{equation*}
	t'Z^n = t_{12}'Z_{12}^n + t_4'Z_4^n,
\end{equation*}
with $Z_4^n := vec\left((Z_{\k}^n)_{\k\in\K_4(\overline{k})}\right) \in \Rb^{\overline{k}(\overline{k}+1)^2}$. We first remark that by a simple application of the continuous mapping in distribution
\begin{equation}
	\label{eq:weak_conv_first_part}
	t_{12}'Z_{12}^n \convD \mathcal{N}(0,||t_{12}||^2).
\end{equation}
If $t_4$ is the null vector, \eqref{eq:weak_conv_first_part} is enough to get~\eqref{eq:weak_conv_final_result}. For the remaining of the proof, we thus focus on the situation when $t_4$ is different from the null vector, and we go through the steps of the proof of Theorem 2 in \cite{ChenRao2007}. We let $K_{12,n} := t_{12}'Z_{12}^n$, $K_{4,n} := t_4'Z_4^n$ and $\mathcal{B}_n$ be the sigma-algebra generated by $\left((U_{i0}^n)_{1\le i \le C_1}, (U_{0j}^n)_{1 \le j \le C_2}\right)$. By construction, $K_{4,n}$ is an i.n.i.d sum of bounded random variables conditional on $\mathcal{B}_n$. We also have
\begin{align*}
	V\left( K_{4,n} \mid \mathcal{B}_n \right) & = \frac{1}{C_1C_2}\sum_{i=1}^{C_1}\sum_{j=1}^{C_2}\sum_{1 \le k_1,k_2,k_1',k_2',k_3 \le \overline{k} : k_3>0}t_{4,k_1k_2k_3}t_{4,k_1'k_2'k_3}\psi_{k_1}(U_{i0}^n)\psi_{k_1'}(U_{i0}^n)\psi_{k_2}(U_{0j}^n)\psi_{k_2'}(U_{0j}^n) \\
	& \convAS ||t_4||^2.
\end{align*}
A conditional version of the Lindeberg-Feller CLT immediately yields
\begin{equation}
	\label{eq:as_conv_to_0_cond_clt}
	\sup_{x \in \Rb} \left| P\left( K_{4,n} \leq x \mid \mathcal{B}_n \right) - \Phi(x/||t_4||) \right| \convAS 0.
\end{equation}
Obtaining~\eqref{eq:weak_conv_final_result} is equivalent to proving for every $x \in \Rb$
\begin{equation}
	\label{eq:rewriting_weak_conv}
	\lim_{n \to \infty} P\left( K_{4,n} + K_{12,n} \leq x \right) = \Phi(x/||t||).
\end{equation}
We can write
\begin{align*}
	P\left( K_{4,n} + K_{12,n} \leq x \right) & = E\left[\Phi\left(\frac{x-K_{12,n}}{||t_4||}\right)\right] + E\left[ P\left( K_{4,n} \leq x - K_{12,n} \mid \mathcal{B}_n \right) - \Phi\left(\frac{x-K_{12,n}}{||t_4||}\right) \right] \\
	& \le E\left[\Phi\left(\frac{x-K_{12,n}}{||t_4||}\right)\right] + E\left[ \sup_{y \in \Rb} \left| P\left( K_{4,n} \leq y \mid \mathcal{B}_n \right) - \Phi\left(\frac{y}{||t_4||}\right) \right| \right].
\end{align*}
The random variable $\sup_{y \in \Rb} \left| P\left( K_{4,n} \leq y \mid \mathcal{B}_n \right) - \Phi\left(\frac{y}{||t_4||}\right) \right|$ is almost surely bounded and converges to 0 almost surely as well by~\eqref{eq:as_conv_to_0_cond_clt}. We can therefore apply the dominated convergence theorem and claim
\begin{equation}
	\label{eq:dct}
	\lim_{n \to \infty} E\left[ \sup_{y \in \Rb} \left| P\left( K_{4,n} \leq y \mid \mathcal{B}_n \right) - \Phi\left(\frac{y}{||t_4||}\right) \right| \right] = 0.
\end{equation}
Let $Z_{12}$ and $Z_4$ be two independent random variables that satisfy $Z_{12} \sim \mathcal{N}(0,||t_{12}||^2)$ and $Z_4 \sim \mathcal{N}(0,||t_4||^2)$. Using the fact that $v \mapsto \Phi\left((x-v)/||t_4||\right)$ is a bounded and continuous function, the weak convergence result~\eqref{eq:weak_conv_first_part} implies
\begin{align}
	\label{eq:weak_conv_equivalence_criterion}
	\lim_{n \to \infty} E\left[\Phi\left(\frac{x-K_{12,n}}{||t_4||}\right)\right] & = E\left[\Phi\left(\frac{x-Z_{12}}{||t_4||}\right)\right] \nonumber \\
	& = E\left[P\left( Z_4 \leq x - Z_{12} \mid Z_{12}\right)\right] \nonumber \\
	& = P\left(Z_{12}+Z_4 \le x \right) = \Phi(x/||t||).
\end{align}
Combining~\eqref{eq:dct} and~\eqref{eq:weak_conv_equivalence_criterion} yields~\eqref{eq:rewriting_weak_conv} and the final result. $\Box$


\end{document}